\documentclass[%
preprintnumbers,
amsmath,amssymb,
aps,
prfluids,
longbibliography,
notitlepage,
floatfix
]{revtex4-1}

\usepackage[utf8]{inputenc}

\usepackage{graphicx}
\usepackage[subrefformat=parens,labelformat=parens,caption=false]{subfig} 
\usepackage{bm} 
\usepackage{color}
\usepackage{dcolumn} 
\newcolumntype{d}[1]{D{.}{.}{#1}}
\usepackage{adjustbox}
\usepackage{placeins} 

\usepackage{hyperref}
\hypersetup{
    colorlinks=true,
    linkcolor=blue,
    filecolor=magenta,      
    urlcolor=cyan,
    breaklinks=true
}

\DeclareMathOperator{\real}{Re}
\DeclareMathOperator{\imag}{Im}
\DeclareMathOperator{\diag}{diag}

\begin{document}


\title{Unsteady fluid--structure interactions
in a soft-walled microchannel:\\ A one-dimensional lubrication model for finite Reynolds number}

\author{Tanmay C.\ Inamdar}%
\author{Xiaojia Wang}%
\author{Ivan C.\ Christov}\thanks{Author to whom correspondence should be addressed.}%
\email{christov@purdue.edu}%
\homepage{http://tmnt-lab.org}%
\affiliation{School of Mechanical Engineering, Purdue University, West Lafayette, Indiana 47907, USA}%

\date{\today}

\begin{abstract}
We develop a one-dimensional model for the unsteady fluid--structure interaction (FSI) between a soft-walled microchannel and viscous fluid flow within it. A beam equation, which accounts for both transverse bending rigidity and nonlinear axial tension, is coupled to a one-dimensional fluid model obtained from depth-averaging  the two-dimensional incompressible Navier--Stokes equations across the channel height. Specifically, the Navier--Stokes equations are scaled in the viscous lubrication limit relevant to microfluidics. The resulting set of coupled nonlinear partial differential equations is solved numerically through a segregated approach employing fully-implicit time stepping and second-order finite-difference discretizations. Internal FSI iterations and under-relaxation are employed to handle the stiff nonlinear algebraic problems within each time step. Then, we explore both the static and dynamic FSI behavior of this example microchannel system by varying a reduced Reynolds number $Re$, which necessarily changes the Strouhal number $St$, while we keep the geometry and a modified dimensionless Young's modulus $\Sigma$ fixed. At steady state, an order-of-magnitude analysis (balancing argument) shows that the axially-averaged pressure in the flow, $\langle P\rangle$, exhibits two different scaling regimes, while the maximum deformation of the top wall of the channel, $H_{\mathrm{max}}$, can fall into four different regimes, depending on the magnitudes of $Re$ and $\Sigma$. These regimes are physically explained as resulting from the competition between the inertial and viscous forces in the fluid flow as well as the bending resistance and tension in the elastic wall. Finally, the linear stability of the steady inflated microchannel shape is assessed via a modal analysis, showing the existence of many {highly oscillatory but} stable modes, which further highlights the computational challenge of simulating unsteady FSIs.
\end{abstract}

\maketitle

\section{Introduction}
\label{sec:intro}

Microfluidics is the part of fluid mechanics that deals with flows at small scales, on the order of microns, wherein the small dimensions, and the resulting confinement of the system, start to affect the flow physics~\cite{NW06}. Microfluidic flows are characterized by small geometric scales (say, $h_{0f}$ transverse to the flow), which under normal flow conditions (velocity scale $U$, kinematic viscosity $\nu$ and density $\varrho_f$) correspond to a small Reynolds number, $Re \equiv U h_{0f}/\nu \ll 1$. In principle, aside from giving rise to attractive academic research problems, microdevices are important as they can be used in micro total analysis systems ($\mu$TAS) designed to perform the tests and assays (currently performed with macroscale laboratory tools) using smaller amount of working fluids at lower costs, more efficiently, and, eventually, with higher accuracy \cite{RDIAM02,AIRM02,BKLF04}. Therefore, microfluidics is transforming many applications, including biomedical devices, chemical processing, and thermal cooling, to name a few. For example, in the field of medical technology, microfludics has enabled the development of a whole new field of science and technology known as \emph{lab-on-a-chip}~\cite{C13_book}. The latter has, in the last decade, given rise to \emph{organ-on-a-chip} technologies \cite{Huhetal10} with the advent of biocompatible materials for microfluidic applications. Consequently, there is now  significant interest in understanding unsteady fluid--structure interactions (FSIs) between low $Re$ flows and compliant (elastic) boundaries in these contexts \cite{KCC18}. Similarly, the dynamics of lubricated elastic sheets \cite{HM04,HBDB14} have broad relevance to micro-electro-mechanical systems
(MEMS) design \cite{HT98} and the study of complex mechanical instabilities \cite{JPPH18}. {Recently, by harnessing viscous FSIs, passive fuses have been built \cite{GMV17}, the efficiency of micropumps has been analyzed \cite{BBUGS18}, and soft robotic actuators have been designed \cite{MEG17,BGB18,SGOG20}.}

Several methods of manufacturing microfluidic devices, such as soft lithography \cite{XW98} and additive or subtractive manufacturing (e.g., 3D printing) \cite{TWL03,KRSDC12} have emerged over the last few decades \cite{NW06}. The availability of new polymer-based materials and processes, such as ink jet printing~\cite{SCFT16}, has made it possible to manufacture complex geometries with relative ease, at low cost, and with high through-put~\cite{B08}. For example, polydimethylsiloxane (PDMS) is a silicon-based polymeric material that is often used in manufacturing of microdevices \cite{MW02}. PDMS can be cured in layers, which allows for manufacturing of complicated geometries via casting \cite{FY10}. The rheological properties of PDMS can be controlled by mixing different concentrations of the constituent polymeric substances \cite{JMTT14}, which allows for control of the compliance of soft microchannels \cite{VK13}.

Microfluidic devices handle very small volumes of fluids, in the range of nano to microliters \cite{SQ05}. Various  inventive techniques such as imposing an electric field, acoustic streaming, capillary forces, and fluid--structure interactions are required to transport fluids at the microscale~\cite{SSA04}. However, when a fluid is pumped through a microfluidic channel, the viscous stresses at such small geometric scales result in significant pressure drops, even for very low flow rates \cite{panton}. Consequently, when this pressure drop acts against a compliant wall, it causes appreciable deformation of the microchannel \cite{HKBC03,GEGJ06,RDC17,CCSS17}. This deformation in turn affects the flow profile and the pressure drop, which again changes the deformation of the wall, and this cycle continues thus forming the feed-back loop of FSI \footnote{See, e.g., Fung's illustration \cite[Figure~3.4:2]{F97} for a schematic visual representation of this FSI feed-back loop in hemoelastic system.}. At vanishing $Re$, microchannels deform to a well-defined steady state. At finite $Re$, however, the coupled problem can destabilize, leading to high-frequency vibrations of the soft wall and subsequent laminar-to-turbulent transition of the flow \cite{VK13,SK17,VK15}. Likewise, in physiological contexts, a weakened portion of a blood vessel (e.g., an artery, see also \cite{F97}) is known to exhibit lateral wall vibrations \cite{P80}, which can lead to deadly aortic dissections \cite{GJ04}. 

The widespread use of soft polymeric materials in microfluidics makes the task of understanding the transient response of a compliant channel wall due to the fluid flow underneath (and its subsequent effect on the flow itself) accessible experimentally. In fact, this inherent softness has been recently exploited to create mechanically active heart-on-a-chip devices being capable of mimicking physiological FSI \cite{Lindetal17}. At the same time, if most lab-on-a-chip devices are fabricated from PDMS, it is important to understand the fundamental physics of  unsteady FSIs in order to be able to effectively design microfluidic systems, though few studies have done so. For example, Dendukuri et al.~\cite{DGPHD07} studied the unsteady FSI problem that might arise during \emph{stop-flow lithography}. Specifically, they were interested in the dependence of the characteristic response time $\mathcal{T}_r$, due to sudden initiation or stoppage of flow, of a soft microchannel wall on the various system parameters, following the scaling arguments from \cite{GEGJ06}. This approach is essentially two-dimensional (2D) and does not consider the deformation in the cross-section. Others have incorporated electro-osmotic flow \cite{MCC13} and finite-size ionic effects \cite{NCC17} into this problem. More recently, Mart\'{i}nez-Calvo et al.~\cite{MCSPS19} re-analyzed the start-up problem for flow in a soft microchannel by considering the unsteady version of the three-dimensional (3D) coupled problem posed in \cite{CCSS17}. Specifically, they obtained a reduced model  showing that the approach to the steady state is controlled by the ratio of the viscous flow time scale to the elastic inertial time scale. It was further shown that the scaling of $\mathcal{T}_r$ with the various system parameters is significantly different under 3D thin (plate-like) elastic models as in \cite{CCSS17,MCSPS19} compared to 2D thick elastic structures as in \cite{GEGJ06,DGPHD07,MCC13,NCC17}.

In parallel, the effect of instabilities in such coupled flow--compliant wall problems has been studied extensively in the high-Reynolds-number regime \cite{RGHM88,G96,Pai03}.  { Reduced-order formulations (2D, or even 1D) of FSI in such systems have made stability analysis tractable from the point of view of theory (asymptotics). A membrane-like top wall (again, assuming the solid and fluid layer aspect ratios are small) has been one popular solid model. For example, it was found that flutter instabilities can arise and be superimposed on the tension-induced low-frequency instabilities if the wall inertia is not negligible \cite{LP98}. A rich phenomenology of FSI instabilities has been constructed by also including  bending, pre-tension and stretching, in solid mechanics problem  \cite{CL03}. It has been shown that the boundary conditions (either pressure drop or fixed upstream flux) also have a profound effect on the onset of instabilities \cite{XBJ13}. However, these questions, coupling mechanics and stability analyses have not been carried out} in the low-Reynolds-number regime, leaving the latter class of FSIs relatively underdeveloped \cite{DS16}. {Thus, we are motivated to formulate reduced-order (specifically, 1D) models of FSI in low-Reynolds-number flows. Importantly, this line of research might pave the way for new microfluidics applications. For example, previous work has shown that} wall-mode instabilities in the coupled problem can lead to efficient mixing in microchannels \cite{VK13,SK17,VK15}. This effect can have significant implications for microfluidic technologies, as previously only diffusion and certain features of periodic laminar flows (``chaotic mixing'' \cite{O89}) were thought to achieve mixing in viscous flows at the microscale \cite{OW04,SSA04}. { Specifically, it is worth mentioning that} this instability-driven mixing takes place at Reynolds numbers that are \emph{orders of magnitude smaller} than the Reynolds number for  transitional/turbulent flow in an equivalent rigid geometry \cite{VK13,VK15}. This striking effect generates further interest in unsteady low-Reynolds-number FSIs in soft-walled microchannels, which is the subject of the present work. 

Specifically, this paper is organized as follows. First, we develop a one-dimensional (1D) model of fluid--structure interactions (FSIs) at finite Reynolds number in flexible, soft-walled microchannels (Sect.~\ref{sec:theory}), which can then be made computationally tractable (Appendix~\ref{app:compute}) \footnote{{A less general unsteady 1D model was derived in \cite{BBUGS18}, in parallel to the present work. The latter model does not consider fluid inertia, stretching of the elastic wall, the inflationary nonlinear dynamics, or the stability of the deformed channel shape.}}. {Next, we analyze the static (Sect.~\ref{sec:steady}) and the dynamic (Sect.~\ref{sec:inflate}) response of this model system. Of interest is the fact that microchannel walls (unlike collapsible tubes) have significant bending rigidity, thus steady-state channel deformation profiles are not flat. Importantly, despite  significant dynamic complexity in the transient response of the microchannel, ultimately stable steady states are achieved. To justify this observation from, a linear stability analysis of the nonlinear deformed shape is presented in Sect.~\ref{sec:linear}. Conclusions and avenues for future work are discussed in Sect.~\ref{sec:concl}.}


\section{Derivation of the 1D lubrication model}
\label{sec:theory}

Consider a topologically rectangular fluid channel whose top wall is made from a soft, compliant solid. The length of the channel (in the flow-wise direction) is $\ell$, while $h_{0f}$ and $h_{0s}$ denote the undeformed heights (in the direction perpendicular to the flow) of the fluid channel and solid wall, respectively. The  positive $x$-direction is taken as the flow-wise direction, i.e., the fluid flows from left to right in Fig.~\ref{fig:geo}. Meanwhile, the solid wall can deform in the perpendicular $y$-direction. The solid displacement is assumed to vary only with $x$, while the fluid flow is two-dimensional (2D) having both $x$ and $y$ velocity components each of which might depend on both $x$ and $y$. In the lubrication approach described below, the fluid model will be averaged in $y$ to yield a one-dimensional (1D) model. Thus, at the end of the derivation, $x$ and $t$ will be the independent variables.

\subsection{Solid mechanics: Governing equations}
\label{sec:solid}

\begin{figure}
\includegraphics[width=0.75\textwidth]{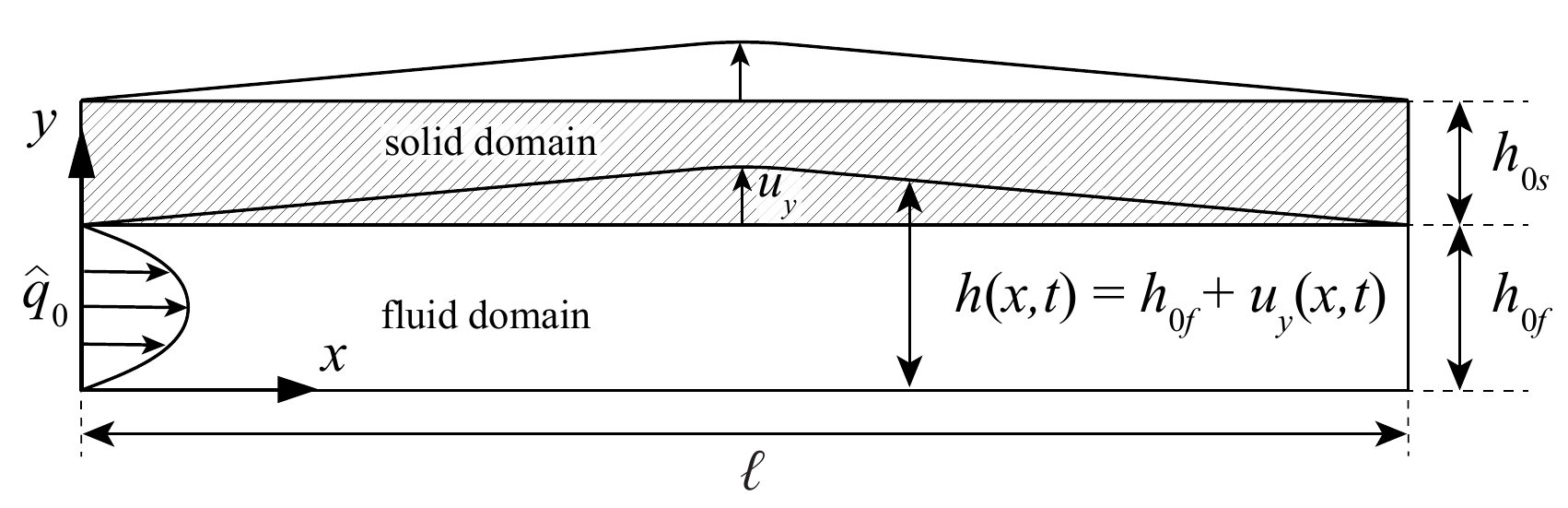}
\caption{Schematic of the geometry of the exemplar  two-dimensional soft-walled microchannel.}
\label{fig:geo}
\end{figure}

To take into account the mass (inertia), bending and stretching of the channel's top wall, we use a nonlinear tension model derived on the basis of von K\'arm\'an strains \cite{reddy07}. The equilibrium equations for a beam, along with the constitutive equations, which relate stress resultants to strains, are simplified by making the assumptions of no axial load, negligible axial displacement, uniform cross section and constant solid properties. {Considering a 2D problem, i.e., a beam with unit width (out of the page in Fig.~\ref{fig:geo}),} the following governing equation for the solid's vertical displacement $u_y=u_y(x,t)$ takes can be derived: 
\begin{equation}\label{solidmodel1}
\hat{\varrho}_s \frac{\partial^2 u_y}{\partial t^2} + \frac{\partial^2}{\partial x^2} \left(E\hat{I} \frac{\partial^2 u_y}{\partial x^2} \right) - \frac{3}{2}E\hat{A} \left(\frac{\partial u_y}{\partial x}\right)^2 \frac{\partial^2 u_y}{\partial x^2} = \hat{\mathcal{L}}[u_y(x,t),x,t],
\end{equation}
{Here, $\hat{A}=h_{0s}$ denotes the cross sectional area of the beam (per unit length), $\hat{\varrho}_s$ is mass per unit area of the solid, $E$ is the Young's modulus, and $\hat{I}=h_{0s}^3/12$ is the second moment of area per unit width}. {Henceforth, hats over quantities denote they are the 2D versions (e.g., per unit width) of the otherwise 3D quantities.} The product $E\hat{I}$ is termed the \emph{bending rigidity} of the beam. The load (per unit width) acting on the beam in positive $y$-direction (i.e., from the fluid side) is denoted by the functional $\hat{\mathcal{L}}$, which depends on the hydrodynamics under the wall through its flow-induced deformation.

Previous studies of FSI in 3D microchannels \cite{GMV17,CCSS17} have shown that deformation is bending rather than stretching dominated. In the present 1D context, however, {we expect large deformations, thus, stretching of the beam due to the rotations of the transverse normals is not negligible. To this end, Eq.~\eqref{solidmodel1} relaxes the infinitesimal strain constraint on the ``classical'' Euler--Bernoulli beam that has been used in previous works \cite{KGV17,HBDB14}, and thus, leads to non-uniform axial tension in our model, which allows it to handle large displacements.} Finally, in this microfluidics context, the weight of the solid is assumed insignificant and gravitational forces are neglected. 

The load on the solid is the result of the forces exerted by the fluid and constitutes one part of the FSI coupling. In a more general (e.g., 3D model), a full traction boundary condition is required at the fluid--solid interface. Since we are developing a 1D model, we assume that the hydrodynamic pressure $p$ is the only force contributing to the load on the beam. (The reason for neglecting shear stresses at the fluid--solid interface is made clear in Sect.~\ref{sec:fluid} after making the problem dimensionless.) {Then, the load per unit width is $\hat{\mathcal{L}} = p$.}

Next, Eq.~\eqref{solidmodel1} can be made dimensionless by choosing the following dimensionless variables: 
\begin{equation}
X=x/\ell,\qquad T=t\Big/\sqrt{\hat{\varrho}_s \ell^4/E\hat{I}},\qquad P=p/p_0,\qquad U_Y=u_y/u'_y,
\label{nd_vars_solid}
\end{equation}
where $p_0$ and $u'_y$ are ``dummy'' scales for the pressure and displacement that will be determined self-consistently as a part of the analysis. Substituting the dimensionless variables from Eq.~\eqref{nd_vars_solid} into Eq.~\eqref{solidmodel1} and using {$\hat{\mathcal{L}} =  p$} results in
\begin{equation}\label{solidmodel2}
\frac{\partial^2 U_Y}{\partial T^2} + \frac{\partial^4 U_Y}{\partial X^4} - \frac{3}{2}\frac{h_{0s}(u'_y)^2}{\hat{I}}\left(\frac{\partial U_Y}{\partial X}\right)^2\frac{\partial^2 U_Y}{\partial X^2} = \frac{p_0\ell^4}{E\hat{I}u'_y}P.
\end{equation}
In order to couple the fluid and solid mechanics, the right-hand side of Eq.~\eqref{solidmodel2} must be $\mathcal{O}(1)$, which allows us to determine the characteristic vertical displacement scale self-consistently as
\begin{equation}\label{uydef}
u'_y = \frac{p_0\ell^4}{E\hat{I}} = {\frac{12p_0\ell^4}{E{h_{0s}^3}}}.
\end{equation}
Substituting Eq.~\eqref{uydef} into Eq.~\eqref{solidmodel2}, we arrive at the dimensionless governing equation for the solid mechanics problem:
\begin{equation}\label{solidmodel3}
\frac{\partial^2 U_Y}{\partial T^2} + \frac{\partial^4 U_Y}{\partial X^4} - \alpha \left(\frac{\partial U_Y}{\partial X}\right)^2\frac{\partial^2 U_Y}{\partial X^2} = P,
\end{equation}
where $\alpha$ is a dimensionless tension given by
\begin{equation}\label{alphadef}
\alpha=\frac{3}{2}\frac{h_{0s}(u'_y)^2}{\hat{I}}=\frac{3}{2}\frac{h_{0s}p_0^2\ell^8}{E^2\hat{I}^3}  = {2592\left(\frac{p_0}{E}\right)^2\left(\frac{\ell}{h_{0s}}\right)^8}.
\end{equation}
Note that the ultimate definition of the characteristic displacement scale $u'_y$ will depend on the choice of fluid model and how it is nondimensionalized, through the form of $p_0$ to be substituted into Eq.~\eqref{uydef}. 

The top wall is assumed to be clamped at both ends (the entry and exit planes of the microchannel). Hence, the relevant boundary conditions for Eq.~\eqref{solidmodel3} are 
\begin{equation}\label{solidbc}
U_Y|_{X=0} = \left.\frac{\partial U_Y}{\partial X}\right|_{X=0} = 0,\qquad U_Y|_{X=1} = \left.\frac{\partial U_Y}{\partial X}\right|_{X=1} = 0.
\end{equation}
To ensure two-way coupling, we must also take into consideration the changing fluid domain. The deformed channel height is thus scaled by the undeformed channel height, and using the definition of $u'_y$ from Eq.~\eqref{uydef}, we obtain
\begin{equation}\label{betaeq}
H = \frac{h}{h_{0f}} = \frac{h_{0f}+u_y}{h_{0f}}=1+\frac{u_y}{h_{0f}}=1+\frac{u'_y}{h_{0f}} U_Y = 1 + \left(\frac{p_0\ell^4}{E\hat{I}\,h_{0f}}\right)U_Y.
\end{equation}

\subsection{Fluid mechanics: Governing equations}
\label{sec:fluid}

To derive the fluid model, we start with the 2D incompressible continuity and Navier--Stokes equations \cite{panton}:
\begin{subequations}\begin{align}
\frac{\partial v_x}{\partial x}+\frac{\partial v_y}{\partial y} &= 0,\label{cont1}\\
\frac{\partial v_x}{\partial t}+v_x\frac{\partial v_x}{\partial x}+v_y\frac{\partial v_x}{\partial y} &= -\frac{1}{\varrho_f}\frac{\partial p}{\partial x}+\nu \left(\frac{\partial^2 v_x}{\partial x^2}+\frac{\partial^2 v_x}{\partial y^2}\right), \label{xmom1}\\
\frac{\partial v_y}{\partial t}+v_x\frac{\partial v_y}{\partial x}+v_y\frac{\partial v_y}{\partial y} &= -\frac{1}{\varrho_f}\frac{\partial p}{\partial y}+\nu \left(\frac{\partial^2 v_y}{\partial x^2}+\frac{\partial^2 v_y}{\partial y^2}\right), \label{ymom1}
\end{align}\label{eq:ins}\end{subequations}
where $\varrho_f$ is the fluid's density, and $\nu$ is its kinematic viscosity. The planar velocity field is denoted $\bm{v} = (v_x,v_y)$, where both $v_x$ and $v_y$ can depend on $x$, $y$ and $t$. Next, we introduce the following dimensionless variables:
\begin{equation} 
X=x/\ell,\qquad Y=y/h_{0f},\qquad T=t\Big/\sqrt{\hat{\varrho}_s\ell^4/E\hat{I}},\qquad V_X=v_x\big/(\hat{q}_0/h_{0f}),\qquad V_Y=v_y\big/(\epsilon \hat{q}_0/h_{0f}),\qquad P=p/p_0,
\label{nd_vars_fluid}
\end{equation}
where $\epsilon\equiv h_{0f}/\ell$ is the aspect ratio of the fluid region, and $\hat{q}_0$ is the inlet area flow rate. The scales chosen for $x$, $t$ and $p$ are consistent with the ones used for the solid model. Specifically, we must use the same time scale for both the fluid and solid model to ensure a two-way coupled FSI system. The scales chosen for the fluid velocity are necessary to maintain the leading-order balance in the continuity equation. Substituting the dimensionless variables from Eq.~\eqref{nd_vars_fluid} into Eqs.~\eqref{eq:ins} results in
\begin{subequations}\begin{align}
\frac{\partial V_X}{\partial X}+\frac{\partial V_Y}{\partial Y} &= 0,\label{cont3}\\
\underbrace{\frac{h_{0f}^2}{\nu \sqrt{\hat{\varrho}_s\ell^4/E\hat{I}}}}_{=\epsilon Re^* St} \frac{\partial V_X}{\partial T} + \underbrace{\frac{\hat{q}_0 h_{0f}}{\nu \ell}}_{=\epsilon Re^*} V_X\frac{\partial V_X}{\partial X} + \underbrace{\frac{\epsilon \hat{q}_0}{\nu}}_{=\epsilon Re^*} V_Y\frac{\partial V_X}{\partial Y} &=
-\underbrace{\frac{p_0h_{0f}^3}{\varrho_f \nu \hat{q}_0\ell}}_{=\mathcal{O}(1)} \frac{\partial P}{\partial X} + \Bigg(\underbrace{\frac{h_{0f}^2}{\ell^2}}_{=\epsilon^2} \frac{\partial^2 V_X}{\partial X^2}+ \frac{\partial^2 V_X}{\partial Y^2}\Bigg),\label{xmom3}\\
\underbrace{\frac{\epsilon h_{0f}^2}{\nu \sqrt{\hat{\varrho}_s\ell^4/E\hat{I}}}}_{=\epsilon^2 Re^* St} \frac{\partial V_Y}{\partial T} + \underbrace{\frac{\epsilon \hat{q}_0 h_{0f}}{\nu \ell}}_{=\epsilon^2 Re^*} V_X\frac{\partial V_Y}{\partial X} + \underbrace{\frac{\epsilon^2 \hat{q}_0}{\nu}}_{=\epsilon^2 Re^*} V_Y\frac{\partial V_Y}{\partial Y} &=
- \underbrace{\frac{p_0 h_{0f}^2}{\varrho_f \nu \hat{q}_0}}_{=\mathcal{O}(\epsilon^{-1})} \frac{\partial P}{\partial Y}+ \Bigg(\underbrace{\frac{\epsilon h_{0f}^2}{\ell^2}}_{=\epsilon^3} \frac{\partial^2 V_Y}{\partial X^2}+\epsilon \frac{\partial^2 V_Y}{\partial Y^2}\Bigg). \label{ymom3}
\end{align}\label{eq:ins_ND}\end{subequations}

Now, two key dimensionless groups arise from inspection. First {is $Re\equiv\epsilon Re^*$, which is introduced as the ``reduced'' Reynolds number, with the regular Reynolds number, $Re^* \equiv \hat{q}_0/\nu$ defined based on the inlet flow rate.} Second is the Strouhal number $St \equiv \epsilon \sqrt{E\hat{I}/\hat{\varrho}_s \hat{q}_0^2}$, which multiplies the unsteady terms of Eqs.~\eqref{xmom3} and \eqref{ymom3}. While the Reynolds number quantifies the balance between inertial and viscous forces, {the Strouhal number (see, e.g., \cite[p.~351]{panton}) is the ratio of a characteristic fluid time scale ($\tau_f \sim \ell h_{0f}/\hat{q}_0$) to a characteristic solid time scale ($\tau_s \sim \sqrt{\hat{\varrho}_s\ell^4/E\hat{I}}$)}.
To make the pressure gradient in Eq.~\eqref{xmom3} a $\mathcal{O}(1)$ term, we must set
\begin{equation}\label{pscale}
p_0=\frac{\varrho_f \nu \hat{q}_0\ell}{h_{0f}^3}.
\end{equation}
Thus, our approach for the nondimensionalization of Eqs.~\eqref{eq:ins} is similar to the one outlined by Stewart et al.~\cite{SWJ09}, \emph{however}, we have used a low-Reynolds formulation [viscous pressure scale, i.e., Eq.~\eqref{pscale}] while Stewart et al.~\cite{SWJ09} used a high-Reynolds number nondimensionalization (inertial pressure scale). 

Next, as it is typical of microchannels, we assume a long and shallow geometry: {$h_{0f} \ll w \ll \ell$ (see, e.g., the discussion in \cite{GEGJ06,CCSS17})}. In other words, $\epsilon \equiv h_{0f}/\ell \ll 1$. Thus, all higher powers of $\epsilon$ can be dropped in the dimensionless Navier--Stokes equations. Nevertheless, we do {\emph{not} need to assume that $Re^*$ is small. Therefore, terms of order $Re = \epsilon Re^*$ are allowed to be $\mathcal{O}(1)$}, as in standard lubrication theory \cite{panton,B08}. Consequently, our dimensionless governing equations [i.e., Eqs.~\eqref{eq:ins_ND}] for the fluid become
\begin{subequations}\begin{align}
\frac{\partial V_X}{\partial X}+\frac{\partial V_Y}{\partial Y} &= 0, \label{cont5}\\
{ Re} St \frac{\partial V_X}{\partial T}+{ Re} V_X\frac{\partial V_X}{\partial X}+{ Re} V_Y\frac{\partial V_X}{\partial Y} &= - \frac{\partial P}{\partial X}+ \frac{\partial^2 V_X}{\partial Y^2}, \label{xmom5}\\
\frac{\partial P}{\partial Y} &= 0. \label{ymom5}
\end{align}\end{subequations}

Finally, note that in our 2D Newtonian fluid model, the only non-trivial component of the shear stress is $\tau_{xy}$ \cite{panton}, which can be made dimensionless to yield a scale for the shear stress: $\tau_0={\mu \hat{q}_0}/{h_{0f}^2}=\epsilon p_0$. Therefore, we draw the usual conclusion that shear forces from the fluid onto the solid can be neglected in comparison to the pressure load.

\subsection{Coupled fluid--solid model}
\label{sec:fscouple}

The no-slip boundary condition is  enforced on both the top and bottom walls of the microchannel. In addition, a no penetration boundary condition is imposed at the bottom wall. Since the top wall moves, a kinematic boundary condition is required there \cite{panton}, which takes the form 
\begin{equation}\label{dynBC}
 St \frac{\partial H}{\partial T}=V_Y|_{Y=H}.
\end{equation}
Equation~\eqref{dynBC} ensures that the vertical velocity of the fluid in contact with the moving wall is equal to the vertical velocity of the wall. {The horizontal motion of the elastic wall is negligible within the beam model from Sect.~\ref{sec:solid}.}

Since our goal is to obtain a 1D model, the continuity Eq.~\eqref{cont5} and the $x$-momentum Eq.~\eqref{xmom5} are integrated over the channel height. For the continuity equation, we immediately obtain 
\begin{equation}\label{cont7}
\frac{\partial Q}{\partial X}+ St \frac{\partial H}{\partial T}=0,
\end{equation}
after defining the dimensionless area flow rate, $Q = \int^H_0 V_X \,\mathrm{d}Y$, and using Eq.~\eqref{dynBC} to obtain the value of $V_Y$ at $Y=H$. Next, the non-convective terms in the $x$-momentum equation are re-cast in conservative form and integrated over $Y$ to obtain
\begin{equation}\label{xmom7}
{Re} St \frac{\partial Q}{\partial T} + {Re} \frac{\partial }{\partial X} \int^H_0 V_X^2 \,\mathrm{d}Y =- H\frac{\partial P}{\partial X} + \left.\frac{\partial V_X}{\partial Y}\right|^{Y=H}_{Y=0},
\end{equation}
having assumed that $V_X$ is a continuous function (so that we can switch the order of operation between derivative and integral), applying the no slip boundary condition, and using the reduced $y$-momentum Eq.~\eqref{ymom5} to deduce that $P=P(X;T)$ (so that $\partial P/\partial X$ can be treated as constant in the integration over $Y$).

The final step in the process of averaging over $Y$ is to invoke the \emph{von K\'arm\'an--Polhausen approximation} \cite[p.~541]{panton}. That is, we assume a parabolic velocity profile at each cross section in the flow \cite{SWJ09}, specifically the 2D Poiseuille profile with horizontal component $v_x=6qy(h-y)/h^3$ (in dimensional variables), where $h = h_{0f} + u_y$ is the height of deformed microchannel, as above. After nondimensionalization, we have $V_X=6QY(H-Y)/H^3$ [the corresponding $V_Y$ can be found via Eq.~\eqref{cont5}], and this expression can be used to evaluate the integral on the left-hand side and the last term on the right-hand side of Eq.~\eqref{xmom7} to yield
\begin{equation}\label{xmom8}
{Re} St \frac{\partial Q}{\partial T} + {Re} \frac{6}{5} \frac{\partial }{\partial X} \left(\frac{Q^2}{H}\right) = - H\frac{\partial P}{\partial X} - \frac{12Q}{H^2}.
\end{equation}
Thus, Eqs.~\eqref{cont7} and \eqref{xmom8} are the final dimensionless governing equations of the fluid mechanics problem.

As mentioned above, to ensure two-way fluid--solid coupling, one final equation is required to close the problem. To this end, substituting the pressure scale from Eq.~\eqref{pscale} into the dimensionless deformed channel height in Eq.~\eqref{betaeq}, we obtain
\begin{equation}\label{betaeq2}
H = 1 + {\left(\frac{\varrho_f \nu \hat{q}_0\ell^5}{E\hat{I}h_{0f}^4}\right)U_Y = 1+\beta U_Y,}
\end{equation}
where $\beta$ can be termed the \emph{FSI parameter} because it combines \emph{all} the fluid, solid and geometrical properties of the given setup. Specifically, we have defined
\begin{equation}\label{betadef}
\beta \equiv \frac{\varrho_f \nu \hat{q}_0\ell^5}{E\hat{I}h_{0f}^4}={\frac{\hat{q}_0/\nu}{E\hat{I}/(\varrho_f \nu^2 h_{0f})}\frac{\ell^5}{h_{0f}^5}=\frac{Re^*}{\Sigma^*}\frac{\ell^5}{h_{0f}^5}=\frac{Re}{\epsilon^6\Sigma^*}=\frac{Re}{\Sigma}.}
\end{equation}
Note that, in Eq.~\eqref{betadef}, we were able to re-write the FSI parameter $\beta$ in terms of {the reduced Reynolds number $Re \equiv \epsilon \hat{q}_0/\nu$ (representing the fluid's contribution), the reduced dimensionless bending rigidity $\Sigma\equiv\epsilon^6\Sigma^*$, where $\Sigma^* \equiv E\hat{I}/(\varrho_f \nu^2 h_{0f})$ (representing the solid's contribution)} \footnote{We use the notation $\Sigma^*$, which coincides with the notation in \cite{VK13} and the references therein, because our definition is similar to the dimensionless solid mechanics parameter therein.}. Equation~\eqref{betaeq2} achieves the second part of the two-way FSI coupling by transferring solid displacements into the change of shape of the microchannel.

\subsection{Summary of the 1D model}

Equations \eqref{solidmodel3}, \eqref{cont7}, \eqref{xmom8} and \eqref{betaeq2} all-together form the coupled system of governing equations for our 1D viscous FSI problem. Additionally, initial and boundary conditions are required to fully specify the problem. 

The initial conditions are those of uniform flow under an undeformed wall:
\begin{equation}
Q|_{T=0} = 1, \qquad U_Y|_{T=0} = 0 \quad \Leftrightarrow\quad H|_{T=0} = 1. 
\label{eq:IC}
\end{equation}
An initial inlet flow rate must be imposed, which is used to define the characteristic scales. Therefore, consistent with the von K\'arm\'an--Polhausen approximation, the flow rate everywhere in the channel ($0\le X\le 1$) is initially (at time $T=0$) set equal to the dimensionless inlet flow rate, leading to the first condition in Eq.~\eqref{eq:IC}. No initial conditions can be imposed on the pressure, as usual. 

The boundary conditions on the solid mechanics problem are those of clamping at $X=0$ and $X=1$ as given in Eq.~\eqref{solidbc}. The boundary conditions on the fluid mechanics problem are the imposed inlet flow rate and the outlet pressure set to gauge:
\begin{equation}
Q|_{X=0} = 1,\qquad P|_{X=1} = 0.
\label{eq:fluid_bc}
\end{equation}

The key dimensionless groups of the 1D model are
\begin{equation}\label{finalparaeq}
\epsilon=\frac{h_{0f}}{\ell},\qquad {Re=\frac{\epsilon \hat{q}_0}{\nu}},\qquad St=\epsilon \sqrt{\frac{E\hat{I}}{\hat{\varrho}_s \hat{q}_0^2}},\qquad \Sigma = \frac{\epsilon^6E\hat{I}}{\varrho_f \nu^2 h_{0f}},\qquad {\alpha=18\beta^2\left(\frac{h_{0f}}{h_{0s}}\right)^2},\qquad \beta=\frac{Re}{\Sigma}.
\end{equation}
The last expression for $\alpha$ is derived from Eq.~\eqref{alphadef} by substituting Eq.~\eqref{pscale} into it. {In the following discussions, $Re$ and $\Sigma$ will be varied independently to investigate the steady and dynamic responses of the system. We will also consider the cases $\alpha=0$ and $\alpha\ne0$, but we will not study the parametric variation of $\alpha$. Meanwhile, $\beta$ is a useful dimensionless quantity to gauge the ``strength'' of FSI in the system, but it is not an independent parameter.}

{Table~\ref{table:dim} lists the typical values of dimensional system parameters and the corresponding dimensionless numbers of the FSI model. The geometrical parameters of the microchannel are chosen so that the assumption $h_{0f}\ll \ell$ is satisfied. The thickness of the top wall, $h_{0s}$, is simply taken the same as the fluid domain's thickness, $h_{0f}$; i.e., we are not restricted to the limit of either a thin ($h_{0s}\ll h_{0f}$) or thick ($h_{0s}\gg h_{0f}$) structure. As for the mechanical properties, we are interested in microchannels fabricated from polymeric (soft) materials, whose density is usually comparable to that of water, and the magnitude of their Young's modulus ranges from several MPa to several GPa \cite{SMMC11}. Water is chosen as the working fluid because its small viscosity allows a relatively large range of Reynolds numbers, compared with fluids of higher viscosity (such as glycerol used in \cite{GMV17}). 

For these example values of the dimensional parameters, we have $Re = 1$ and $St = 1$.} We also observe that the value of $\alpha$ is large compared to the other dimensionless parameters. This observation raises the possibility that terms other than the tension in Eq.~\eqref{solidmodel3} are negligible. However, neglecting terms in comparison to the tension term in Eq.~\eqref{solidmodel3} results in an oversimplified ODE, which admits only the trivial solution $U_Y=0$. This oversimplified ($\alpha\to\infty$) ODE is also decoupled from the pressure, thus it will not allow for any nontrivial deformation. On the other hand, it can be shown that combining Eq.~\eqref{solidmodel3} and Eq.~\eqref{betaeq2} results in a PDE [see, e.g., Eq.~\eqref{st0eq5}] in which {the coefficient multiplying the tension term is $\alpha/\beta^2\equiv 18(h_{0f}/h_{0s})^2 = \mathcal{O}(1)$,} while the coefficient multiplying $P$ is $\beta$, instead of unity. All terms in this latter PDE are now the same order of magnitude. Hence, the fluid pressure has a substantial effect, even for $\alpha\gg1$, and all the terms in Eq.~\eqref{solidmodel3} must be retained (even if not multiplied by $\alpha$).

\begin{table}[t]
\begin{ruledtabular}
\begin{tabular}{clll}
Variable &Name &Experimental value & SI Unit \\
\hline
$\ell$ & channel's length &$5.0\times10^{-3}$ & m  \\
$h_{0s}$ & solid's thickness &$5.0\times10^{-5}$ & m  \\
$E$ & solid's Young's modulus &$4.8\times10^{8}$ & Pa  \\
$\hat{\varrho}_s$ & solid's mass per unit area & $5.0\times10^{-2}$ & kg/m$^2$  \\
$\nu$ & fluid's kinematic viscosity & $1.0\times10^{-6}$ & m$^2$/s \\
$\varrho_f$ & fluid's density &$1.0\times10^{3}$ & kg/m$^3$ \\
$\hat{q}_0$ & inlet area flow rate &$1.0\times10^{-4}$ & m$^2$/s \\
$h_{0f}$ & channel height &$5.0\times10^{-5}$ & m\\
\hline
$\epsilon$ & channel's aspect ratio & $1.0\times 10^{-2}$ &  --\\
$Re$ & Reynolds number & $1.0$ &  --\\
$St$ & Strouhal number &  $1.0$ &  --\\
$\Sigma$ & dimensionless Young's modulus & $1.0\times 10^{-4}$ & --\\
$\beta$ & fluid--structure interaction parameter & $1.0\times 10^4$ &  --\\
$\alpha$ & dimensionless beam tension & $1.8 \times 10^9$ & --
\end{tabular}
\end{ruledtabular}
\caption{Typical values of the model's dimensional and dimensionless [see Eq.~\eqref{finalparaeq} for definitions] parameters used in our example simulations below.}
\label{table:dim}
\end{table}

\section{Shape of the inflated microchannel}
\label{sec:steady}

{In this section, we discuss the microchannel's steady-state characteristics. To this end, we set $St = 0$ (i.e., the characteristic fluid time scale is much shorter than the characteristic solid time scale, precluding any unsteady FSI response). First, we  compute the steady-state shape of the microchannel (Sect.~\ref{subsec:st0}), and discuss the forces required to achieve a particular static response of the system. Second, we determine how the maximum dimensionless channel deformation, $H_\mathrm{max}$, and the axially-average hydrodynamic pressure, $\langle P \rangle$, scale with key dimensionless groups, such as the reduced Reynolds number $Re$ and the reduced dimensionless bending rigidity $\Sigma$ (Sect.~\ref{sec:scalings}).}

\subsection{Steady-state shape of the top wall of the inflated microchannel}
\label{subsec:st0}
{In the limit of $St\to0$}, Eq.~\eqref{cont7} simply states that $Q$ is independent of $X$: ${\partial Q}/{\partial X} = 0$. The flow rate is, thus, simply given by the boundary condition imposed: $Q(X,T)\equiv1$ $\forall X\in[0,1]$, $T\ge0$. Subsequently, Eq.~\eqref{solidmodel3} can be reconstituted as a PDE for $H$ using Eq.~\eqref{betaeq2}. After taking an $X$ derivative of the resulting PDE and dropping the remaining unsteady terms, we obtain a fifth-order PDE:
\begin{equation}\label{st0eq4}
\frac{\partial^5 H}{\partial X^5} -\frac{\alpha}{\beta^2}\frac{\partial}{\partial X}\left[ \left(\frac{\partial H}{\partial X}\right)^2 \frac{\partial^2 H}{\partial X^2} \right] = 
\beta \frac{\partial P}{\partial X}.
\end{equation}
{Observe that, due to pressure loading of the soft wall, $H\equiv 1$ is \emph{not} a steady state, unless $\beta=0$. This feature of the microchannel problems makes it distinct from the collapse vessel problems studied in the literature \cite{SWJ09,XBJ13,S17}.} Next, Eq.~\eqref{xmom8} can be used to solve for $\partial P/\partial X$, the expression for which can then be substituted into Eq.~\eqref{st0eq4}:
\begin{equation}\label{st0eq5}
\frac{\partial^5 H}{\partial X^5}  -{18\left(\frac{h_{0f}}{h_{0s}}\right)^2}\frac{\partial}{\partial X}\left[ \left(\frac{\partial H}{\partial X}\right)^2 \frac{\partial^2 H}{\partial X^2}\right] = 
{\frac{Re}{\Sigma}} \left({ Re} \frac{6}{5} \frac{1}{H^3}\frac{\partial H}{\partial X}-\frac{12}{H^3}\right).
\end{equation}
{Here, we have made use of the relations $\alpha/\beta^2\equiv 18(h_{0f}/h_{0s})^2$ and $\beta\equiv Re/\Sigma$ from Eq.~\eqref{finalparaeq}, to make the parametric dependencies in Eq.~\eqref{st0eq5} more explicit.}

This final fifth-order nonlinear PDE \eqref{st0eq5} for $H$ can be compared to Stewart et al.~\cite[Eq.~(2.12a)]{SWJ09}, which was derived in the high-$Re$ context. In the model in \cite{SWJ09}, stretching is the dominant solid mechanics response and bending is neglected by assuming small deformations. Thus, \cite[Eq.~(2.12a)]{SWJ09} differs from Eq.~\eqref{st0eq5} in two principal ways: (i) $Re$ only modifies the fluid inertia term in Eq.~\eqref{st0eq5}, while it modifies \emph{both} the fluid inertia and the nonlinear stretching terms in \cite[Eq.~(2.12a)]{SWJ09}; (ii) the higher-order bending term on the left-hand side of Eq.~\eqref{st0eq5} is not present in \cite[Eq.~(2.12a)]{SWJ09} and, likewise, the nonlinear stretching term on the left-hand side of Eq.~\eqref{st0eq5} is to be contrasted with the linearized tension term in \cite[Eq.~(2.12a)]{SWJ09}. Consequently, we expect that the steady states governed by Eq.~\eqref{st0eq5}, and their linear stability, to differ significantly from those studied in the literature, paving the way to potentially rich new dynamic behaviors in the present viscous FSI model.

To compute the steady state channel shape, denoted $H_0(X)$, we re-interpret  Eq.~\eqref{st0eq5} as a two-point boundary-value problem that can be solve numerically using SciPy's {\tt solve\_bvp} \cite{SciPy}. Specifically, Eq.~\eqref{st0eq5}  is subject to
\begin{equation}
H_0(X=0) = 1,\quad \left.\frac{\partial H_0}{\partial X}\right|_{X=0} =0,\qquad H_0(X=1) = 1,\quad \left.\frac{\partial H_0}{\partial X}\right|_{X=1} = 0,\qquad  \left.\frac{\partial^4 H_0}{\partial X^4}\right|_{X=1} = 0,
\label{ssH0BC}
\end{equation}
where the first four boundary conditions are simply the clamped conditions [see Eq.~\eqref{solidbc}], while the last one is the outlet pressure condition [see Eq.~\eqref{eq:fluid_bc}] rewritten in terms of the steady-state channel height via Eq.~\eqref{solidmodel3}. 

\begin{figure}
\subfloat[\label{fig:H0steady}]{\includegraphics[width=0.5\textwidth]{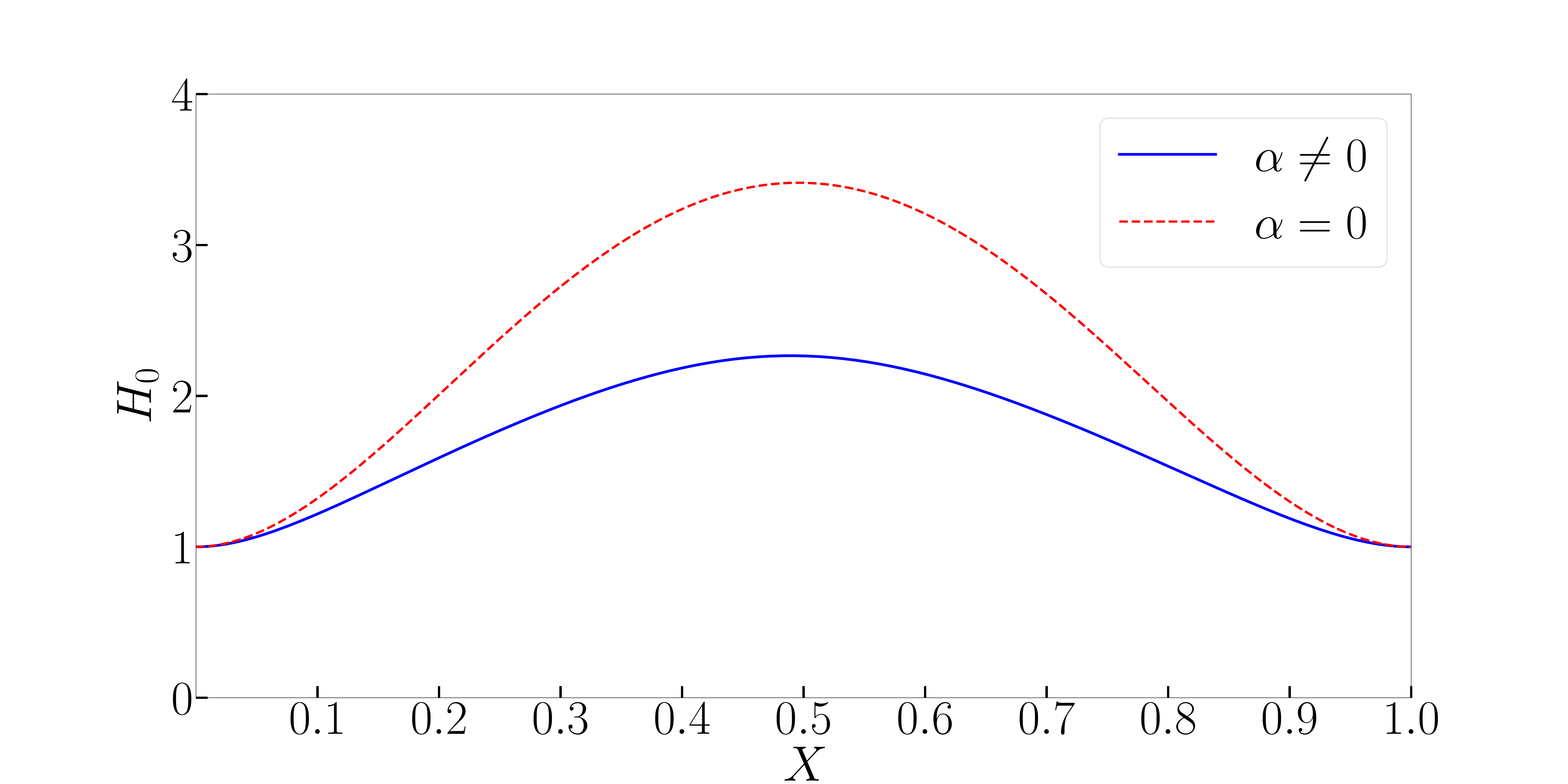}}
\hfill
\subfloat[\label{fig:P0steady}]{\includegraphics[width=0.5\textwidth]{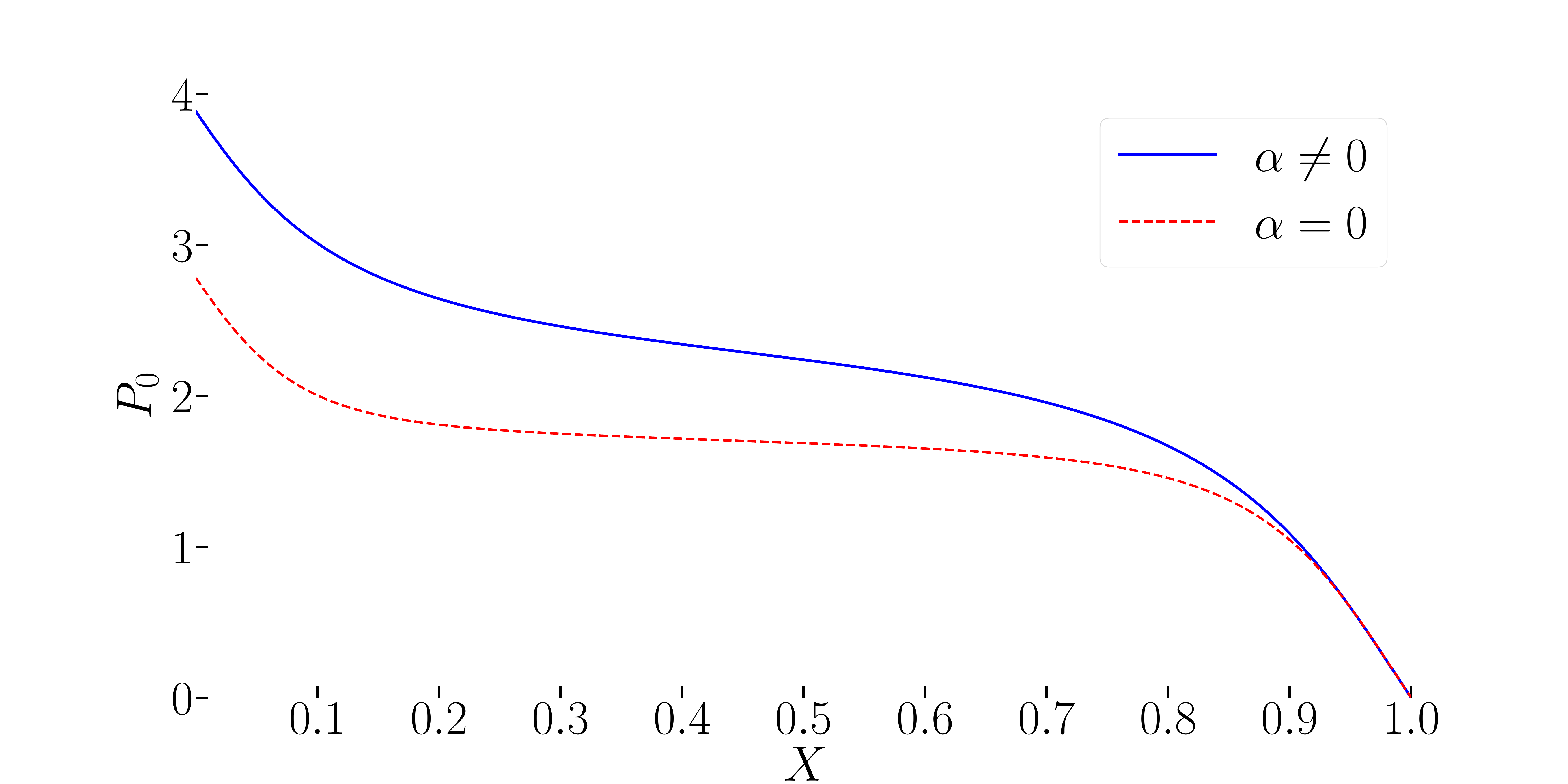}}
\caption{(Color online) {Typical steady-state shapes and pressure distributions inside the soft-walled microchannel. (a) Steady-state deflection, $H_0$, as a function of the flow-wise position $X$. (b) Steady-state pressure distribution, $P_0$, as a function of the flow-wise position $X$. The solid curves represent the results with nonlinear tension included ($\alpha=5.56\times10^6$), while the dashed curves represent results without tension ($\alpha=0$, i.e., pure bending). The remaining dimensionless parameters are $Re=0.5$, $\Sigma=9\times10^{-4}$ and $\beta=5.56\times 10^2$.}}
\label{fig:steadyPlot}
\end{figure}

{Next, we show example plots of the steady-state shape, $H_0(X)$, and the corresponding pressure distribution, $P_0(X)$, along the microchannel. For these examples, we take $\Sigma=9\times10^{-4}$, fix the height ratio at $h_{0f}/h_{0s}=1$, and vary $Re$. Both $\alpha\ne0$ and $\alpha=0$ are considered. The dimensionless parameters used for these computations are not the same as Table \ref{table:dim} but are modified in order to make the deflections (with and without tension) within a similar range, which makes the plots easier to interpret.

First, in Fig.~\ref{fig:steadyPlot}, we consider $Re=0.5$. Whether tension is included or not, $P_0$ is a nonlinear function of $X$ due to FSI, as can be seen in panel (b). However, note that for $\alpha\ne0$, the microchannel displays much smaller deformation for a larger pressure drop, $P_0(0)-P_0(1)$. The reason for this observation is that tension in the beam restricts the deflection of the top wall, resulting in larger flow velocity at fixed flow rate ($Q=1$) and, thus, causes larger pressure losses due to viscosity. Furthermore, $P_0$ is a decreasing function of $X$, because the inertial effects of the flow are negligible in this case ($Re$ is small), thus viscous effects dominate and $\mathrm{d} P_0/\mathrm{d} X \sim -12/H_0(X)^3<0$, consistent with lubrication theory.}

\begin{figure}
\subfloat[\label{fig:H0steady2}]{\includegraphics[width=0.5\textwidth]{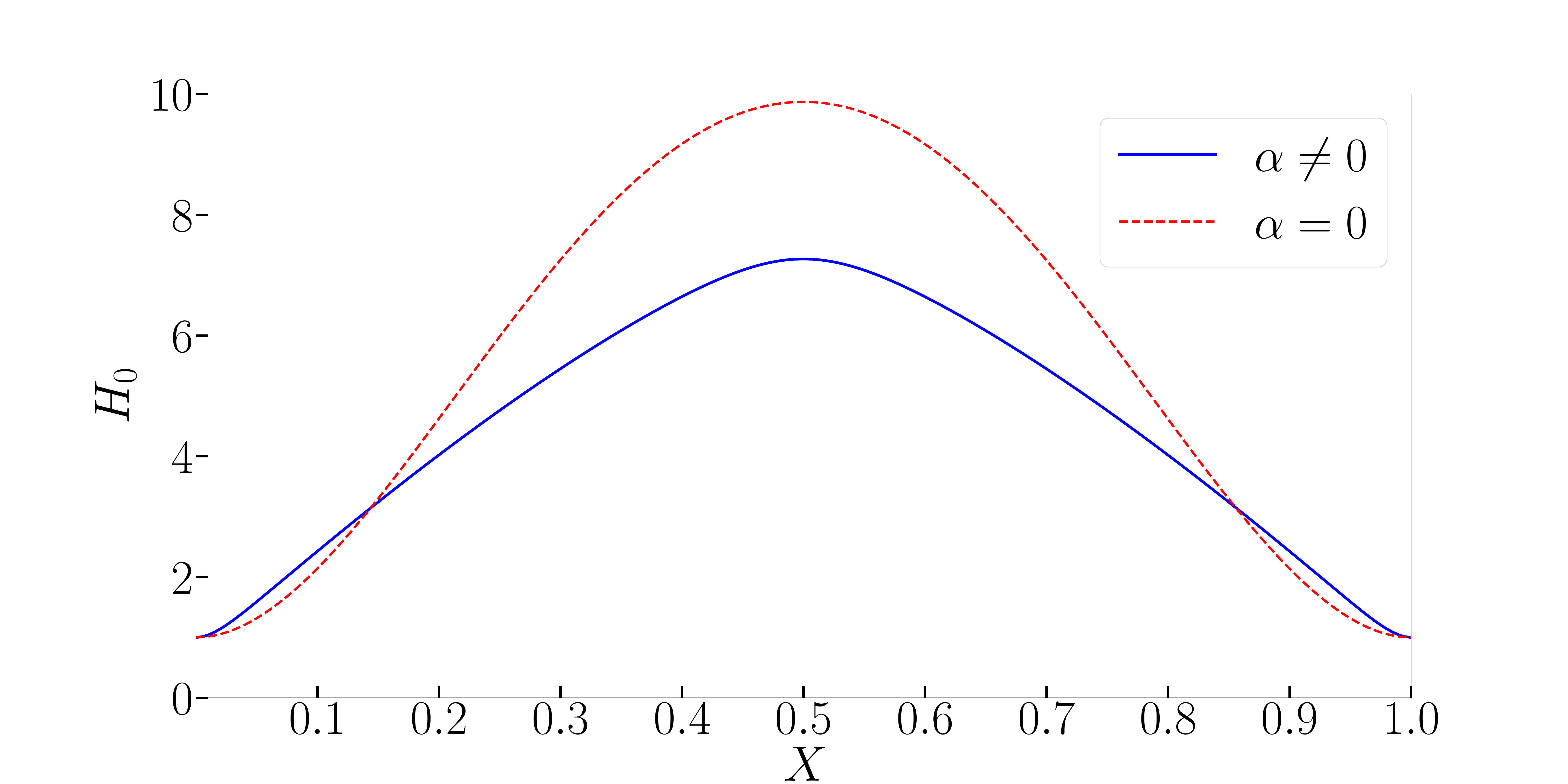}}
\hfill
\subfloat[\label{fig:P0steady2}]{\includegraphics[width=0.5\textwidth]{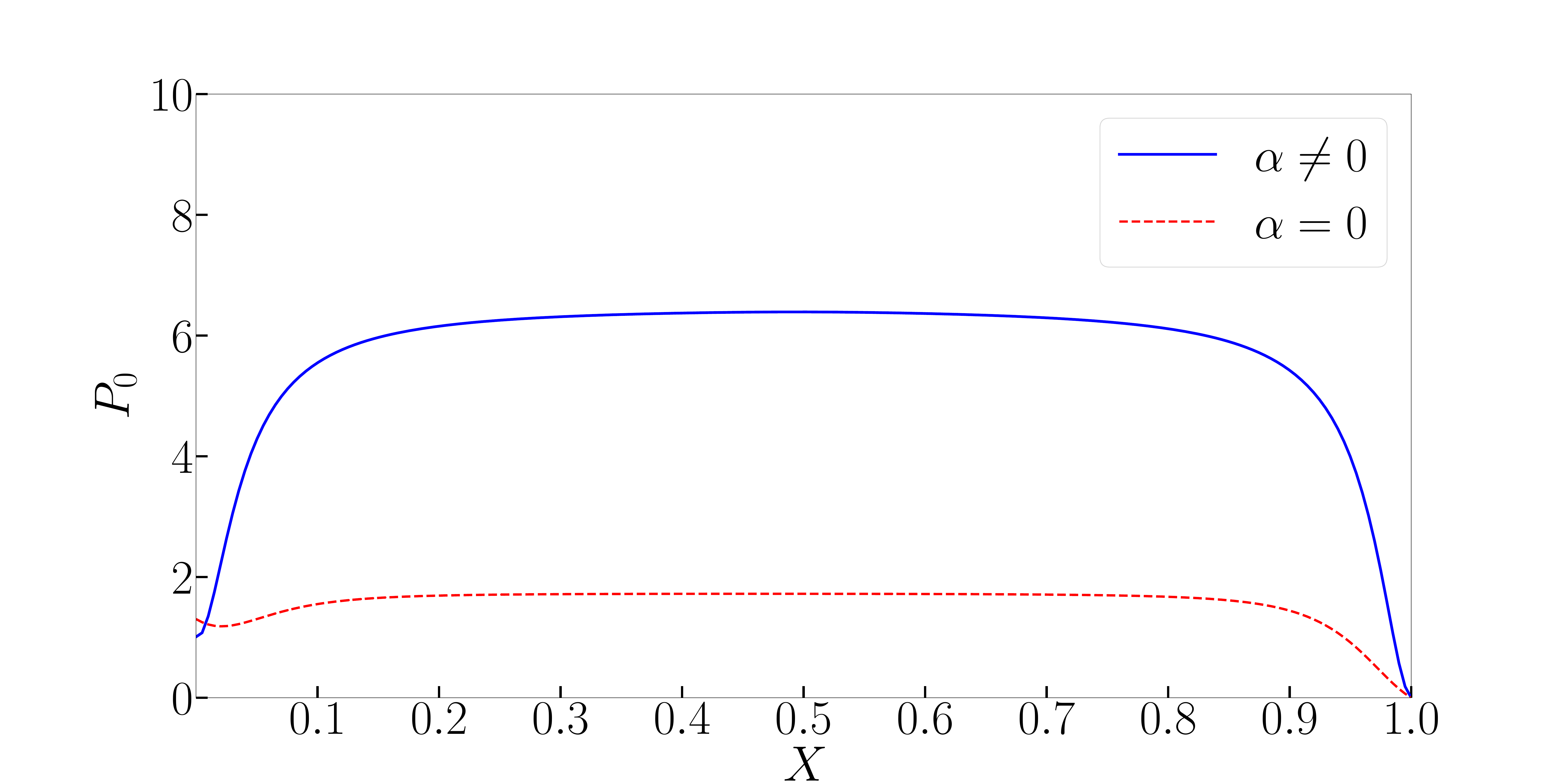}}
\caption{(Color online) {More ``exotic'' steady-state shapes, (a) $H_0(X)$, and pressure distributions, (b) $P_0(X)$, inside the soft-walled microchannel. The solid curves represent the results with nonlinear tension included ($\alpha=2.22\times10^9$), with $Re=10$ and $\Sigma=9\times 10^{-4}$, while the dashed curves represent results without tension ($\alpha=0$, i.e., pure bending) with $Re=1.8$ and $\Sigma=9\times 10^{-4}$.}}
\label{fig:steadyPlot2}
\end{figure}

{ With the increase of $Re$, it is expected that inertial effects in the flow become prominent. While the top wall of the microchannel will still bulge under the pressure load from the flow, the pressure gradient does not have to remain negative, and $P_0$ will not necessarily be a decreasing function of $X$, as shown in  Fig.~\subref*{fig:P0steady2} in contrast to Fig.~\subref*{fig:P0steady}. This observation can be justified by recognizing that $\mathrm{d} P_0/\mathrm{d} X$ is the consequence of the competition between the convective effects and viscous effects in the flow [see the right-hand side of Eq.~\eqref{st0eq5}]. Since the top wall is clamped at both ends, its bulging leads to its slope, $\mathrm{d} H_0/\mathrm{d} X$, increasing near the inlet ($X=0$) and decreasing near the outlet ($X=1$). If inertia is dominant in the flow, a positive pressure gradient can be expected, for $Re$ large enough. As shown in Fig.~\ref{fig:steadyPlot2}, this positive pressure gradient is observed upstream. Note that we have chosen a smaller $Re$ value for the pure bending case (compared to the case with tension), in order to ensure that the deformation is within a reasonable range. Since $Re$ is much larger in the case of $\alpha\ne 0$, it is not surprising that the positive pressure gradient is much more prominent. Interestingly, the pressure profiles are almost flat in the middle part of the channel, with or without tension, indicating a negligible pressure gradient in this region. Also observe that the deformations in both cases are large. Since the flow rate is fixed, the fluid's  velocity has to decrease rapidly along the flow-wise direction in the expanding section of the deformed microchannel, and the positive pressure gradient will help decelerate the flow.}

\subsection{Deformation and pressure scalings at steady state}
\label{sec:scalings}

In this subsection, we address {the different scaling regimes of deformation and hydrodynamics with respect to key dimensionless groups of the problem. To frame the discussion, we define the maximum deformation $H_\mathrm{max} = \max_{0\le X\le 1} H_0(X)$ and the axially-average pressure $\langle P\rangle = \int_0^1 P_0(X) \,\mathrm{d}X$. We seek to establish how each of these scalar quantities scales with $Re$ and $\Sigma$, as we encounter different regimes of physics: e.g., bending- or tension-dominated deformation, inertia- or viscosity-dominated pressure profile, and so on.}

\subsubsection{Scaling of $\langle P \rangle$}\label{Pscaling}

{ Before we start our analysis, it is worth mentioning the reason for choosing $\langle P\rangle$ as the quantity of interest instead of, say, the total pressure drop $P_0(0)-P_0(1)$, which is more commonly discussed in microchannel studies. First, $\langle P\rangle$ better captures the pressure variation, compared to $P_0(0)$, especially when the inertial forces in the flow are dominant, and a positive pressure gradient is observed upstream. In this case, the pressure in the middle part of the channel is larger than the total pressure drop (see Fig.~\ref{fig:steadyPlot2}) and using $P_0(0)$ to infer the characteristic load on the structure will underestimate the deformation. Second, using $P_0(0)$ renders the inertial flow effects difficult to analyze. It is easy to show that $P_0(0)=\int_0^1 12/H_0(X)^3 \,\mathrm{d}X$ by integrating the right-hand side of Eq.~\eqref{st0eq5} and applying the clamped boundary conditions. This expression does not necessarily mean that the inertia of the fluid is not important, rather it ``hides'' this effect in the shape of the channel, $H_0(X)$, which further serves to complicate the scaling analysis.}

{ From Sect.~\ref{subsec:st0}, we already know that the pressure gradient in the flow, $\mathrm{d}P_0/\mathrm{d}X$, is the outcome of the competition between the inertial and viscous forces in the flow. Then, it is natural to investigate the two limits, i.e., the viscosity-dominated and inertia-dominated regimes, respectively, and explore how $\langle P\rangle$ scales in each regimes.

\paragraph{Case 1:} Viscous effects are dominant in the flow. In this case, $\mathrm{d} P_0/\mathrm{d} X \sim -12/H_0(X)^3<0$ and $P_0(X)$ is a decreasing function of $X$ with a relatively flat middle part, as in Fig.~\subref*{fig:P0steady}. Observing that the deformation profile is almost symmetric in Fig.~\subref*{fig:H0steady}, we assume that the pressure in this flat region is a good estimate of $\langle P\rangle$. 

To proceed, define a critical value of the deformed channel height as $H_c$ such that $H_c = H_\mathrm{max}$ for small deformation and $H_c<H_\mathrm{max}$ with $1/H_c^3\ll 1$ for large deformation. We want to find the flow-wise position, $X_c$, at which $H_0(X_c) = H_c$ and also, $P_0(X_c)\sim \langle P \rangle$, per our assumption. Then, with a linear approximation of the deformation profile, we have ${H_c}/H_\mathrm{max}\sim 2(1-X_c)$. Here, we have made use of the (almost) symmetry of the deformation profile. The deformation profile near the outlet is written as a linear function, $H_0(X)\approx -(H_c-1)(X-1)/(1-X_c)+1$, then
\begin{equation}\label{pviscous}
    \langle P\rangle\sim P_0(X_c) \sim -\int_{X_c}^1 \frac{\partial P_0}{\partial X} \,\mathrm{d}X \sim \int_{X_c}^1 \frac{12}{[-(H_c-1)(X-1)/(1-X_c)+1]^3}\,\mathrm{d}X
    =\frac{6(1-X_c)(1+H_c)}{H_c^2}\sim \frac{1}{H_\mathrm{max}}.
\end{equation}
Note that the actual value of $H_c$ is not important in the scaling analysis.

\paragraph{Case 2:} Convective (inertial) effects are dominant in the flow. In this case, as shown in Fig.~\ref{fig:steadyPlot2}, the deformation of the microchannel is usually large and, thus the profile $P_0(X)$ displays a flatter middle part. Again, assume that the pressure in this portion of the microchannel is still a good estimate of $\langle P\rangle$. Following a similar procedure to Case 1 above, but choosing $H_c$ as $1/H_c^2\ll 1$, further balancing the convective term and the pressure gradient in Eq.~\eqref{xmom8}, we can estimate
\begin{equation}\label{pinertial}
    \langle P\rangle \sim -\int_{X_c}^1 \frac{\partial P_0}{\partial X} \,\mathrm{d}X \sim \int_{X_c}^1 \frac{3}{5}Re\frac{\partial}{\partial X}\left(\frac{1}{H_0^2}\right)\,\mathrm{d}X = \frac{3}{5}Re\left(1-\frac{1}{H_c^2}\right)\sim Re .
\end{equation}

\subsubsection{Scaling of  $H_\mathrm{max}$: Pure bending}\label{bendingscaling}
Now, we are ready to analyze the solid mechanics problem to obtain the scaling of $H_\mathrm{max}$. First, we consider pure bending ($\alpha= 0$). In this case, the governing equation of the solid mechanics is the classic Euler--Bernoulli beam equation, $\mathrm{d}^4H_0/\mathrm{d}X^4=\beta P_0={Re P_0/\Sigma}$, which implies that $H_{\mathrm{max}} \sim {Re\langle P\rangle/\Sigma}$. Furthermore, in order for the beam theory to apply, we have strictly restricted the maximum deformation of the top wall to be no greater than 10\% of the length of the channel, corresponding to $H_{\mathrm{max}}\leq 10$ with the aspect ratio $\epsilon=0.01$. This restriction applies to all of the following discussion.

If viscous effects are dominant in the flow, using Eq.~\eqref{pviscous}, we obtain the scalings $H_{\mathrm{max}}\sim {(Re/\Sigma)}^{1/2}$ and $\langle P\rangle\sim {(Re/\Sigma)}^{-1/2}$, which are clearly observed in the numerical data in Fig.~\ref{fig:viscous0alpha}. However, this figure also shows another regime in which the deformation is very small, with $H_{\mathrm{max}}\sim {(Re/\Sigma)}^{1/4}$ and $\langle P\rangle\sim {(Re/\Sigma)}^{-1/4}$. Note that Eq.~\eqref{pviscous} is still valid. The key difference, in this case, is that, $P_0(X)$ is nearly linear with an almost constant gradient given by the lubrication approximation, $\mathrm{d} P_0/ \mathrm{d} X\sim 1/H^3_{\mathrm{max}}$, since the deformation is very small, i.e., $H_{\mathrm{max}}\approx 1$. Then, a more appropriate scaling is obtained by considering $\mathrm{d}^5 H_0/\mathrm{d}X^5={(Re/\Sigma)}\mathrm{d}P_0/\mathrm{d}X$, indicating that $H_{\mathrm{max}}\sim {(Re/\Sigma)}/H^3_{\mathrm{max}} $ and thus yielding $H_{\mathrm{max}}\sim {(Re/\Sigma)}^{1/4}$. 

In Fig.~\ref{fig:viscous0alpha}, we also observe outliers for the last two cases with $\Sigma=1.0\times 10^{-3}$ and $\Sigma=1.0\times 10^{-2}$ at large $\beta$ because, in these cases, $Re$ can be large (even under the restrictions on the maximum deformation) so that the response of the system deviates from the viscosity--bending force balance/regime. Specifically, as we will show next, the outliers in the case of the most rigid microchannel actually belong to an inertia--bending force balance/regime.

If the inertial effects are dominant in the flow, $H_{\mathrm{max}} \sim Re^2/\Sigma$ since $\langle P\rangle\sim Re$. However, for the parameters chosen, which cover six orders of $\Sigma$ (and thus we believe should cover a significant number of actual microchannel systems), only the last set of data with $\Sigma=1.0\times 10^{-2}$ reaches this regime, Meanwhile, the cases of $\Sigma=1.0\times 10^{-3}$ are more likely to be in the transitional stage at relatively high $Re$, as shown in Fig.~\ref{fig:inertia0alpha}. 

\begin{figure}
    \centering
    \subfloat[]{\includegraphics[width=0.5\textwidth]{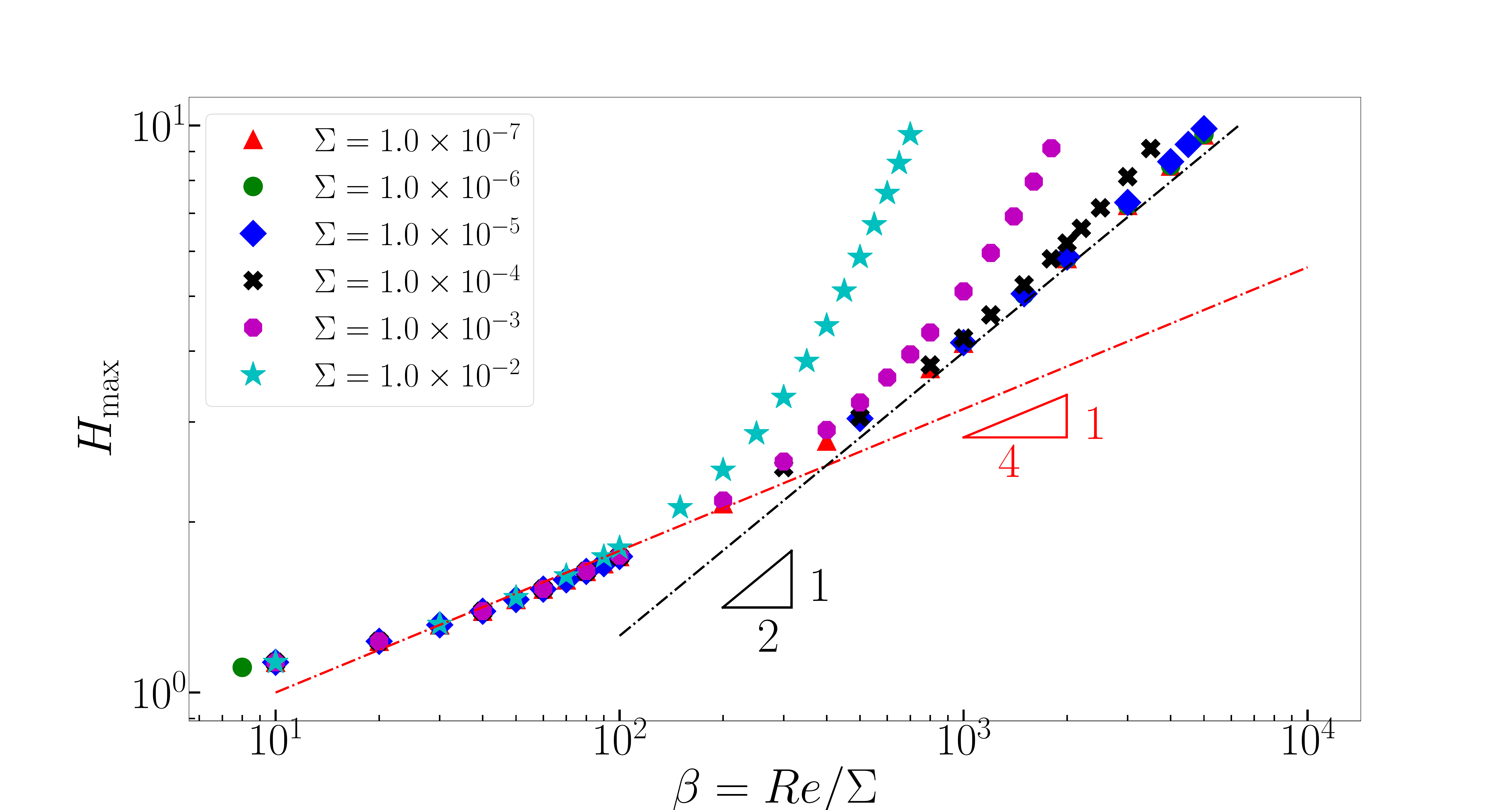}}
    \subfloat[]{\includegraphics[width=0.5\textwidth]{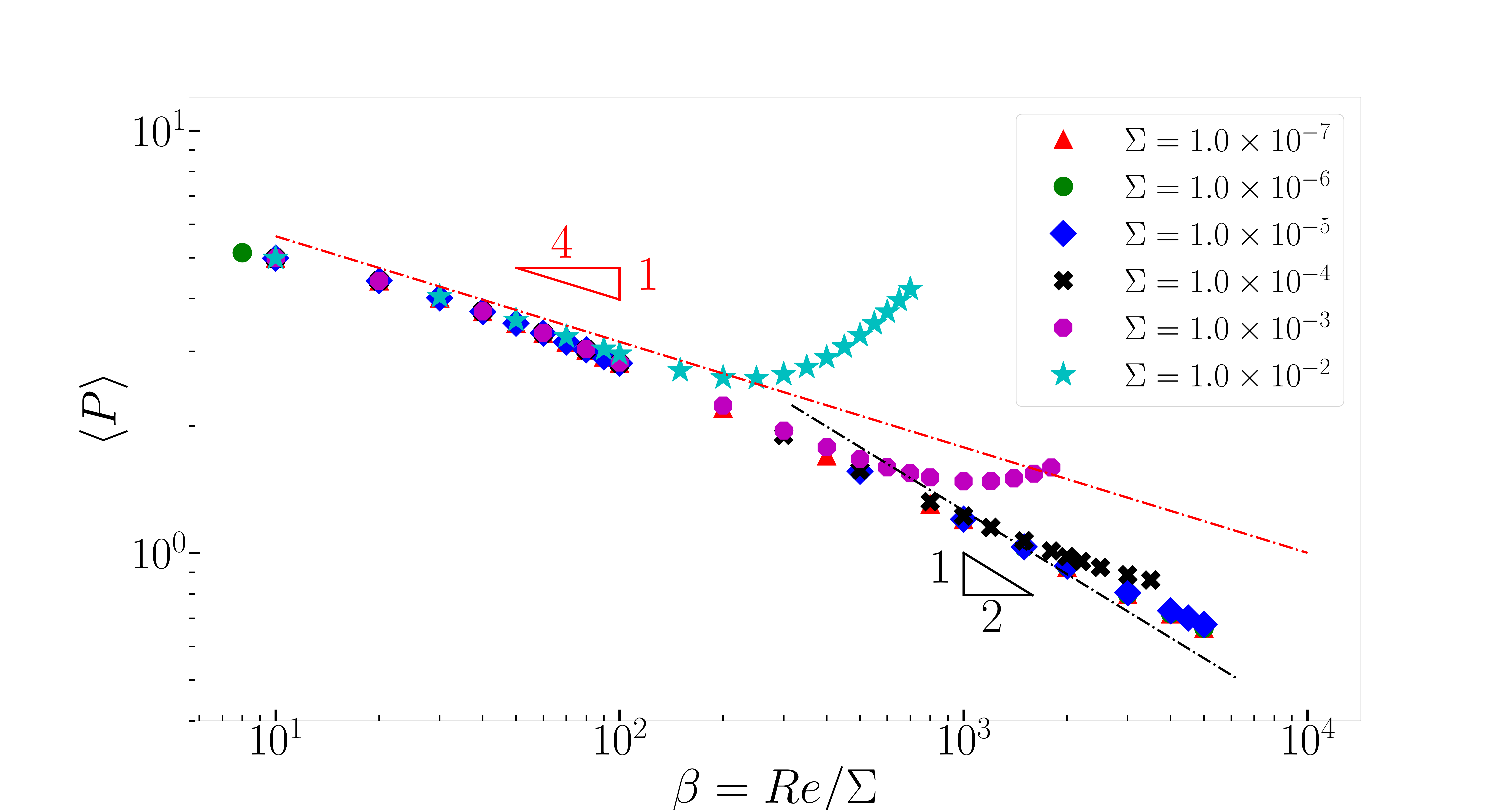}}
    \caption{(Color online) Scaling of (a) $H_{\mathrm{max}}$ and (b) $\langle P\rangle$ for the case of viscous--bending force balance. All regimes can be described in terms of the FSI parameter $\beta=Re/\Sigma$ in this case. The dash-dotted lines represent different slopes as shown. {The range of $Re$ is adjusted for each value of $\Sigma$ to keep $H_{\mathrm{max}}\leq 10$.}}
    \label{fig:viscous0alpha}
\end{figure}

\begin{figure}
    \centering
    \subfloat[]{\includegraphics[width=0.5\textwidth]{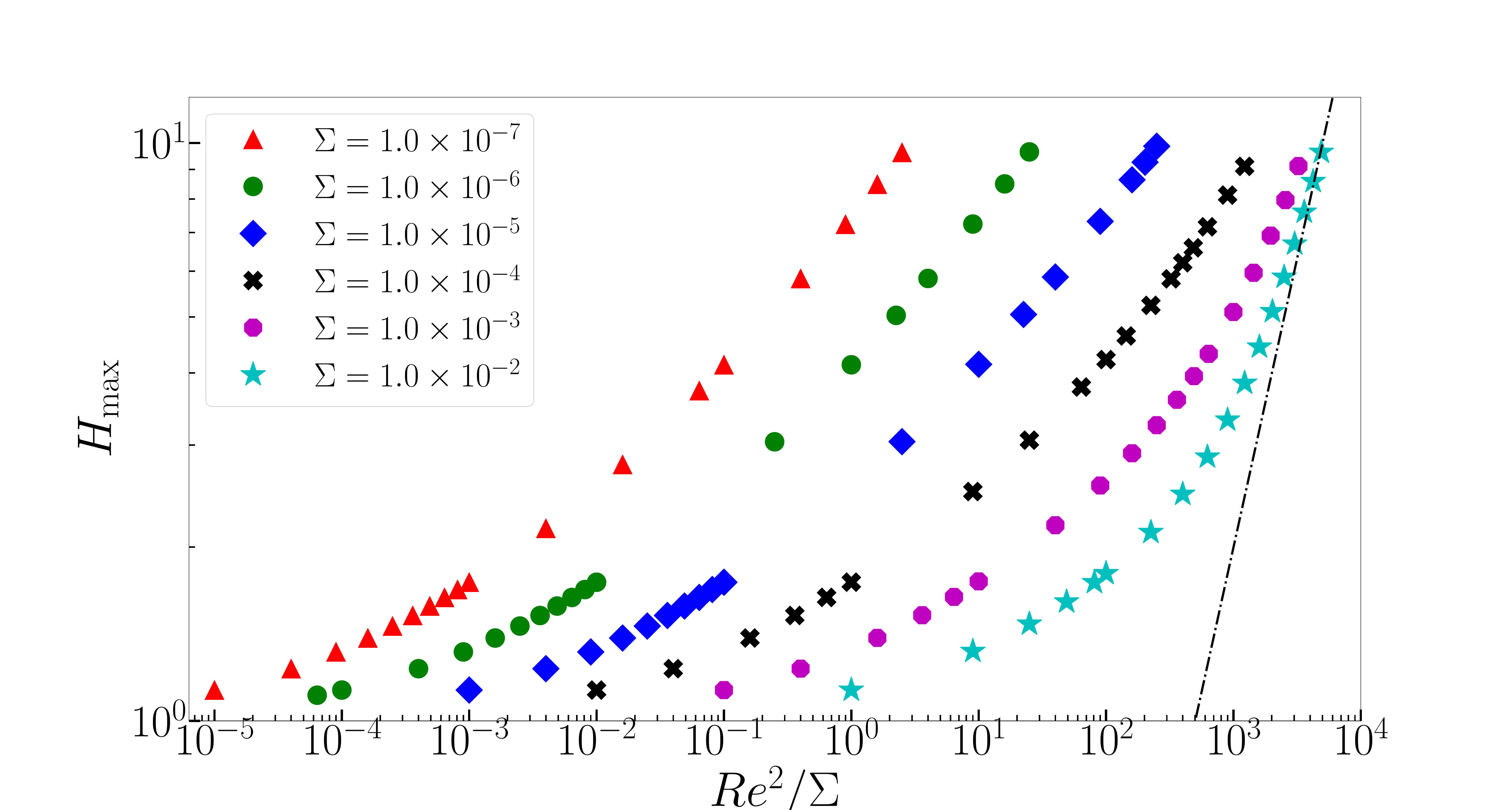}}
    \subfloat[]{\includegraphics[width=0.5\textwidth]{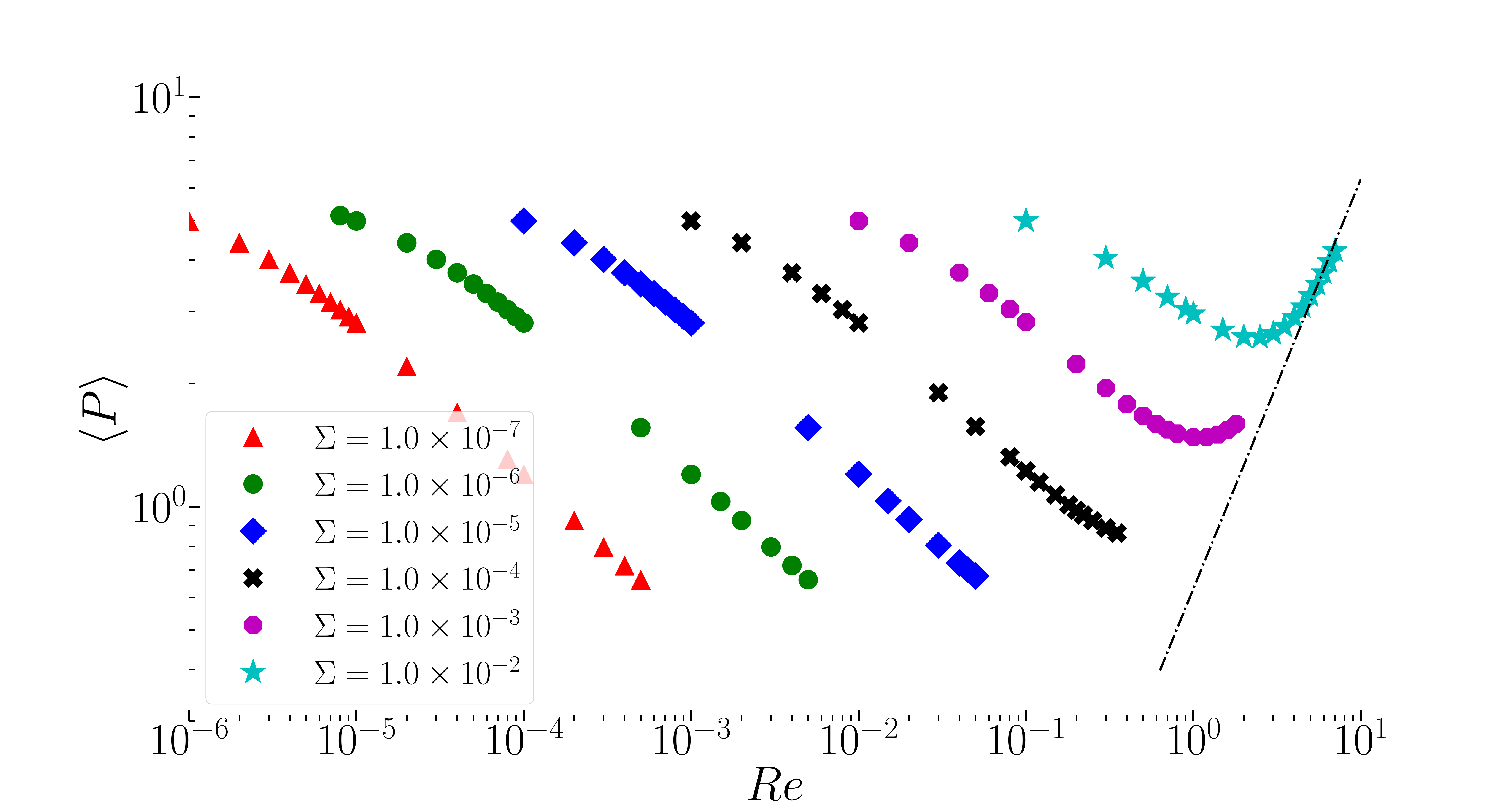}}
    \caption{(Color online) Scaling of (a) $H_{\mathrm{max}}$ and (b) $\langle P \rangle$ for the case of inertia--bending force balance. The dash-dotted lines represent a slope of 1. {The range of $Re$ is adjusted for each value of $\Sigma$ to keep $H_{\mathrm{max}}\leq 10$.}}
    \label{fig:inertia0alpha}
\end{figure}

\subsubsection{Scaling of $H_\mathrm{max}$: Bending and tension}\label{tensionscaling}

Now, we consider the beam equation \eqref{st0eq4} with bending and tension ($\alpha \ne 0$), which are expressed by the first and second term, respectively, and scale as $H_{\mathrm{max}}$ and $\alpha H_{\mathrm{max}}^3/\beta^2$ (recall that $H_{\mathrm{max}}\geq 1$), with $\alpha/\beta^2=18(h_{0f}/h_{0s})^2$. Varying the height ratio, $h_{0f}/h_{0s}$, will change the tension effects in the beam but it will not affect the classification of different regimes. Specifically, if $h_{0f}^2/h_{0s}^2\ll 1$, then the tension is negligible and the previous discussions for the pure bending case will apply. In the following analysis, we are interested in the tension-dominated regime and thus we fix  $h_{0f}/h_{0s}=1$, yielding $\alpha/\beta^2 = 18$, which ensures that tension is the dominant effect in the elastic response of the top wall.

In the case of the viscosity-dominated flow regime, $H_{\mathrm{max}}^3\sim {Re\langle P\rangle/\Sigma \sim Re H_{\mathrm{max}}^{-1}/\Sigma}$ leads to $H_{\mathrm{max}}\sim {(Re/\Sigma)}^{1/4}$ and $\langle P\rangle\sim {(Re/\Sigma)}^{-1/4}$ according to Eq.~\eqref{pviscous}. However, as discussed in Sect.~\ref{bendingscaling}, if the deformation is small, it is more appropriate to consider $H_{\mathrm{max}}^3\sim{(Re/\Sigma)} \mathrm{d}P/\mathrm{d}X\sim {(Re/\Sigma)}/H_{\mathrm{max}}^3$, indicating $H_{\mathrm{max}}\sim {(Re/\Sigma)}^{1/6}$ and $\langle P \rangle\sim {(Re/\Sigma)}^{-1/6}$. In Fig.~\ref{fig:viscous}, we show the results of six sets of data across six orders of magnitude of $\Sigma$. Again, the maximum deformation is strictly restricted to be less than 10\% of the channel length. The two predicted scaling regimes are clearly observed in Fig.~\ref{fig:viscous}. Outliers exist in the cases with relatively large $Re$, for which the viscous effects are no longer dominant.

On the other hand, for the regime with an  inertia--tension force balance, $\langle P \rangle\sim Re$ and $H_{\mathrm{max}}^3\sim {Re \langle P \rangle/\Sigma}\sim Re^2/\Sigma$, yielding $H_{\mathrm{max}}\sim Re^{2/3}/\Sigma^{1/3}$. Thanks to the tension effects suppressing the inflation of the microchannel, we are able to consider a larger range of $Re$ than the bending-dominated cases so that more cases are observed to reach this regime. As shown in Fig.~\ref{fig:inertial}, the last three data sets all collapse along the line of slope 1, as predicted by the proposed $H_{\mathrm{max}}$ and $\langle P \rangle$ scalings.

\begin{figure}
    \centering
    \subfloat[]{\includegraphics[width=0.5\textwidth]{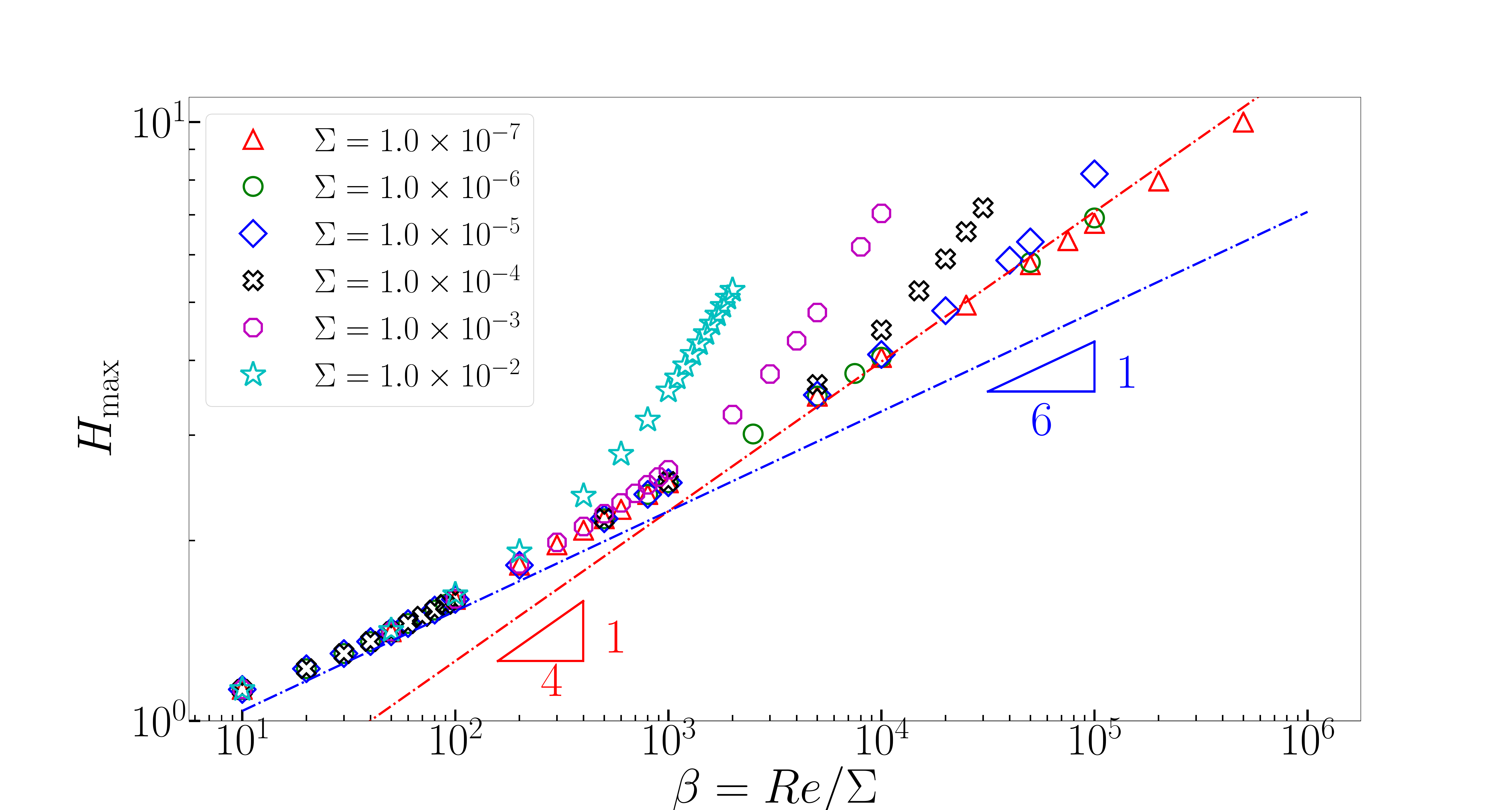}}
    \subfloat[]{\includegraphics[width=0.5\textwidth]{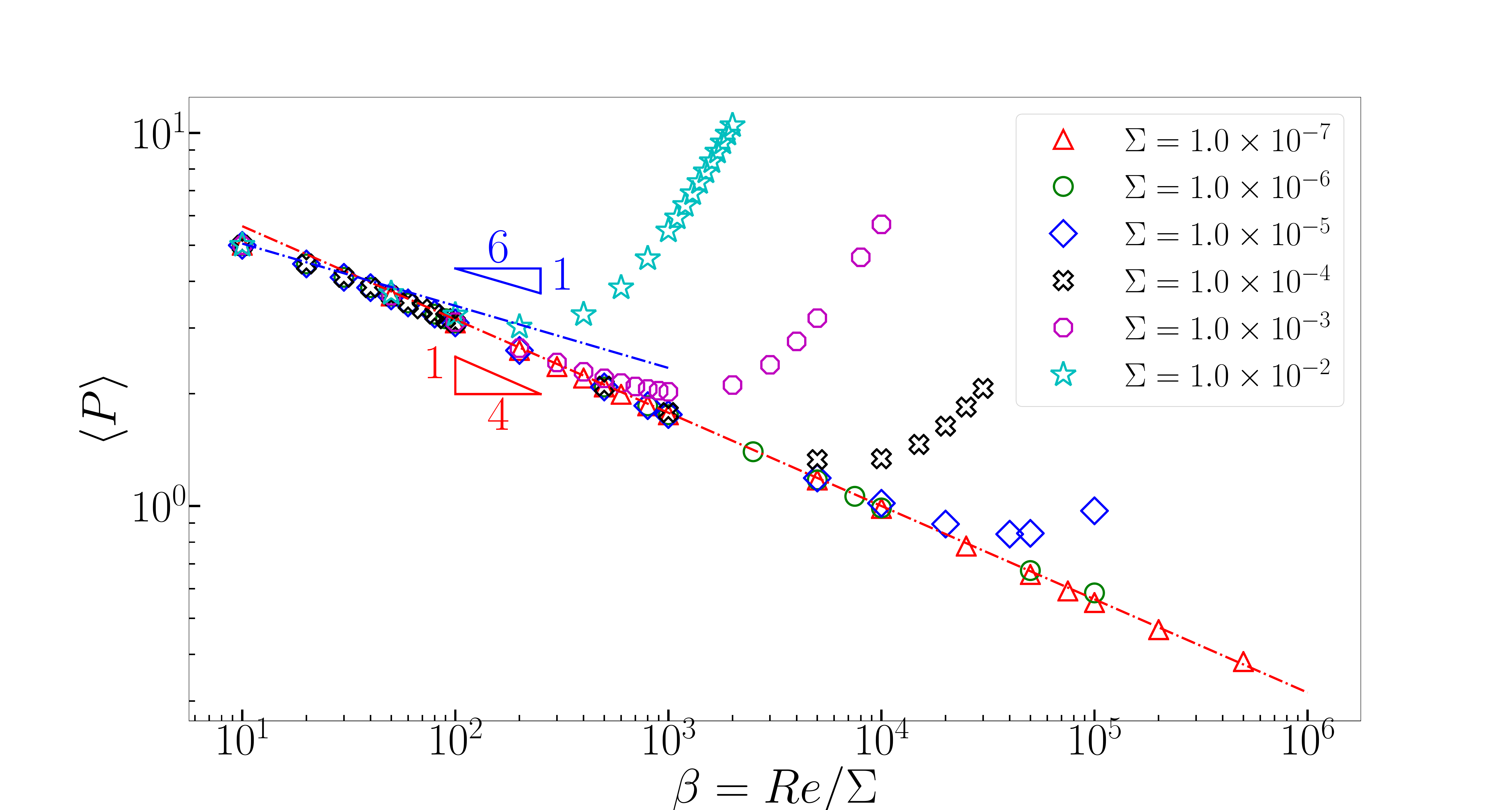}}
    \caption{(Color online) Scaling of (a) $H_{\mathrm{max}}$ and (b) $\langle P \rangle$ for the case of viscous--tension force balance. All regimes can be described in terms of the FSI parameter $\beta=Re/\Sigma$ in this case. The dash-dotted lines represent different slopes as shown. { The range of $Re$ is adjusted for each value of $\Sigma$ to keep $H_{\mathrm{max}}\leq 10$.}}
    \label{fig:viscous}
\end{figure}

\begin{figure}
    \centering
    \subfloat[]{\includegraphics[width=0.5\textwidth]{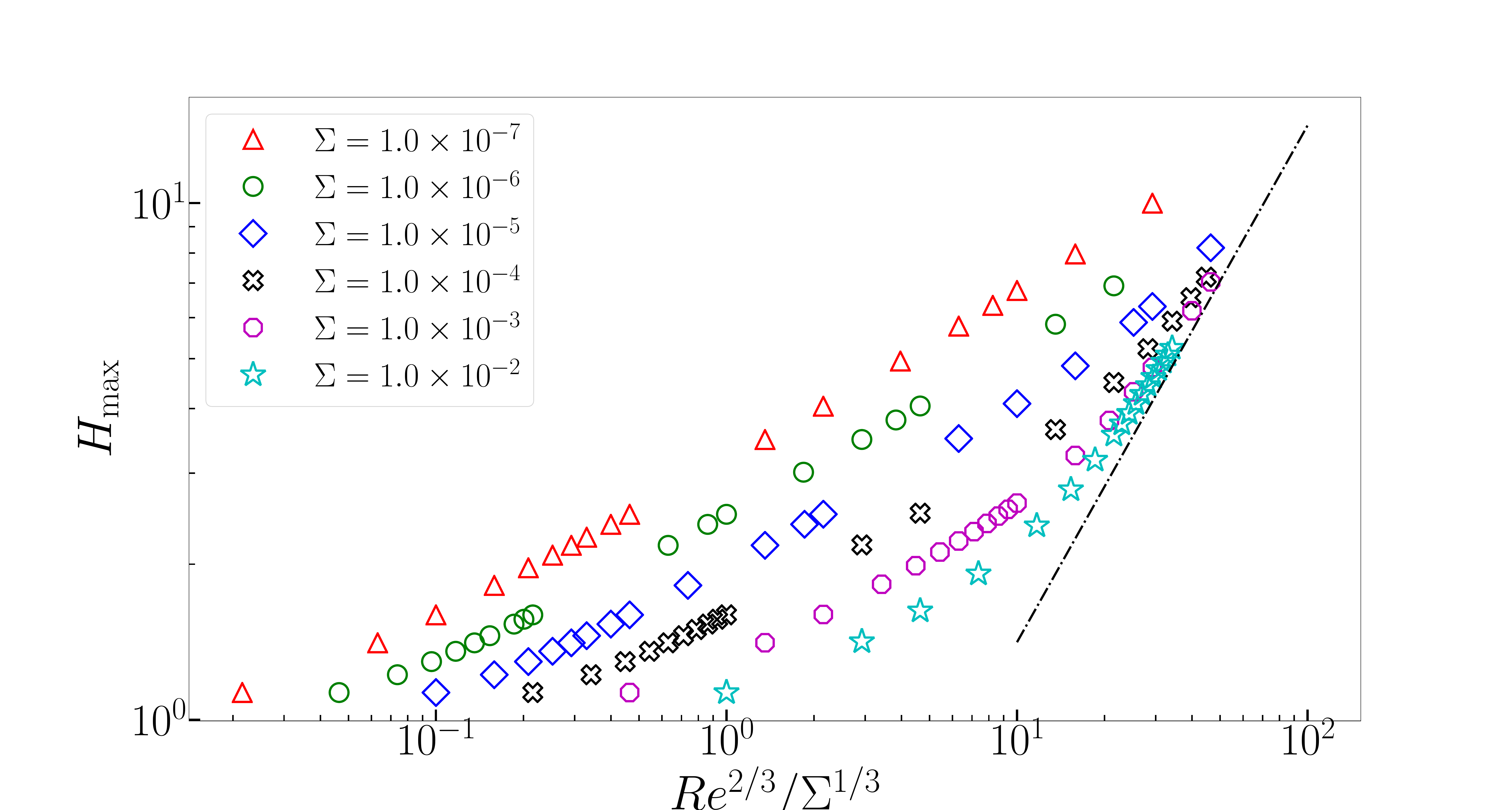}}
    \subfloat[]{\includegraphics[width=0.5\textwidth]{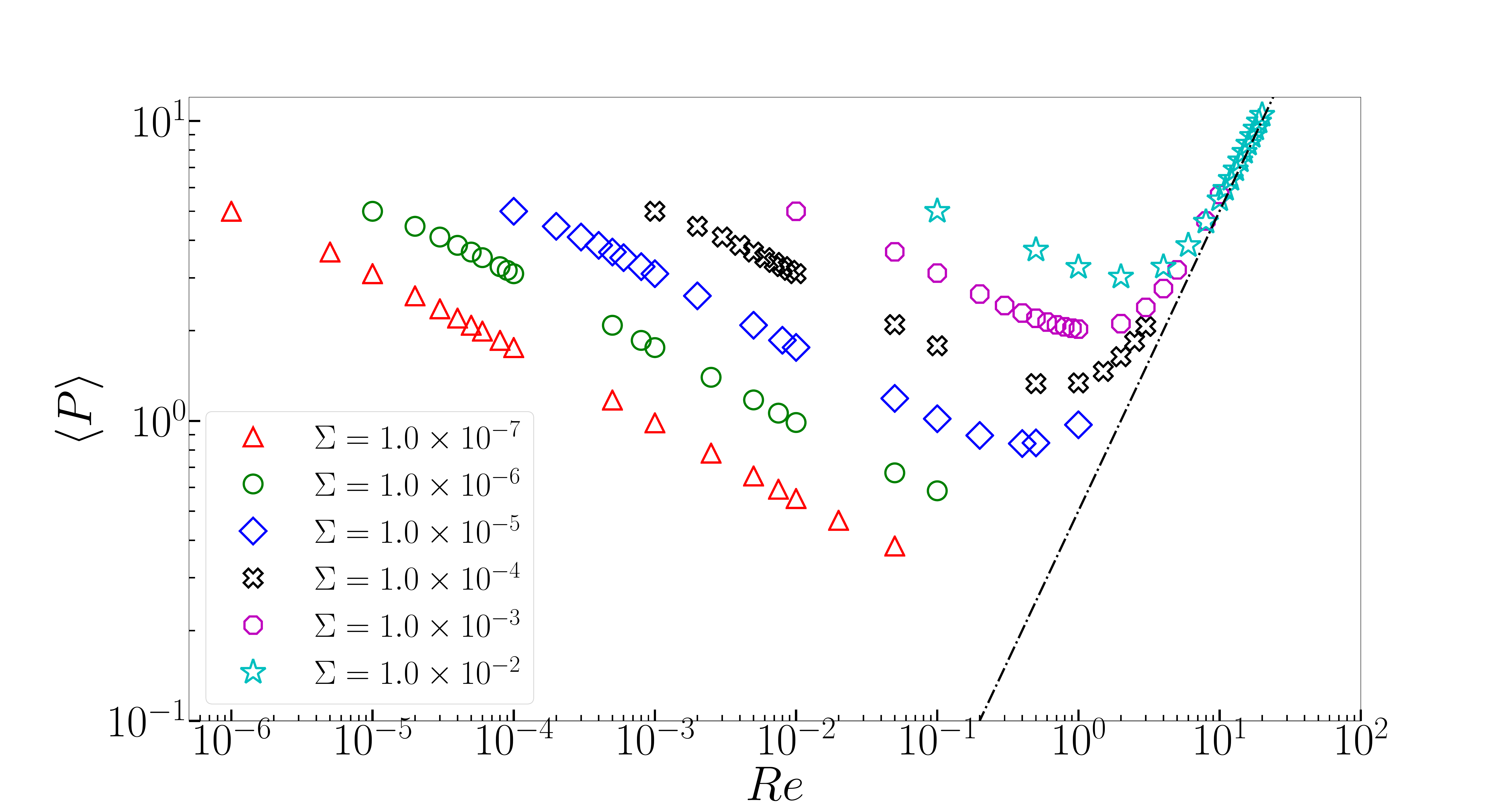}}
    \caption{(Color online) Scaling of (a) $H_{\mathrm{max}}$ and (b) $\langle P \rangle$ for the case of inertia--tension force balance. The dash-dotted lines in each panel indicate a slope of 1. { The range of $Re$ is adjusted for each value of $\Sigma$ to keep $H_{\mathrm{max}}\leq 10$.}}
    \label{fig:inertial}
\end{figure}

}


\section{Inflationary dynamics of the microchannel}
\label{sec:inflate}

{In this section, we discuss example outcomes of unsteady FSI simulations, using the numerical method introduced in Appendix~\ref{app:compute}, of the model derived in Sect.~\ref{sec:theory}. Specifically, as shown in Sect.~\ref{sec:steady}, the initially flat state is not a steady solution, therefore the channel's wall will deform until reaching the stable inflated steady state. The examples below are computed by fixing $\epsilon=0.01$, $\Sigma=9\times 10^{-4}$ and $h_{0f}/h_{0s}=1$ and varying $Re$. In other words, we are studying FSI in microchannel configurations that have in common the same slenderness and bending rigidity of the top wall. Note that changing $Re$ will necessarily lead to different $St$ and $\beta$ values. Thus, the nonlinear dynamics of the inflationary process are expected to depend strongly on the value of $Re$, an influence that we now proceed to interrogate.}

\subsection{Pure bending ($\alpha=0$)}
\label{subsec:bendingresult}

First, consider the case of pure bending, in which tension is neglected by setting $\alpha=0$. {In this subsection, we present simulation results for $Re=0.5$ (corresponding to $St=6.0$) and $Re=1.8$ (corresponding to $St=1.67$). Note that the dimensionless parameters used in this subsection are the same as for steady states without tension presented in Figs.~\ref{fig:steadyPlot} and \ref{fig:steadyPlot2}, respectively.} The results in this section are representative of the unsteady FSI dynamics produced by the model derived in Sect.~\ref{sec:theory} without tension. 

\begin{figure}[ht]
\subfloat[\label{fig:pq_0alphaRe50}]{\includegraphics[width=0.49\textwidth]{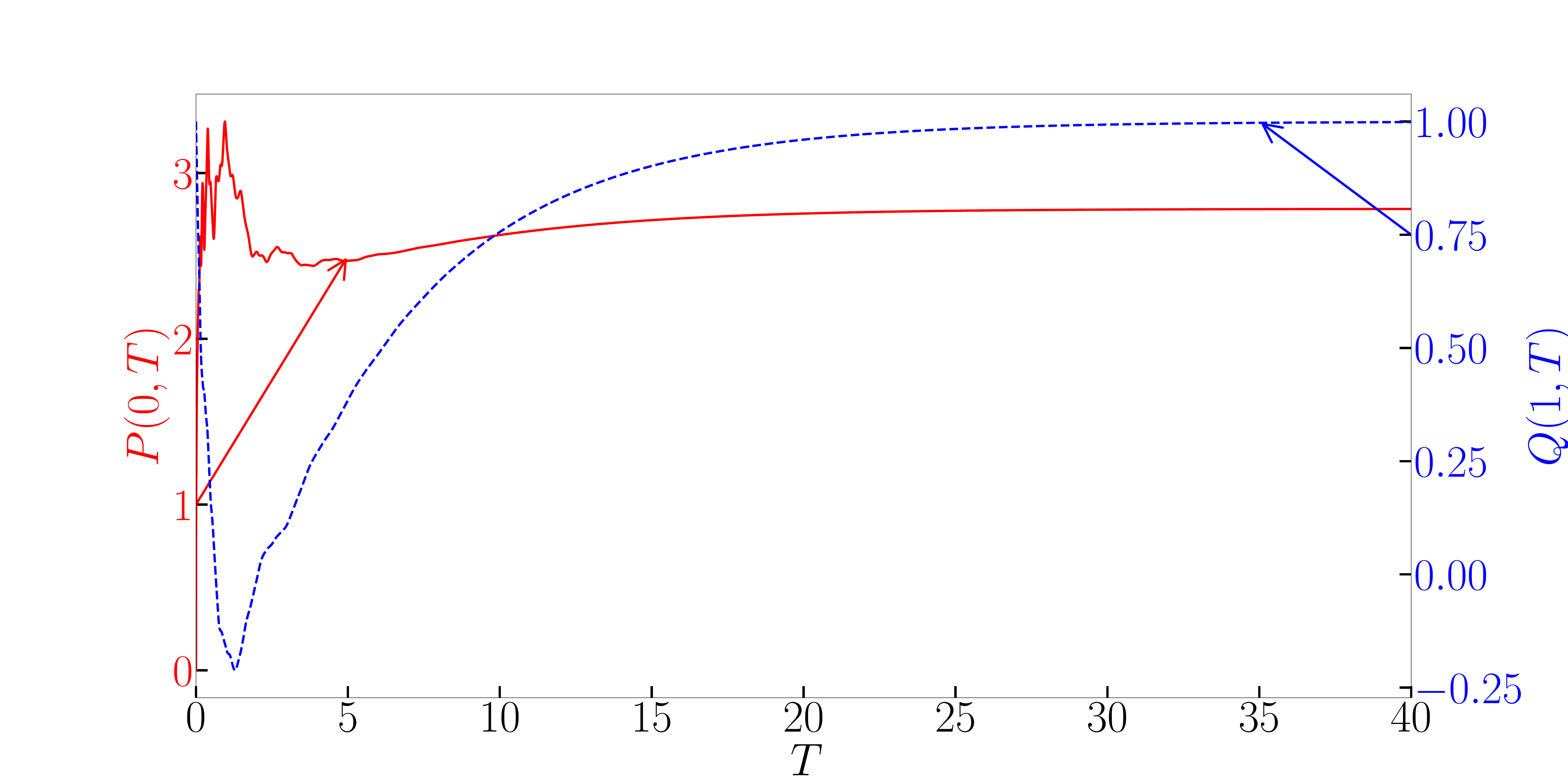}}
\hfill
\subfloat[\label{fig:Havg_0alphaRe50}]{\includegraphics[width=0.49\textwidth]{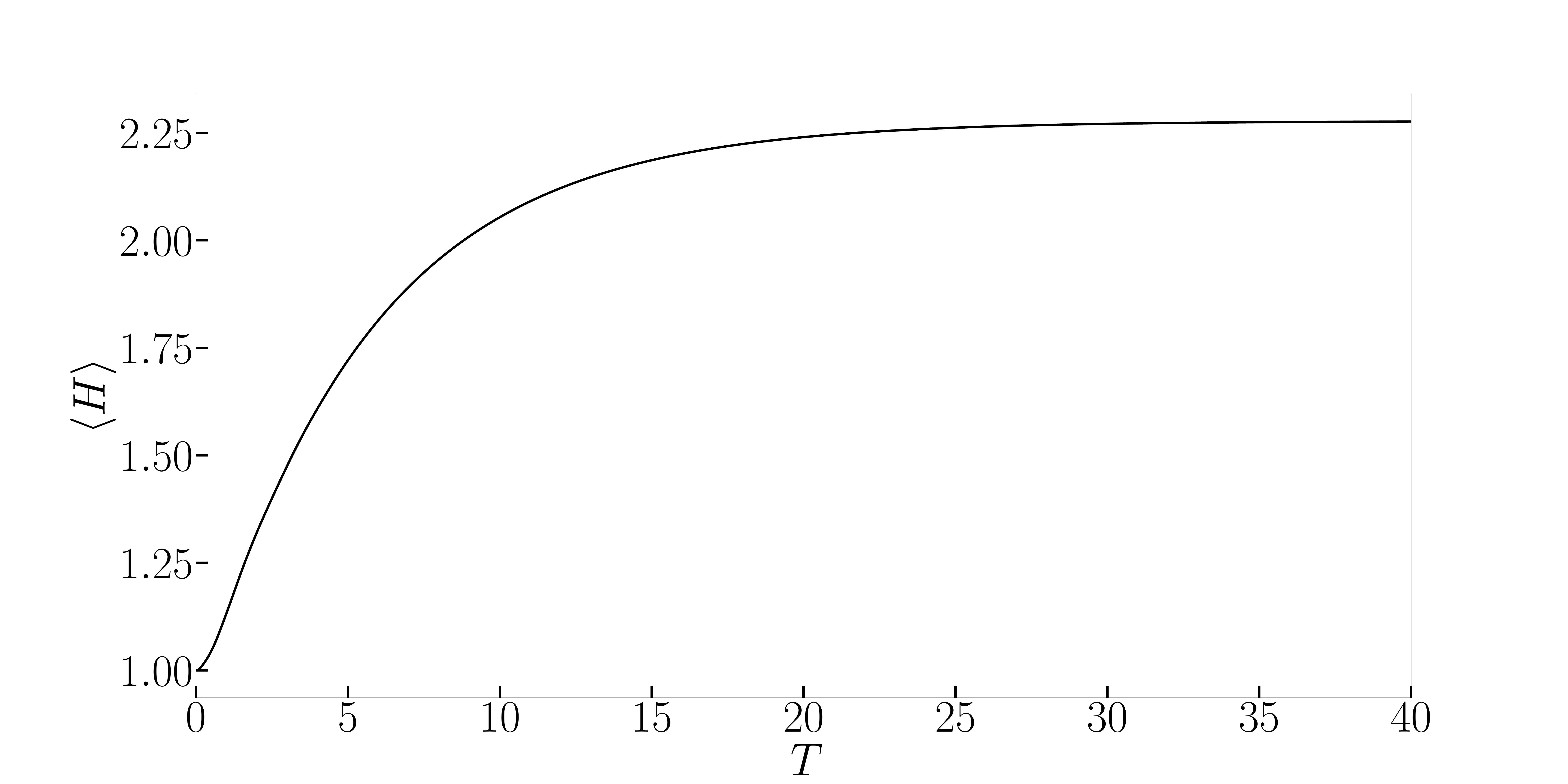}}
\caption{(Color online) Example time histories of (a) the inlet pressure $P(0,T)$ and outlet flow rate $Q(1,T)$, and (b) the axially-averaged deformation, ${\langle H\rangle(T) = \int_0^1 H(X,T)\, \mathrm{d}X}$; $\alpha=0$, ${Re=0.5}$, $St=6.0$, $\Sigma =9.0\times 10^{-4} $ and $\beta = 5.56\times 10^2$. The arrows indicate which curve corresponds to which axis.}
\label{fig:pqu_bendRe50}
\end{figure}

{The results for $Re=0.5$ (``low'' Re) are shown in Fig.~\ref{fig:pqu_bendRe50}. It can be seen that, after a violent initial transient (from $T=0$ to $T\approx 5$) in the fluid, the FSI reaches a steady state gradually.}  The axially-average (over $X\in[0,1]$) height of the deformed channel $\langle H \rangle$, the inlet pressure $P(0,T)$ and the outlet flow rate $Q(1,T)$ all achieve steady values by the end of the simulation at $T=40$. Specifically, the outlet flow rate $Q(1,T)$ reaches 1, which is the imposed inlet boundary condition. As discussed in Sect.~\ref{sec:scalings}, the steady-state values of $\langle H \rangle$ and $P(0,T)$ depend in a nontrivial way on the dimensionless parameters $Re$ and $\Sigma$. A video of the time evolution of the shape of the microchannel (specifically, the top wall), together with a reconstruction of the parabolic velocity profile under the von K\'arm\'an--Polhausen approximation is available in \footnote{See the Supplemental Material at [URL will be inserted by publisher] for the video {\tt notensionRe50.mp4} showing the time evolution of the shape of the microchannel $\alpha=0$ and $Re=0.5$.}.

\begin{figure}
\subfloat[\label{fig:pq_0alphaRe180}]{\includegraphics[width=0.49\textwidth]{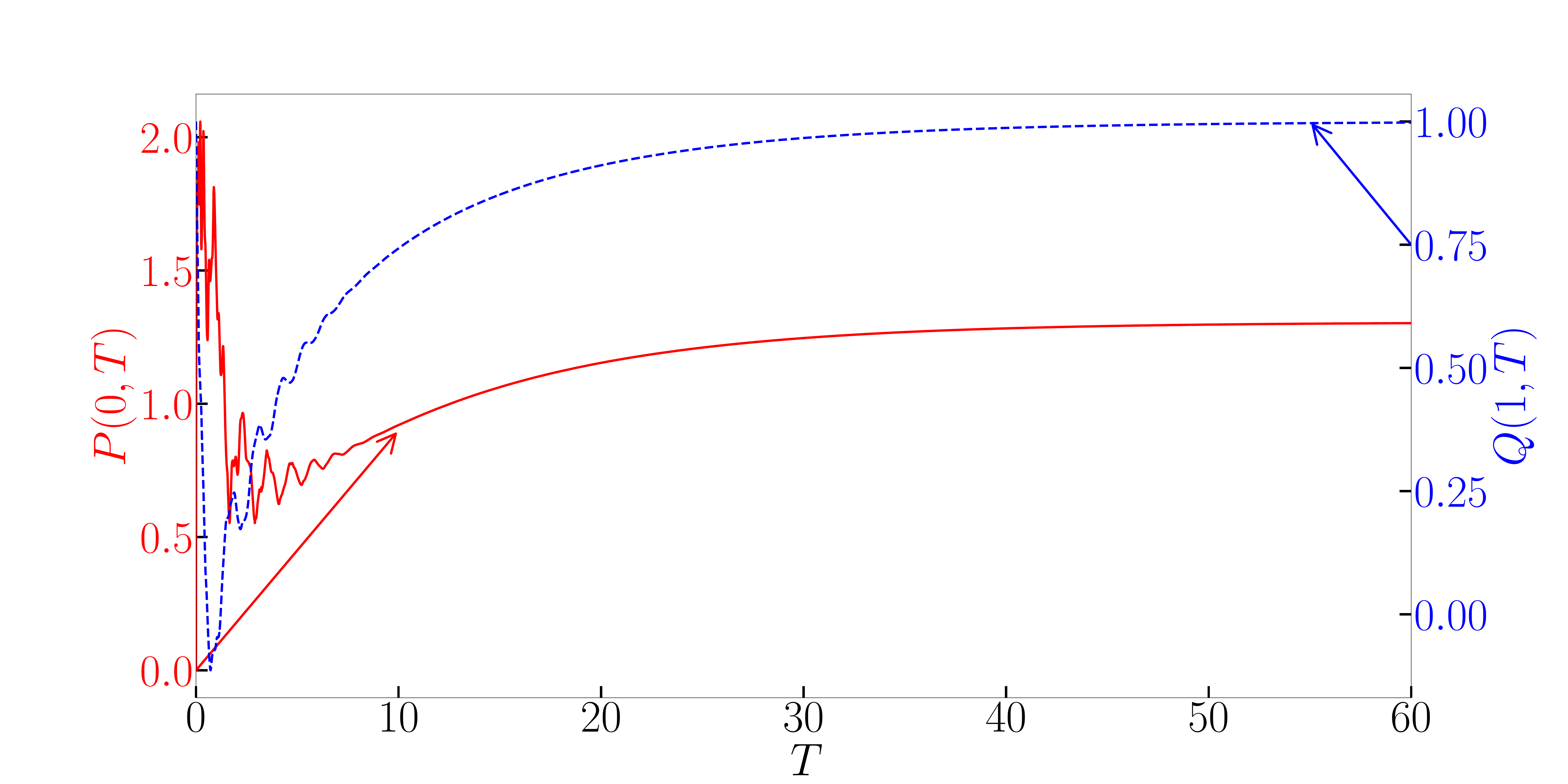}}
\hfill
\subfloat[\label{fig:Havg_0alphaRe180}]{\includegraphics[width=0.49\textwidth]{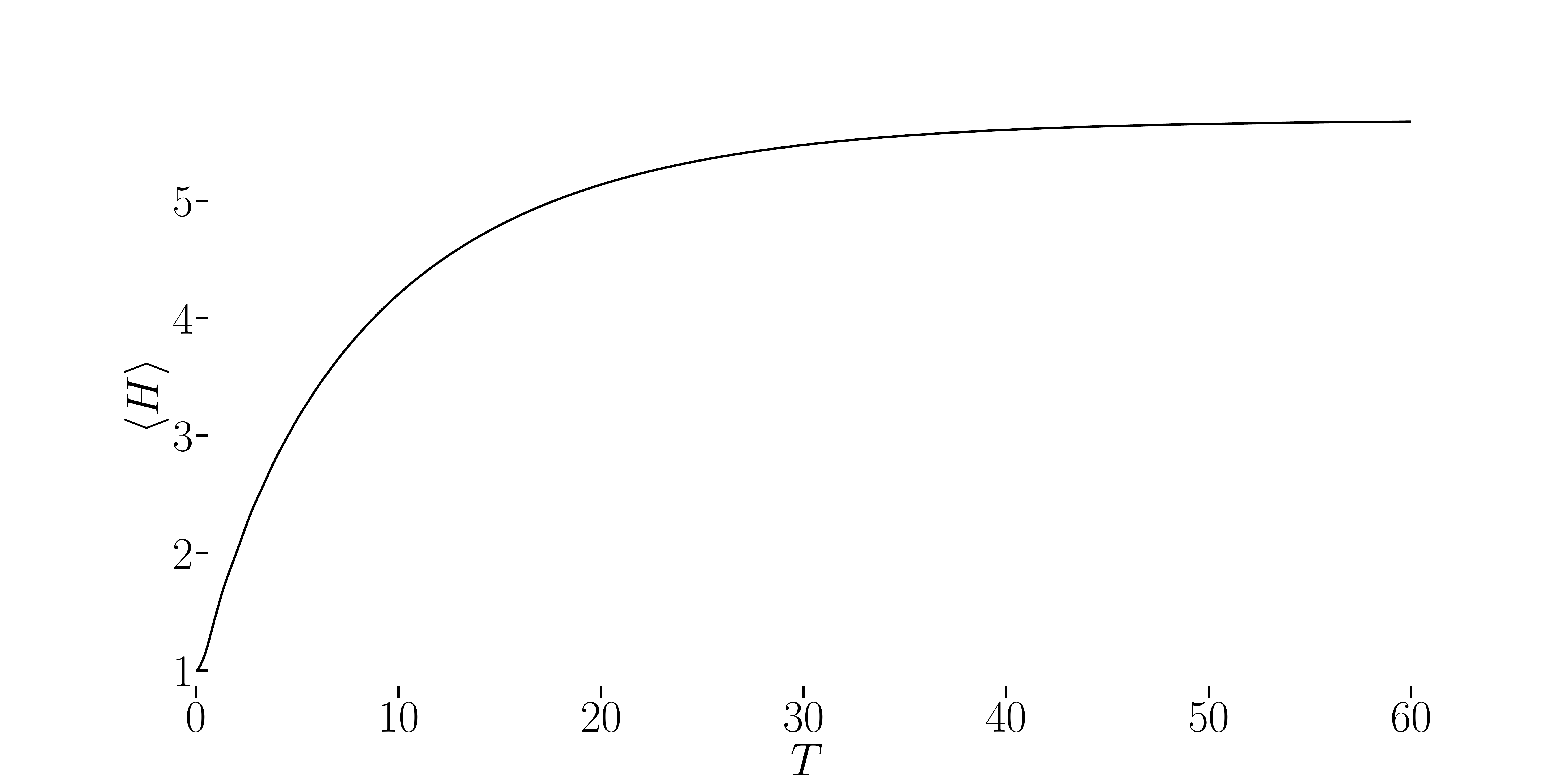}}
\caption{(Color online) Example time histories of (a) the inlet pressure $P(0,T)$ and outlet flow rate $Q(1,T)$, and (b) the axially-averaged deformation, ${\langle H\rangle(T) = \int_0^1 H(X,T)\, \mathrm{d}X}$; $\alpha=0$, ${Re=1.8}$, $St=1.67$, $\Sigma =9.0\times 10^{-4} $ and $\beta = 2.0\times 10^3$.  The arrows indicate which curve corresponds to which axis.}
\label{fig:pqu_bendRe180}
\end{figure}

{ In Fig.~\ref{fig:pqu_bendRe180}, we show the results for $Re=1.8$ (``moderate'' Re) and $St=1.67$. Increasing $Re$ necessarily decreases $St$, if the other parameters are kept fixed. Compared with the ``low'' Re case, it takes longer for the FSI to reach steady state. However, the oscillatory transient response only happens in the initial period, from $T=0$ to $T\approx10$, then the top wall displays a relatively slow inflation until it reaches the steady state. Due to the strong inertial effects (with positive pressure gradient upstream), the top wall is highly inflated and thus a larger average deformation is achieved in Fig.~\subref*{fig:Havg_0alphaRe180} (compared to Fig.~\subref*{fig:Havg_0alphaRe50}). The final steady state shape of the microchannel is exactly the same as shown in Fig.~\ref{fig:steadyPlot2}. A video of the time evolution of the inflation of the top wall is available in \footnote{See the Supplemental Material at [URL will be inserted by publisher] for the video {\tt notensionRe180.mp4} showing the time evolution of the shape of the microchannel $\alpha=0$ and $Re=1.8$.}.}

\subsection{Bending and tension ($\alpha \neq 0$)}

Now consider the full solid model with bending and tension. In this subsection, we discuss the results for { two example simulations with $Re=0.5$ (``low'' $Re$) and $Re=10$ (``moderate'' $Re$) respectively. Again, the parameters chosen here are the same as those of the bending and tension cases in Figs.~\ref{fig:steadyPlot} and \ref{fig:steadyPlot2}, respectively.} 

\begin{figure}
\subfloat[\label{fig:pqRe50}]{\includegraphics[width=0.5\textwidth]{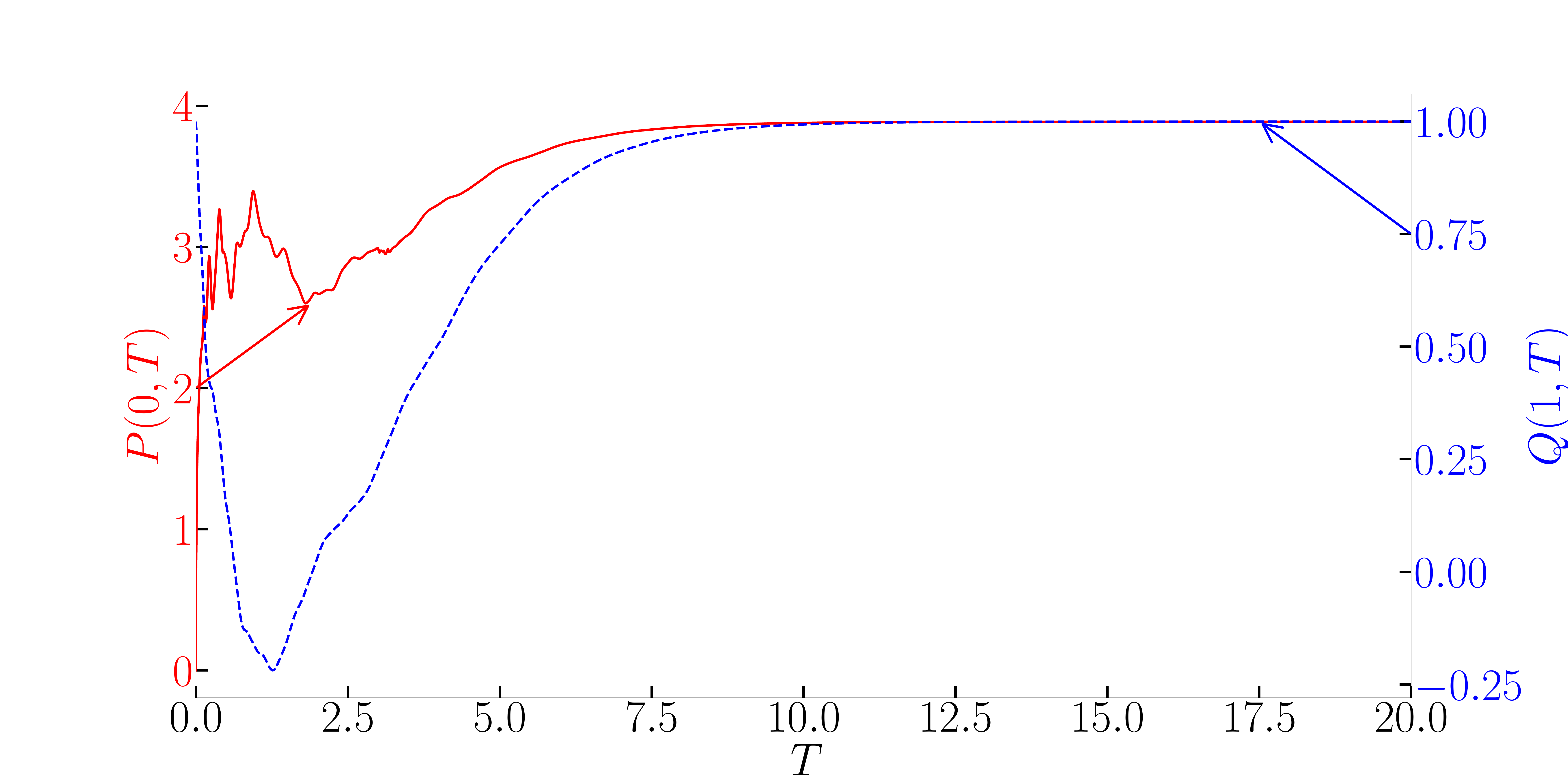}}
\hfill
\subfloat[\label{fig:HavgRe50}]{\includegraphics[width=0.5\textwidth]{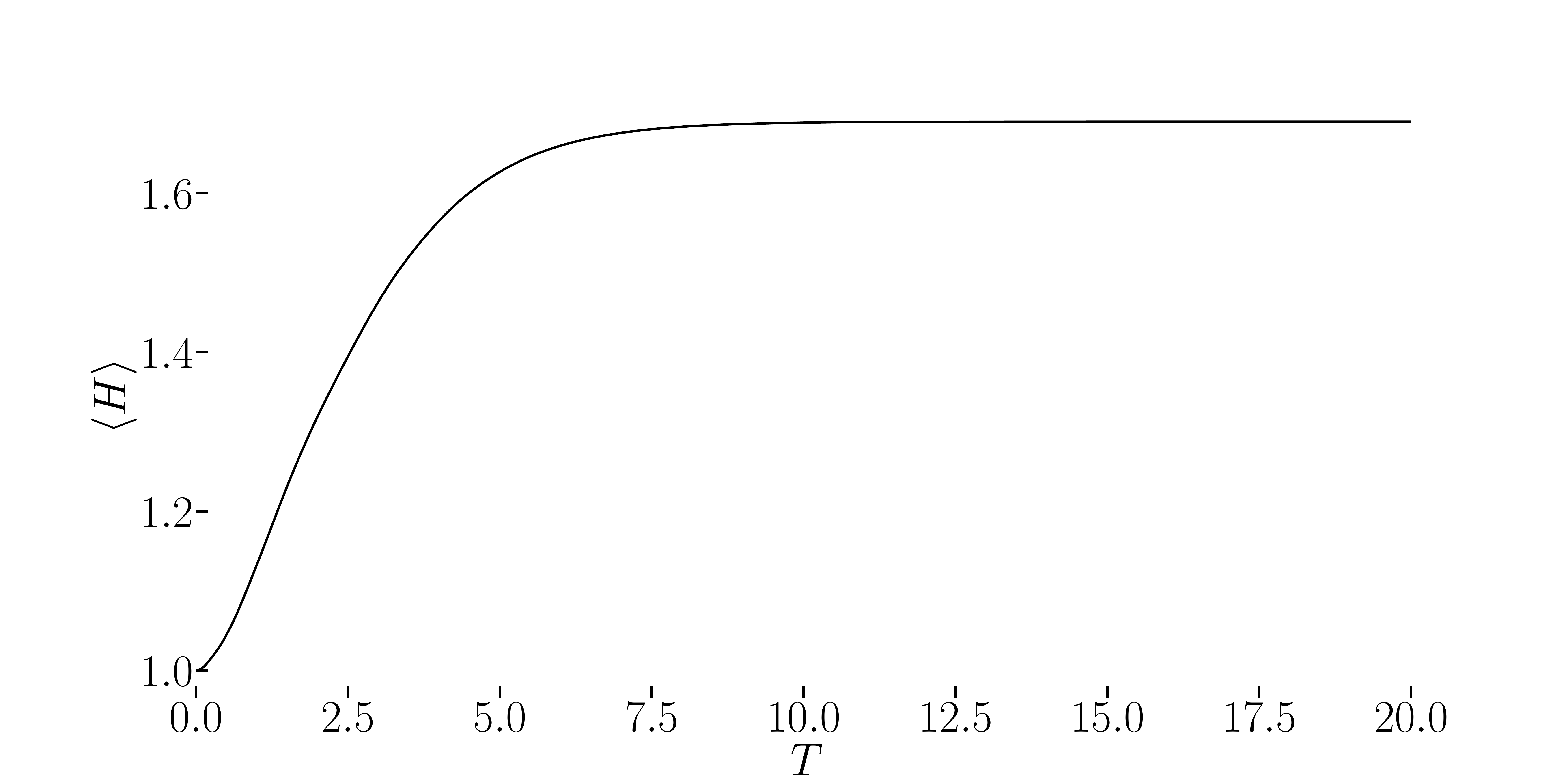}}
\caption{(Color online) Example time histories of (a) the inlet pressure $P(0,T)$ and outlet flow rate $Q(1,T)$, and (b) the axially-averaged deformation, ${\langle H\rangle(T) = \int_0^1 H(X,T)\, \mathrm{d}X}$ ; $Re=0.5$, $St=6$, $\Sigma = 9\times 10^{-4}$, $\alpha=5.56\times 10^6$, and $\beta = 5.56\times 10^2$. The arrows indicate which curve corresponds to which axis.}
\label{fig:pqH_tensionRe50}
\end{figure}

{The results of $Re=0.5$ ($St=6$) are shown in Fig.~\ref{fig:pqH_tensionRe50}.} Similar to the pure-bending case, the FSI reaches a steady state, as can be seen in Fig.~\ref{fig:pqH_tensionRe50}, after a complex initial transient response (from $T=0$ to {$T\approx 10$}) in the fluid. The final average deformation in Fig.~\subref*{fig:HavgRe50} is, however, smaller than the final average deformation in Fig.~\subref*{fig:Havg_0alphaRe50}. This decrease is expected because, now, tension also serves to resist deformation, along with bending. As a result, the steady-state pressure value in Fig.~\subref*{fig:pqRe50} is higher than that in Fig.~\subref*{fig:pq_0alphaRe50} because { the pressure gradient} is inversely proportional to the cube of the channel height in lubrication theory, and tension decreases the deformation (thus height). A video of the inflation of the top wall is available in \footnote{See the Supplemental Material at [URL will be inserted by publisher] for the video {\tt tensionRe50.mp4} showing the time evolution of the shape of the microchannel $\alpha\ne0$ and $Re=0.5$.}.

\begin{figure}
\subfloat[\label{fig:pqRe1000}]{\includegraphics[width=0.5\textwidth]{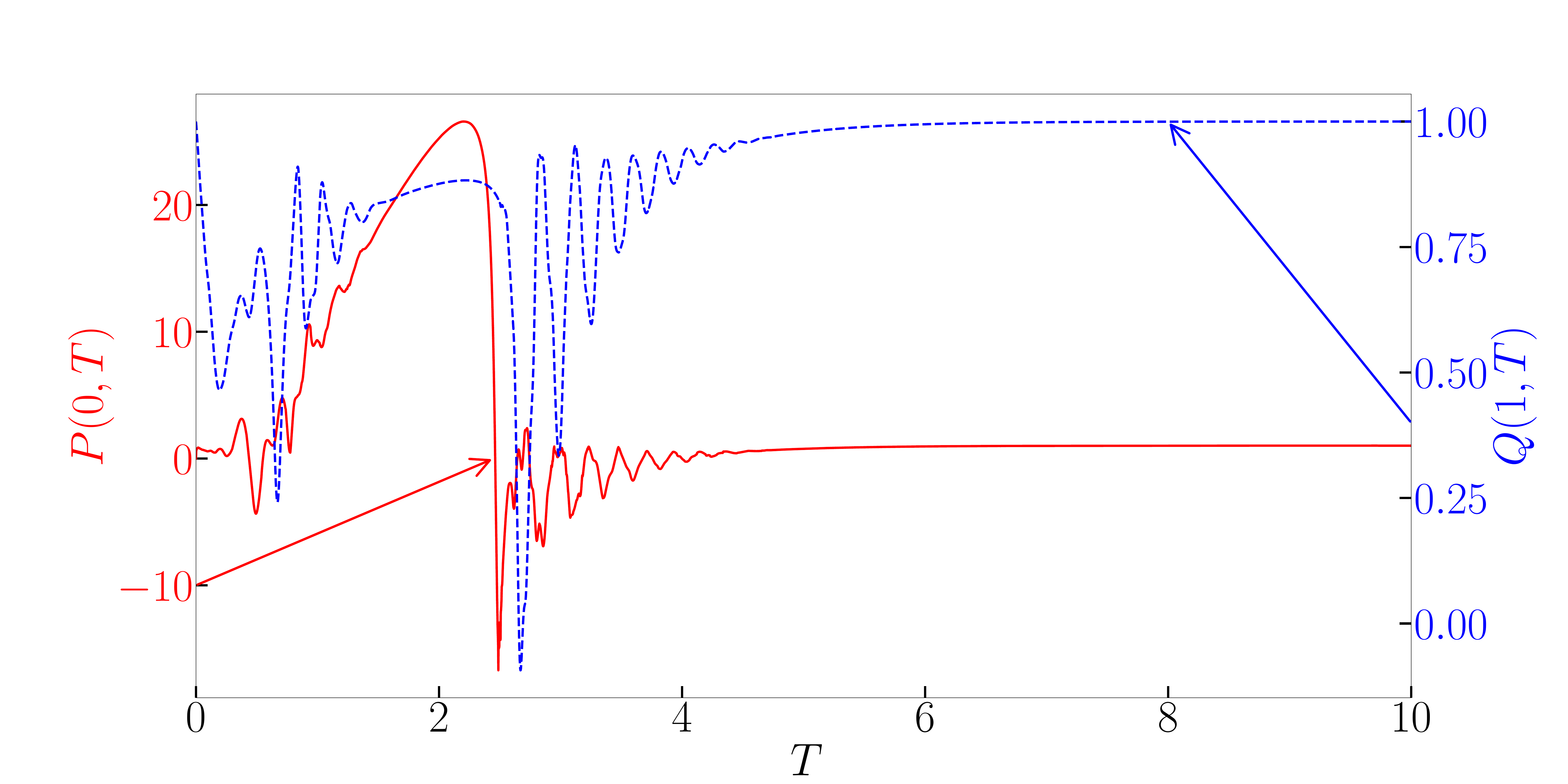}}
\hfill
\subfloat[\label{fig:HavgRe1000}]{\includegraphics[width=0.5\textwidth]{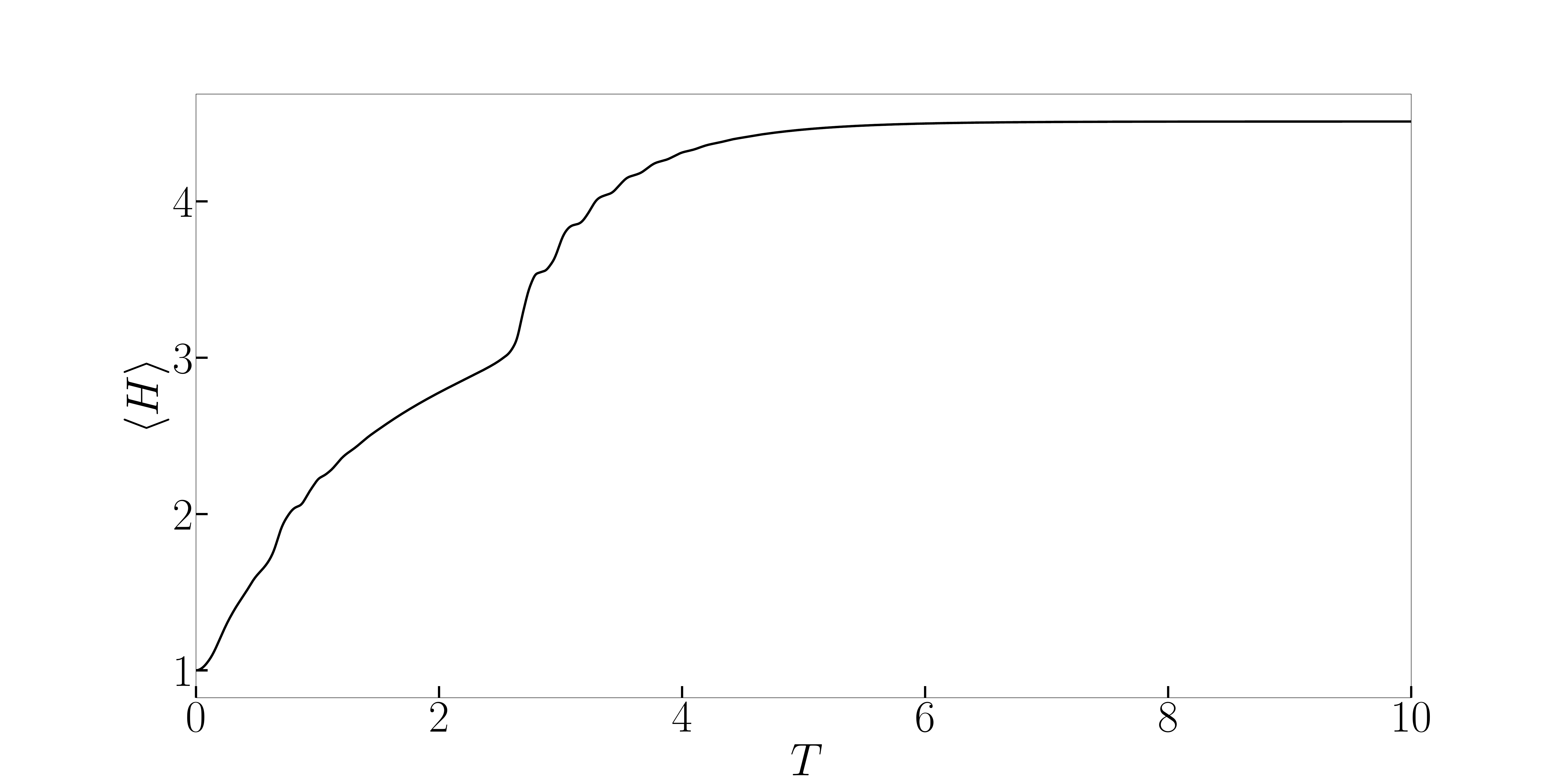}}
\caption{(Color online) Example time histories of (a) $P(0,T)$ and $Q(1,T)$, and (b) ${\langle H\rangle(T) = \int_0^1 H(X,T)\, \mathrm{d}X}$, for bending and tension ($\alpha\ne0$) at ``moderate'' $Re=10$; $St=0.3$, $\Sigma = 9\times 10^{-4}$, $\alpha=2.22\times 10^9$, and $\beta = 1.11\times 10^4$. The arrows indicate which curve corresponds to which axis.}
\label{fig:pqH_tensionRe1000}
\end{figure}

\begin{figure}
\subfloat[$T=2.45$\label{fig:H_Re1000_timeshot1}]{\includegraphics[width=0.5\textwidth]{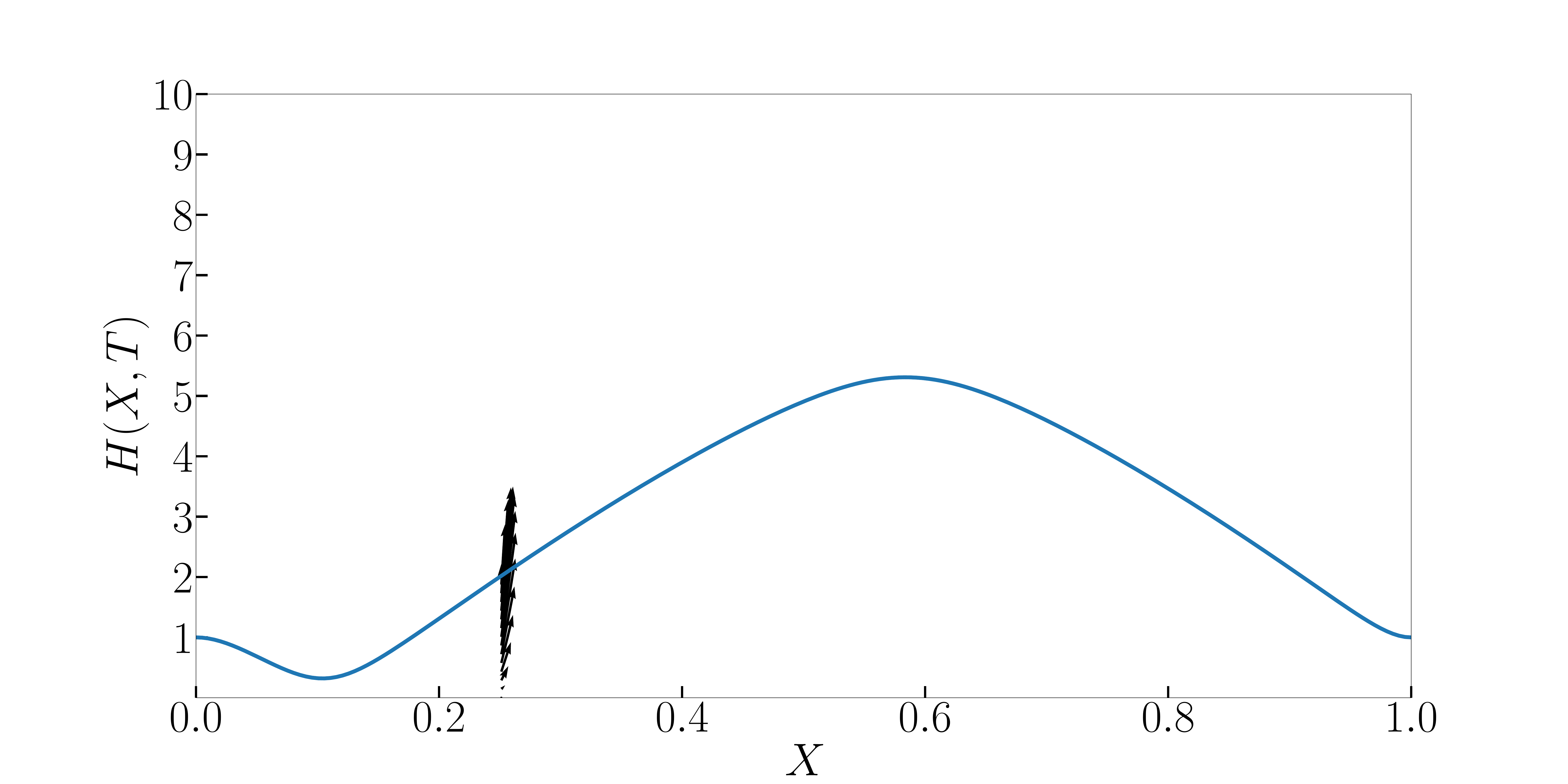}}
\hfill
\subfloat[$T=9.95$\label{fig:H_Re1000_timeshot2}]{\includegraphics[width=0.5\textwidth]{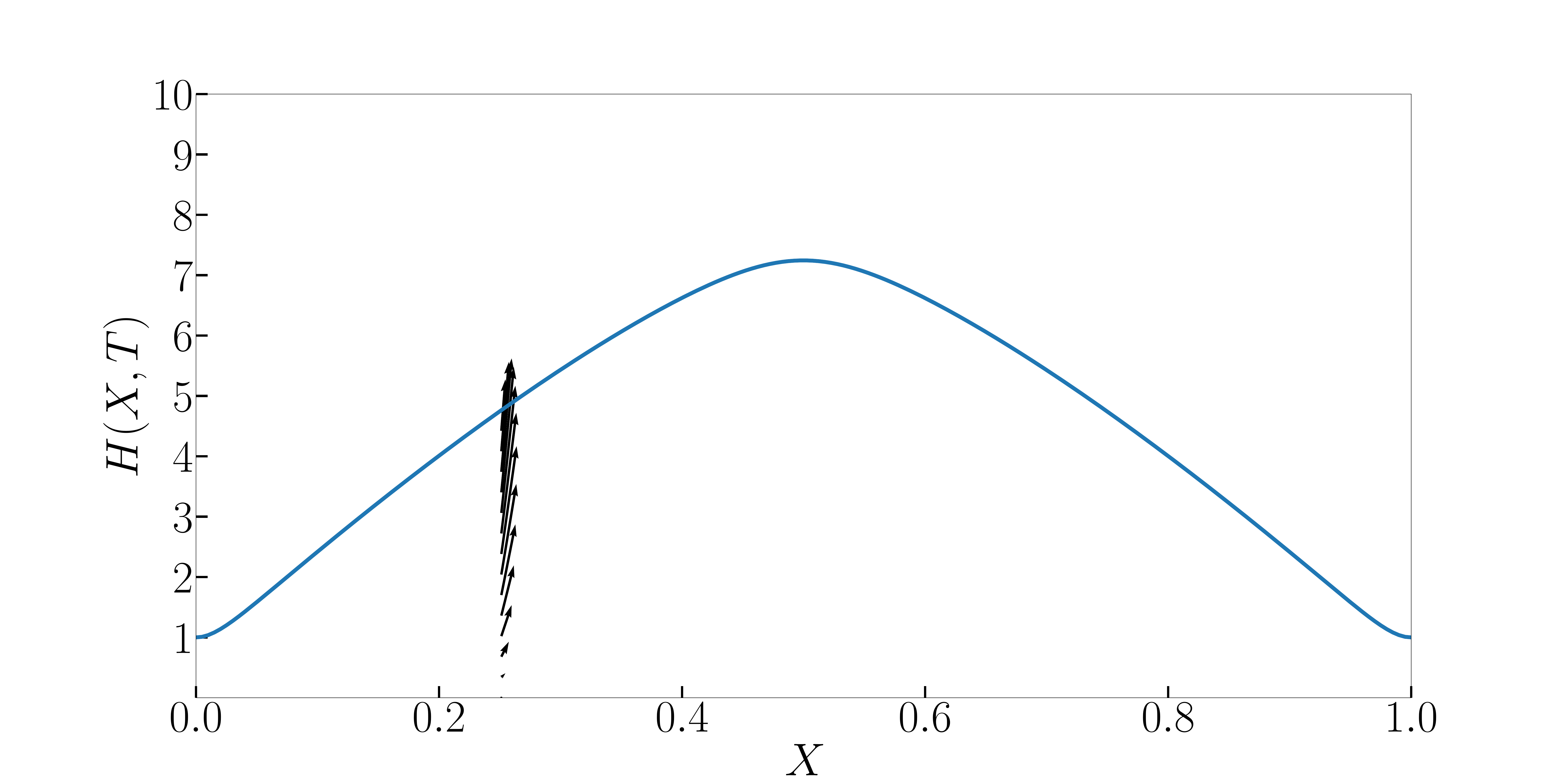}}
\caption{Height of the deformed microchannel and an exemplar flow profile (at a specific cross-section within it, i.e., fixed $X$) for $\alpha \ne 0$ and {$Re=10$}. Panel (a) shows the intermediate ``buckled'' state, while panel (b) shows the final inflated shape. Note that these plots are in \emph{dimensionless} variables, hence the deflection appears exaggerated; the aspect ratio between the dimensional axes is, of course, $\epsilon \ll 1$.}
\label{fig:H_T_tension_1000}
\end{figure}

{ As for the case of $Re=10$ ($St=0.3$), the results are quite different from previous cases, as shown in Fig.~\ref{fig:pqH_tensionRe1000}. First, large amplitude oscillations are observed in both the inlet pressure and the outlet flow rate, indicating more violent initial transient. Second, there is a short ``intermediate'' stage where the inlet pressure reaches a maximum and the outlet flow rate is almost flat, and correspondingly, the growth of the average deformed height, $\langle H\rangle$, slows down. After that, both the inlet pressure and the outlet flow rate drop sharply, followed by oscillations, while the slope of $\langle H\rangle$ increases abruptly before reaching the steady state. 

Figure~\ref{fig:H_T_tension_1000} shows two example snapshots of the time evolution of the shape of fluid domain (the top wall deformation) together with a reconstruction of the parabolic velocity profile under the von K\'arm\'an--Polhausen approximation. Prior to the ``intermediate'' stage mentioned above, a portion of the channel near the inlet is actually collapsed, while the rest of the channel is inflated, as shown in Fig.~\subref*{fig:H_Re1000_timeshot1}. Qualitatively, the ``intermediate'' state resembles a buckling mode of a beam. Interestingly, the position of the collapse does not remain fixed but propagates upstream as the inlet pressure keeps increasing. Then, the solid ``snaps'' into the inflated shape as shown in Fig.~\subref*{fig:H_Re1000_timeshot2}, and the inlet pressure goes down as only the middle portion of the channel is significantly inflated (see Fig.~\subref*{fig:H_Re1000_timeshot2}). A video of the inflation of the top wall is available in \footnote{See the Supplemental Material at [URL will be inserted by publisher] for the video {\tt tensionRe1000.mp4} showing the time evolution of the shape of the microchannel $\alpha\ne0$ and $Re=10$.}.}

\section{Linear stability of the deformed microchannel}
\label{sec:linear}

\subsection{Perturbation about the steady shape}

{As noted above, the flat state $Q,H=const.$, analyzed in a number of prior works on FSI, is \emph{not} relevant to the microchannel problem under consideration because it is not a solution of the steady problem. Indeed, it is easy to show that Eq.~\eqref{st0eq5} has \emph{no} finite constant solutions that also satisfy the boundary conditions in Eq.~\eqref{ssH0BC}. Thus, we are interested in the stability of the \emph{deformed} steady state in the presence of bending and tension of the top wall.} To understand the stability of this non-flat steady state, we perturb about $Q=1$ and $H=H_0(X)$ [i.e., the solution of Eqs.~\eqref{st0eq5} and \eqref{ssH0BC}] as follows:
\begin{subequations}\begin{align}
Q(X,T) &= 1 + \delta {{Q_1(X)\mathrm{e}^{-\mathrm{i}\sigma T}}},\\
H(X,T) &= H_0(X) + \delta {{H_1(X)\mathrm{e}^{-\mathrm{i}\sigma T}}},
\end{align}\label{eq:lin_perturb}\end{subequations}
where $\delta\ll 1$ is the (arbitrary, dimensionless) amplitude of a small perturbation { and $\sigma\in\mathbb{C}$ denotes the growth/decay rate of the perturbations}. { The boundary conditions at both ends are already satisfied by the steady-state solution $(Q_0=1,H_0)^\top$, thus the perturbation $(Q_1,H_1)^\top$  must satisfy \emph{homogeneous} boundary conditions. Specifically, the perturbation should satisfy the boundary conditions from Eq.~\eqref{eq:fluid_bc}, as well as the clamped boundary conditions from Eq.~\eqref{solidbc}. In other words,
\begin{subequations}\label{stability_bc}\begin{align}
    &Q_1|_{X=0} = 0,\qquad\qquad \left.\frac{\mathrm{d} Q_1}{\mathrm{d} X}\right|_{X=0}=\left.\frac{\mathrm{d} Q_1}{\mathrm{d} X}\right|_{X=1} = 0,
    \label{Q1_bc}\\
    &H_1|_{X=0}=\left.\frac{\mathrm{d} H_1}{\mathrm{d} X}\right|_{X=0}=\left.\frac{\mathrm{d} H_1}{\mathrm{d} X}\right|_{X=1}=H_1|_{X=1} = 0, \qquad\qquad \left.\frac{\mathrm{d}^4 H_1}{\mathrm{d} X^4}\right|_{X=1}=0.
\label{H1_bc}
\end{align}\end{subequations}
Note the second relation in Eq.~\eqref{Q1_bc} is the natural consequence of Eq.~\eqref{cont7}, taking into account that the deformation is restricted at both ends of the microchannel by clamping. Meanwhile the last relation in Eq.~\eqref{H1_bc} is the boundary condition that enforces a gauge outlet pressure.}

To determine the growth/decay of the perturbation, we must derive a set of linear evolution equations for $Q_1$ and $H_1$. To this end, we substitute Eqs.~\eqref{eq:lin_perturb} into the governing set of Eqs.~\eqref{solidmodel3}, \eqref{cont7}, \eqref{xmom8} and \eqref{betaeq2}, using the fact that $H_0(X)$ satisfies Eq.~\eqref{st0eq5} and dropping all terms of $\mathcal{O}(\delta^2)$ or higher. The result is two linear evolution equations in which the coefficients depend on the steady-state solution $H_0(X)$ and its derivatives:
\begin{subequations}\begin{align}
&\frac{\mathrm{d}^5H_1}{\mathrm{d} X^5}-\frac{\alpha}{\beta^2}\frac{\mathrm{d}}{\mathrm{d} X}\left[\left(\frac{\mathrm{d} H_0}{\mathrm{d} X}\right)^2\frac{\mathrm{d}^2 H_1}{\mathrm{d} X^2}+2\frac{\mathrm{d}^2H_0}{\mathrm{d} X^2}\frac{\mathrm{d} H_0}{\mathrm{d} X}\frac{\mathrm{d} H_1}{\mathrm{d} X}\right] = { \frac{\sigma}{\mathrm{i} St}\frac{\mathrm{d}^2 Q_1}{\mathrm{d} X^2}} \nonumber\\
&\qquad\qquad -\beta \Bigg\lbrace { -\frac{\mathrm{i} Re St}{H_0}\sigma Q_1} + { Re}\frac{6}{5}\left[\frac{3H_1}{H_0^4}\frac{\mathrm{d} H_0}{\mathrm{d} X}-\frac{1}{H_0^3}\frac{\mathrm{d} H_1}{\mathrm{d} X}
-\frac{2Q_1}{H_0^3}\frac{\mathrm{d} H_0}{\mathrm{d} X}+\frac{2}{H_0^2}\frac{\mathrm{d} Q_1}{\mathrm{d} X}\right] 
+12\left(-\frac{3H_1}{H_0^4}+\frac{Q_1}{H_0^3}\right) \Bigg\rbrace,
\label{eqlin_H1}\\
&\frac{\mathrm{d} Q_1}{\mathrm{d} X}-{\mathrm{i}St\sigma H_1}=0.
\label{eqlin_cont1}
\end{align}\label{eq:linearized}\end{subequations}
{Equations~\eqref{eq:linearized} can be written in the matrix form
\begin{equation}\label{stability_matrix}
\underbrace{\begin{pmatrix}
\displaystyle\frac{\mathrm{d}}{\mathrm{d}X} & 0\\[3mm]
\mathcal{L}_Q & \mathcal{L}_H
\end{pmatrix}}_{=\bm{A}}
\underbrace{\begin{pmatrix}
Q_1\\
H_1
\end{pmatrix}}_{=\bm{\psi}}
=\sigma
\underbrace{\begin{pmatrix}
0 & \mathrm{i} St\\[3mm]
\displaystyle\frac{1}{\mathrm{i} St}\frac{\mathrm{d}^2}{\mathrm{d}X^2}+\frac{\mathrm{i} \beta Re St}{H_0} & 0
\end{pmatrix}}_{=\bm{B}}
\begin{pmatrix}
Q_1\\
H_1
\end{pmatrix}
,
\end{equation}
where we have defined the operators
\begin{subequations}\begin{align}
    \mathcal{L}_H &= \frac{\mathrm{d}^5}{\mathrm{d}X^5}-\frac{\alpha}{\beta^2}\Bigg\lbrace4\frac{\mathrm{d}H_0}{\mathrm{d}X}\frac{\mathrm{d}^2H_0}{\mathrm{d}X^2}\frac{\mathrm{d}^2}{\mathrm{d}X^2}+\left[2\left(\frac{\mathrm{d}^2H_0}{\mathrm{d}X^2}\right)^2+2\frac{\mathrm{d}H_0}{\mathrm{d}X}\frac{\mathrm{d}^3H_0}{\mathrm{d}X^3}\right]\frac{\mathrm{d}}{\mathrm{d}X}+\left(\frac{\mathrm{d}H_0}{\mathrm{d}X}\right)^2\frac{\mathrm{d}^3}{\mathrm{d}X^3}\Bigg\rbrace-\frac{6\beta Re}{5H_0^3}\frac{\mathrm{d}}{\mathrm{d}X}\nonumber\\
    &\phantom{=}-\beta\left(\frac{36}{H_0^4}-\frac{18Re}{5H_0^4}\frac{\mathrm{d}H_0}{\mathrm{d}X}\right)\label{LH}, \\
    \mathcal{L}_Q &= -\beta\left(-\frac{12}{H_0^3}+\frac{12Re}{5H_0^3}\frac{\mathrm{d}H_0}{\mathrm{d}X}-\frac{12Re}{5H_0^2}\frac{\mathrm{d}}{\mathrm{d}X}\right)\label{LQ}.
\end{align}\end{subequations}

Equation~\eqref{stability_matrix} and the boundary conditions in Eqs.~\eqref{stability_bc} constitute a generalized eigenvalue problem $\bm{A}\bm{\psi} = \sigma\bm{B}\bm{\psi}$, with $\sigma$ as the eigenvalue and $\bm{\psi}=(Q_1,H_1)^T$ as the eigenfunction. Note the system in Eq.~\eqref{stability_matrix} has non-constant coefficients due to the non-flat steady-state shape $H_0(X)$ of the microchannel. We say the system is linearly unstable if $\imag(\sigma)>0$, and we proceed to investigate whether this condition holds (or does not hold).}

\subsection{Linear stability of modal perturbations}

{ We use the Chebyshev pseudospectral method \cite{SH01,Boyd00} to numerically solve the generalized eigenvalue problem $\bm{A}\bm{\psi} = \sigma\bm{B}\bm{\psi}$. The details of the numerical approach are presented in Appendix~\ref{app:cheb_stability}. Note} that Eqs.~\eqref{eq:linearized} do \emph{not} give rise to an autonomous system with a self-adjoint matrix operator $\bm{A}$ due to the non-uniform base state (see, e.g., the discussion in \cite{DT03} in the context of thin-film lubrication). Therefore, issues of transient growth and non-modal analysis arise \cite{DT03,SRDM18}. For the present purposes, we are just interested in the \emph{asymptotic} stability of the inflated steady-states, so it suffices to consider the eigenspectrum of $\bm{A}$ for different parameters, and determine the possibility of eigenvalues with positive imaginary part.

\begin{figure}
\subfloat[$Re=0.5$: pure bending.\label{fig:re50bending}]{\includegraphics[width=0.5\textwidth]{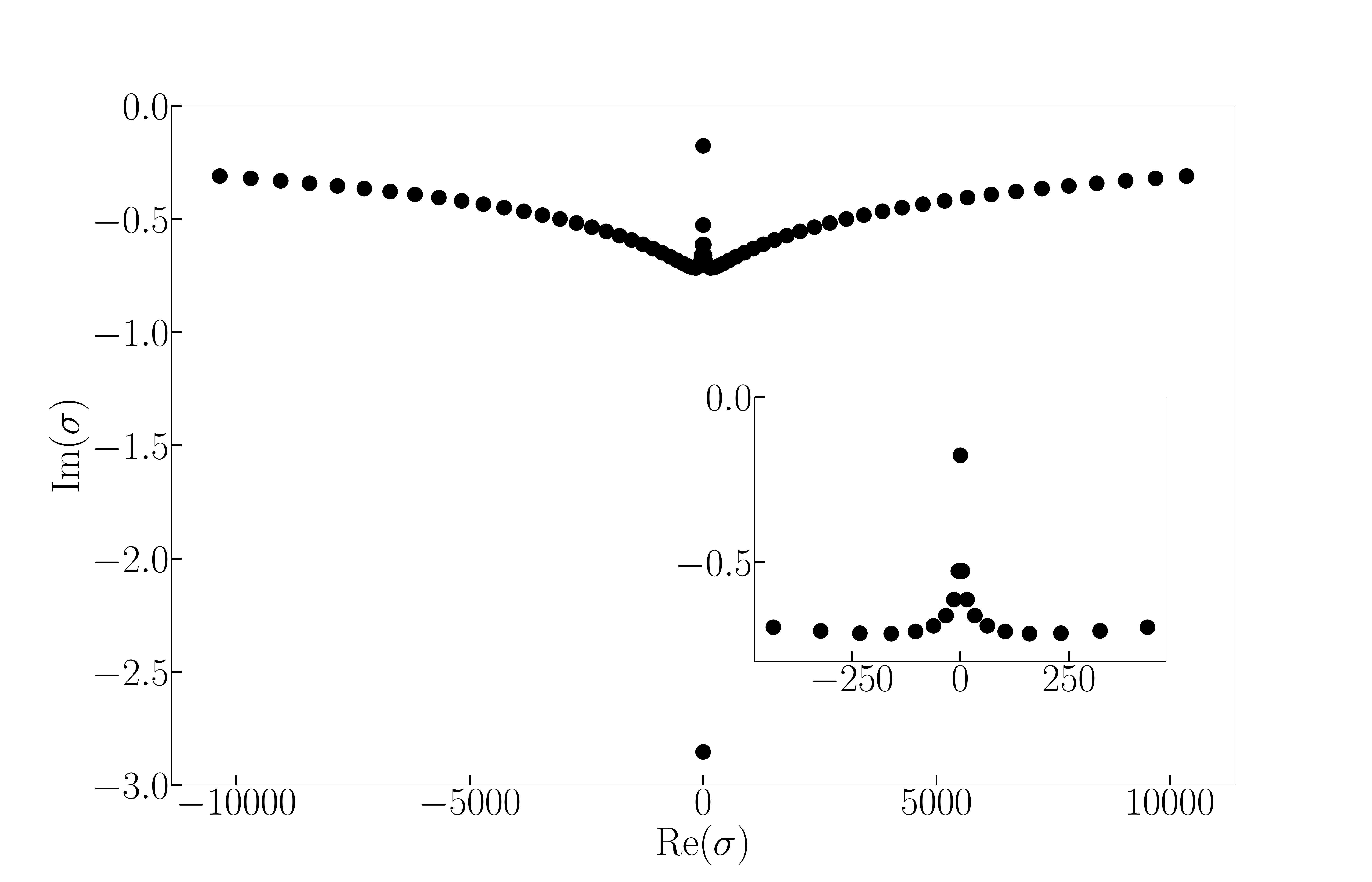}}
\hfill
\subfloat[$Re=0.5$: bending and tension.\label{fig:re50tension}]{\includegraphics[width=0.5\textwidth]{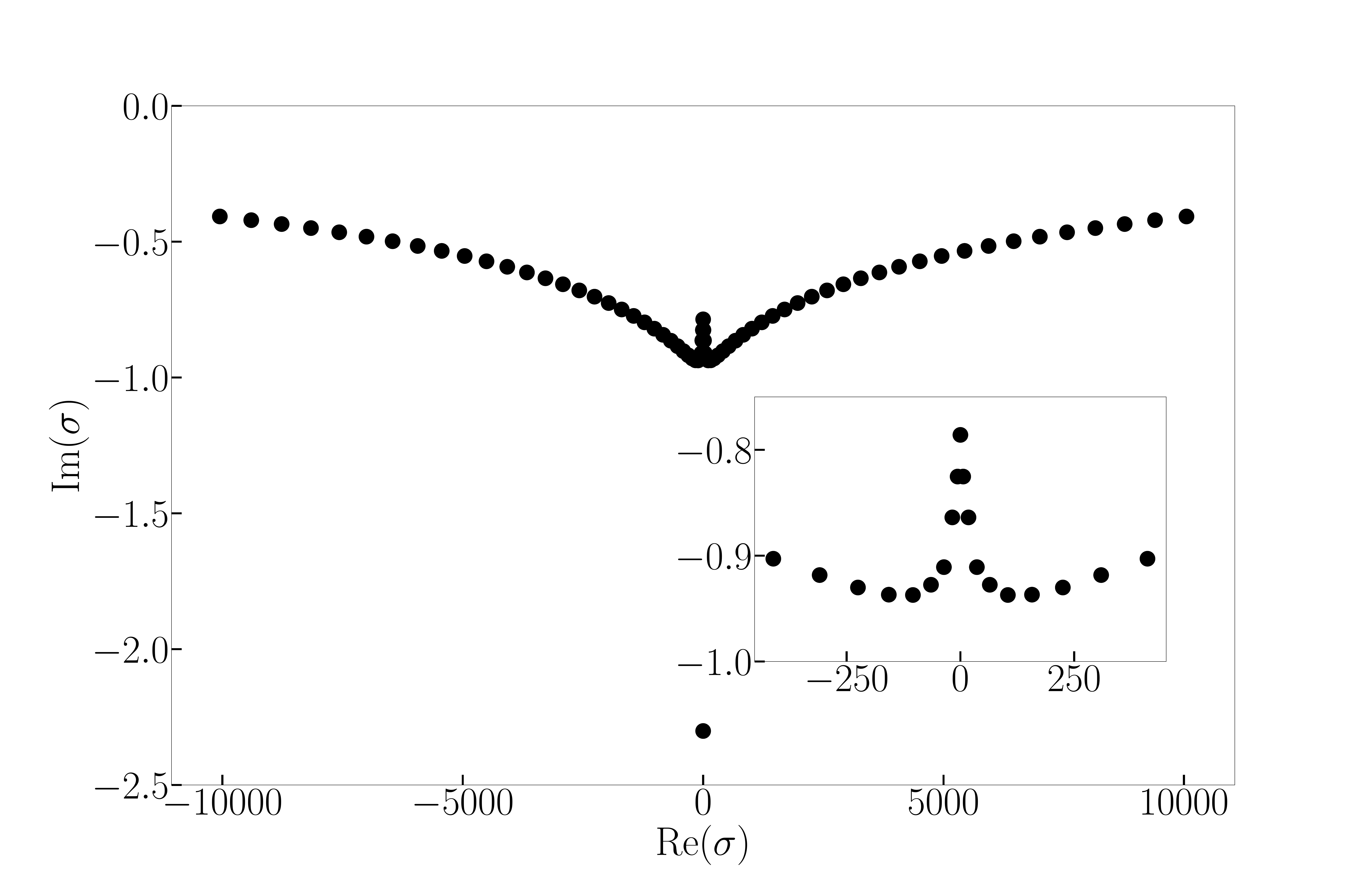}}\\
\subfloat[$Re=1.8$: pure bending.\label{fig:re180bending}]{\includegraphics[width=0.5\textwidth]{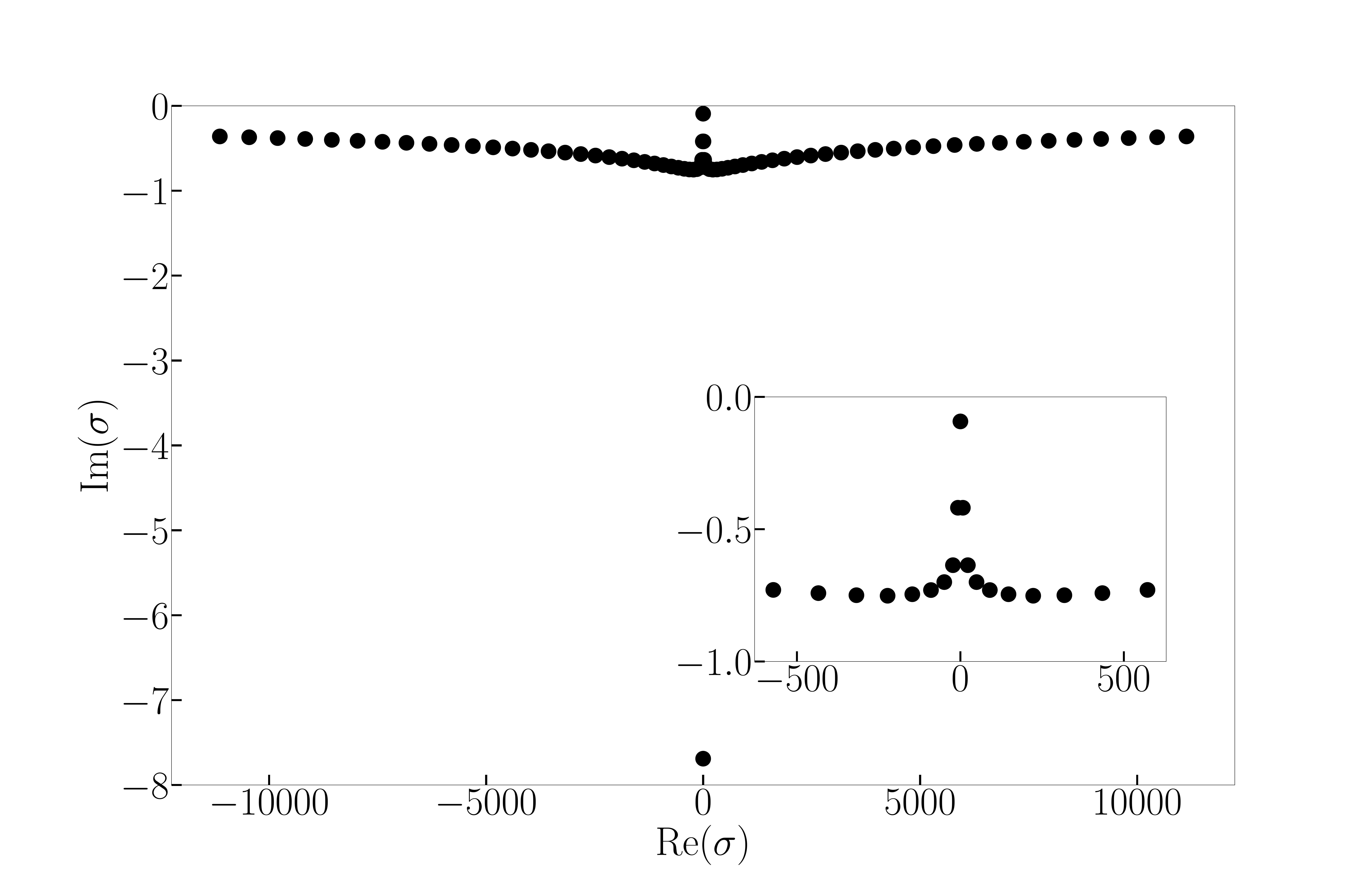}}
\hfill
\subfloat[$Re=10$: bending and tension.\label{fig:re1000tension}]{\includegraphics[width=0.5\textwidth]{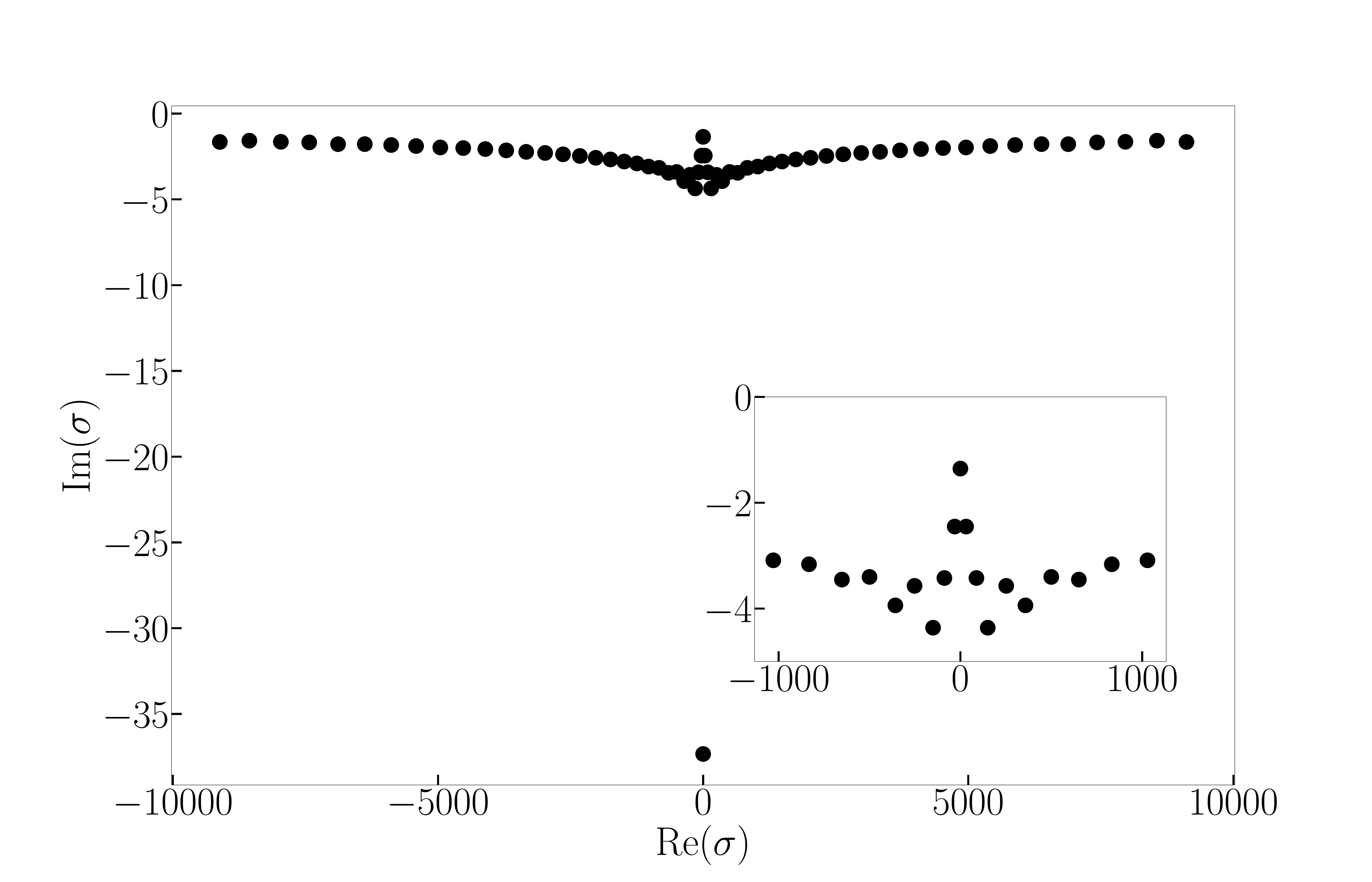}}
\caption{(Color online) Eigenspectra, in the complex plane $\mathbb{C}$, of the discretized linear eigenvalue problem, Eqs.~\eqref{stability_matrix} and \eqref{stability_bc}, governing the linear stability of the deformed microchannel shape. Here, $\epsilon=0.01$, $\Sigma=9\times10^{-4}$ and $h_{0f}/h_{0s}=1$. Two sets of numerical calculations are performed for each set of parameters (i.e., each panel), with $N=60$ (regular grid) and $N=70$ (fine grid), to ensure the accuracy of eigenvalues reported.}
\label{fig:eigenspectra}
\end{figure}

{ Even without solving the generalized eigenvalue problem numerically, we can deduce some salient features. Specifically, note that in Eq.~\eqref{stability_matrix}, the operator $\bm{A}$ is real, while the operator $\bm{B}$ is purely imaginary. This observation indicates that eigenvalues with non-zero real part should come in pairs. In other words, if there exists an eigenvalue $\sigma$ with $\real (\sigma)\neq 0$  and $\bm{\psi}\neq 0$ such that $\bm{A}\bm{\psi} = \sigma\bm{B}\bm{\psi}$, then $\bm{A}\bm{\bar{\psi}} = -\bar{\sigma}\bm{B}\bar{\bm{\psi}}$ is automatically satisfied, meaning $-\bar{\sigma}$ and $\bar{\bm{\psi}}$ are another eigenpair of the problem. Here, $\bar{(\,\cdot\,)}$ denotes the complex conjugate.}

{ Figure~\ref{fig:eigenspectra} shows the first 70  eigenvalues (ordered by magnitudes) for four different combinations of $Re$, $\beta$, $St$ and $\alpha$, corresponding to the cases in Sect.~\ref{sec:inflate} with $\epsilon=0.01$, $\Sigma=9\times10^{-4}$ and $h_{0f}/h_{0s}=1$ fixed. Two Gauss--Lobatto grids are used to compute the eigenspectra and good agreement between the two has been reached for the eigenvalues shown in Fig.~\ref{fig:eigenspectra}. To accurately resolve even higher modes ($>70$), we would have to further increase the number of Gauss--Lobatto grid points, denoted as $N+1$ in Appendix~\ref{app:cheb_stability}, in the Chebyshev pseudospectral method. This increase is impractical because the condition number of the discretized operator matrix $\bm{A}$ grows rapidly with $N$, and finally, round-off error will dominate the calculation \cite{Boyd00}. Therefore, admittedly, our calculation does not lead to any conclusions about the higher-order modes that appear to have an increasing imaginary part. However, we shall report here, that for all the eigenvalues obtained (including those not shown in Fig.~\ref{fig:eigenspectra}), only negative imaginary parts are found, except for two. The latter two are the usual ``spurious'' eigenmodes with magnitudes increasing as $\mathcal{O}(N^5)$; recall that the operator $\bm{A}$ is fifth order, and it has been found that the magnitude of the spurious eigenvalues should grow as $\mathcal{O}(N^m)$, where $m$ is the highest order of the operator \cite{Boyd00}.}

\begin{figure}[ht!]
    \subfloat[First eigenmode: $\imag(\sigma)=-0.7859$. \label{mode0Re50}]{\includegraphics[width=0.5\textwidth]{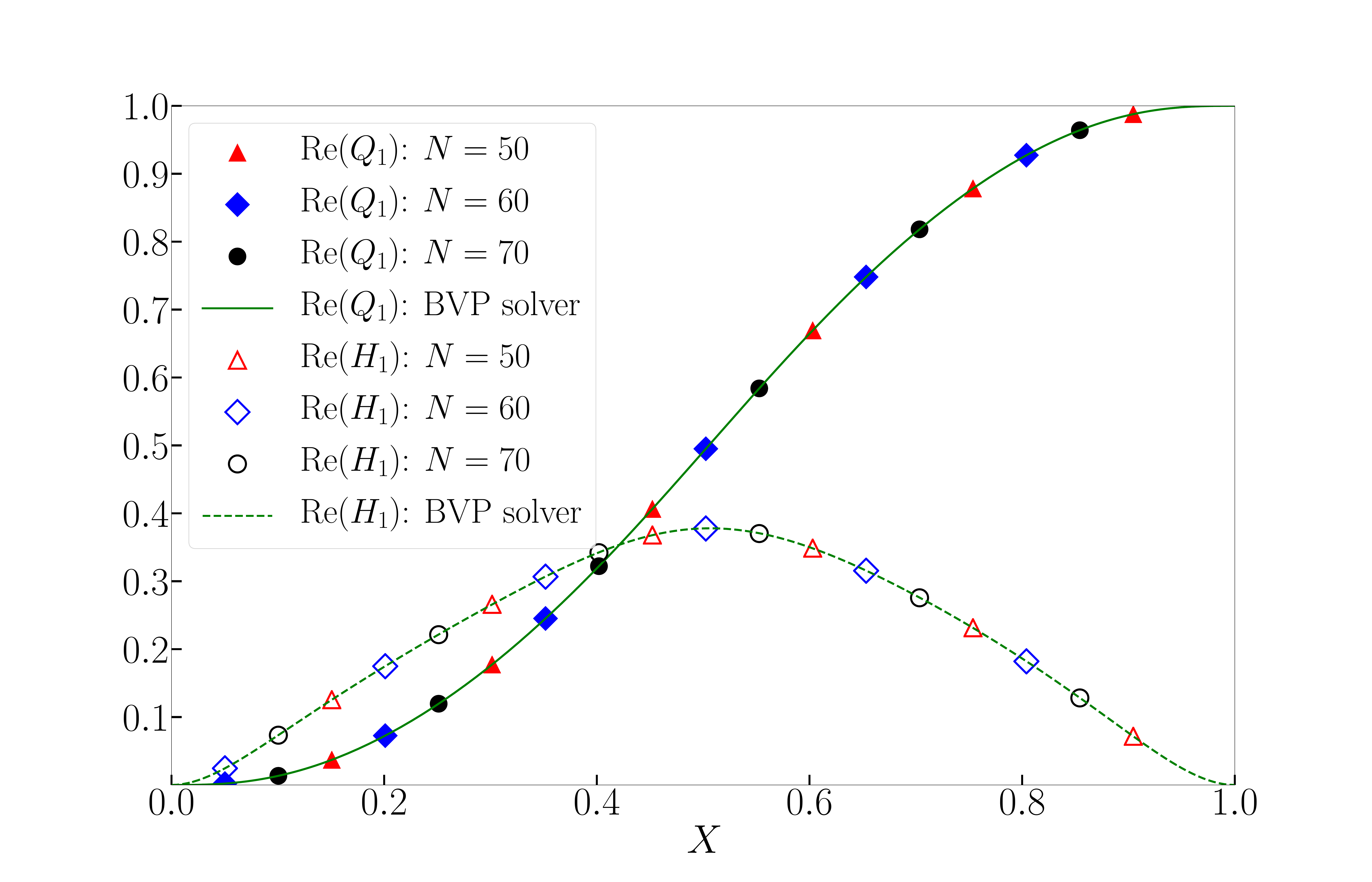}}
    \hfill
    \subfloat[Second eigenmode: $\imag(\sigma)=-2.3013$. \label{mode1Re50}]{\includegraphics[width=0.5\textwidth]{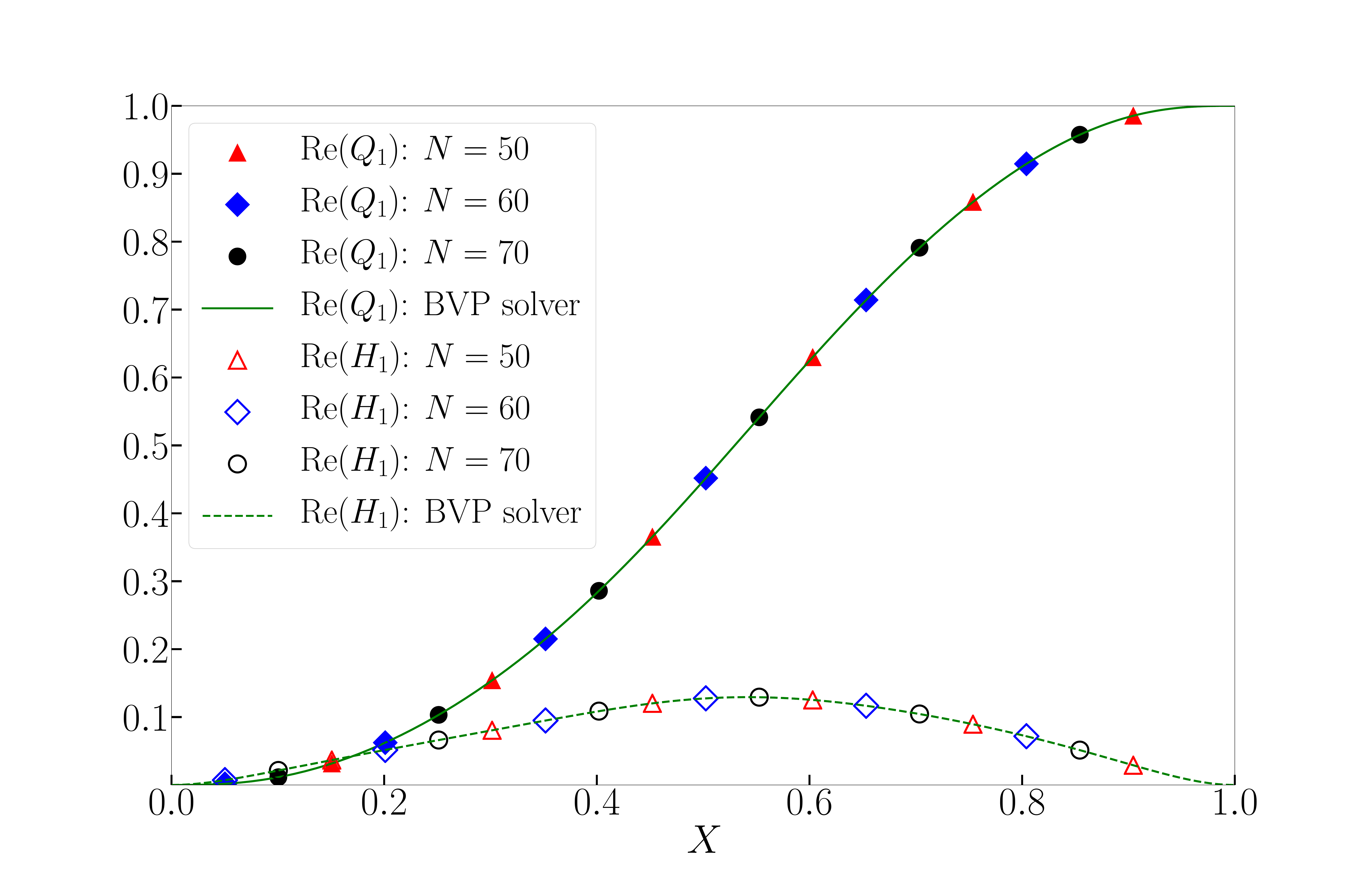}}
    \caption{Comparisons of eigenfunctions, $Q_1(X)$ (filled symbols and solid curve) and $H_1(X)$ (empty symbols and dashed curve), of the first two modes (a,b) using different numbers of Gauss--Lobatto points, as well as a formulation that uses SciPy's {\tt solve\_bvp}, for $Re=0.5$, $St=6$, $\Sigma=9\times 10^{-4}$, and $\alpha=5.56\times 10^5$. The solid curves represent $Q_1$, while the dashed curves represent $H_1$. The eigenvalues calculated from {\tt slove\_bvp} are  $\widetilde{\sigma}=-0.7849$ and $\widetilde{\sigma}=-2.3030$, respectively (recall, $\widetilde{\sigma}=-\mathrm{i}\sigma=\imag(\sigma)$).}
    \label{fig:mode0&1_Re50}
\end{figure}

\begin{figure}[ht!]
    \subfloat[$Re=0.5$: $Q_1(X)$. \label{Re50Q1_0alpha}]{\includegraphics[width=0.49\textwidth]{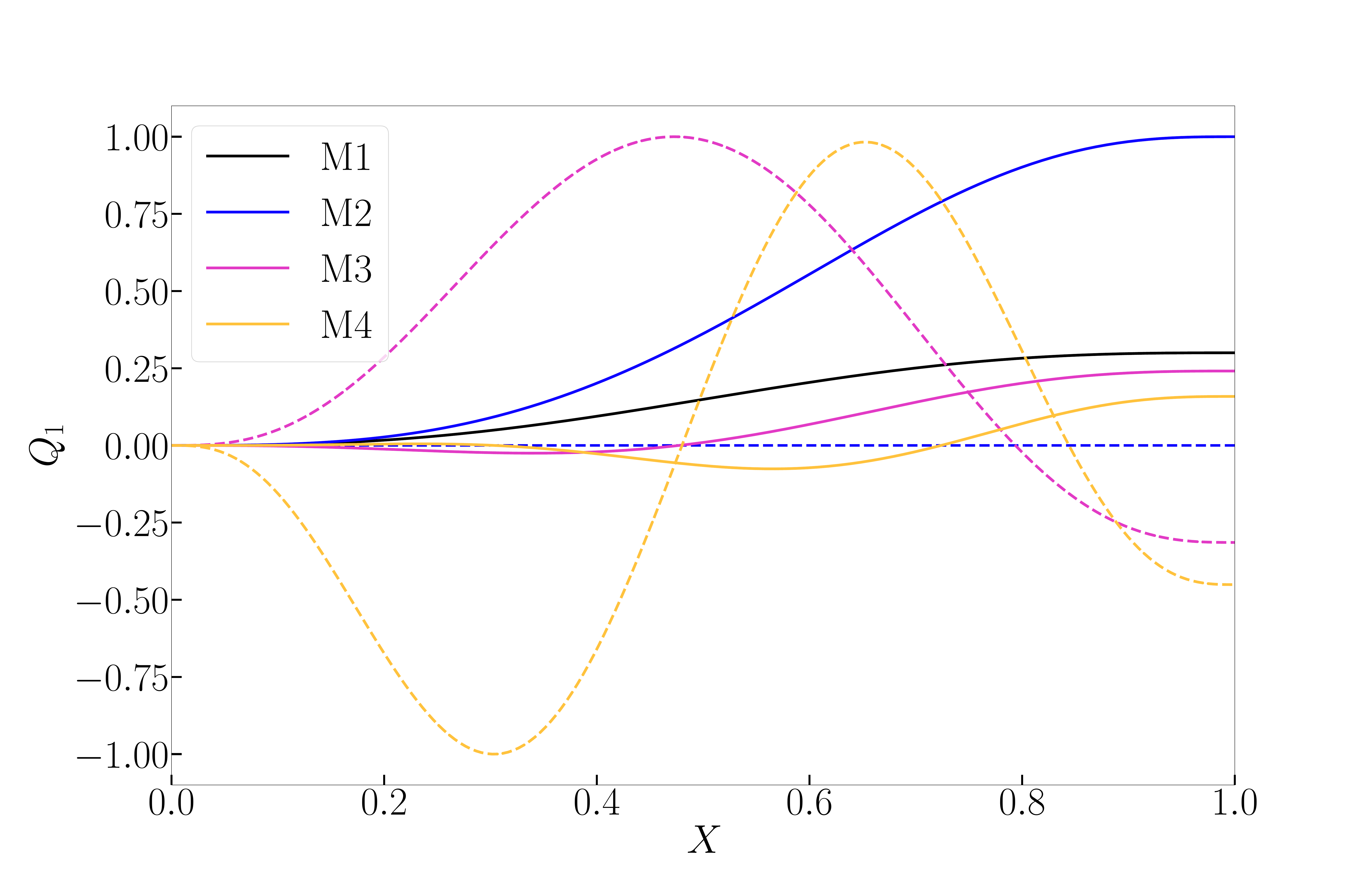}}
    \hfill
    \subfloat[$Re=0.5$: $H_1(X)$. \label{Re50H1_0alpha}]{\includegraphics[width=0.49\textwidth]{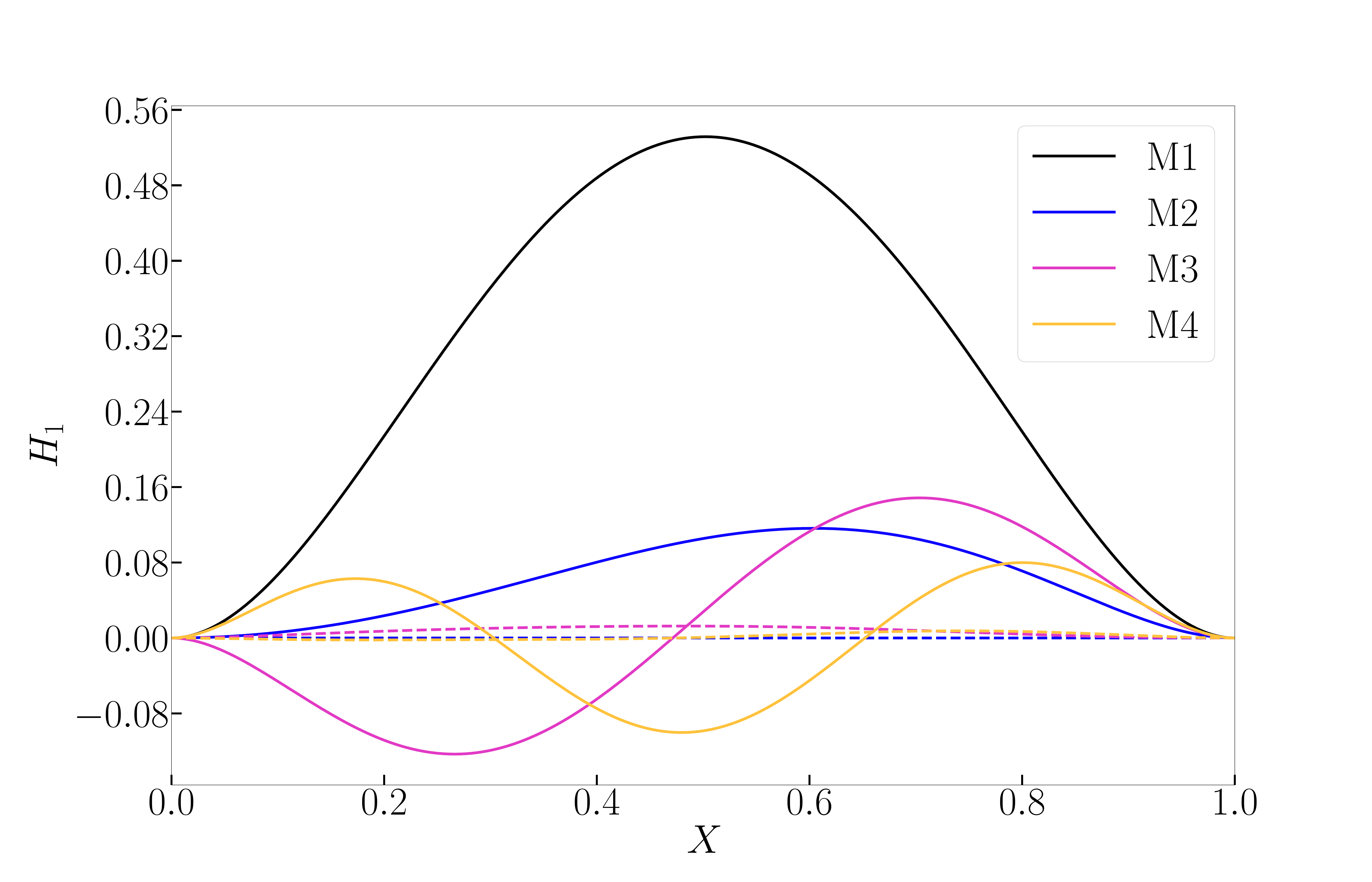}}
    \hfill
    \subfloat[$Re=1.8$: $Q_1(X)$. \label{Re180Q1_0alpha}]{\includegraphics[width=0.49\textwidth]{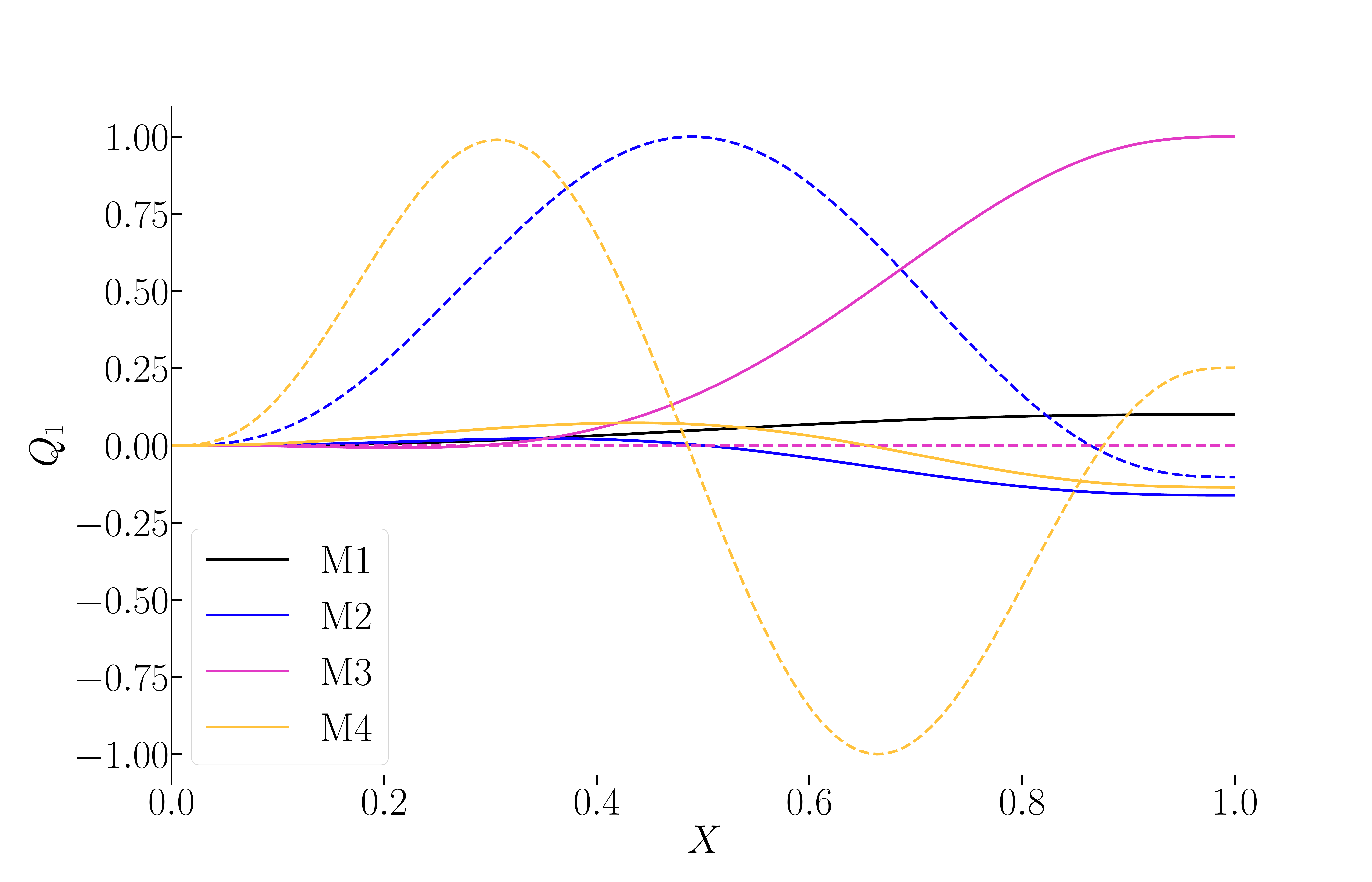}}
    \hfill
    \subfloat[$Re=1.8$: $H_1(X)$. \label{Re180H1_0alpha}]{\includegraphics[width=0.49\textwidth]{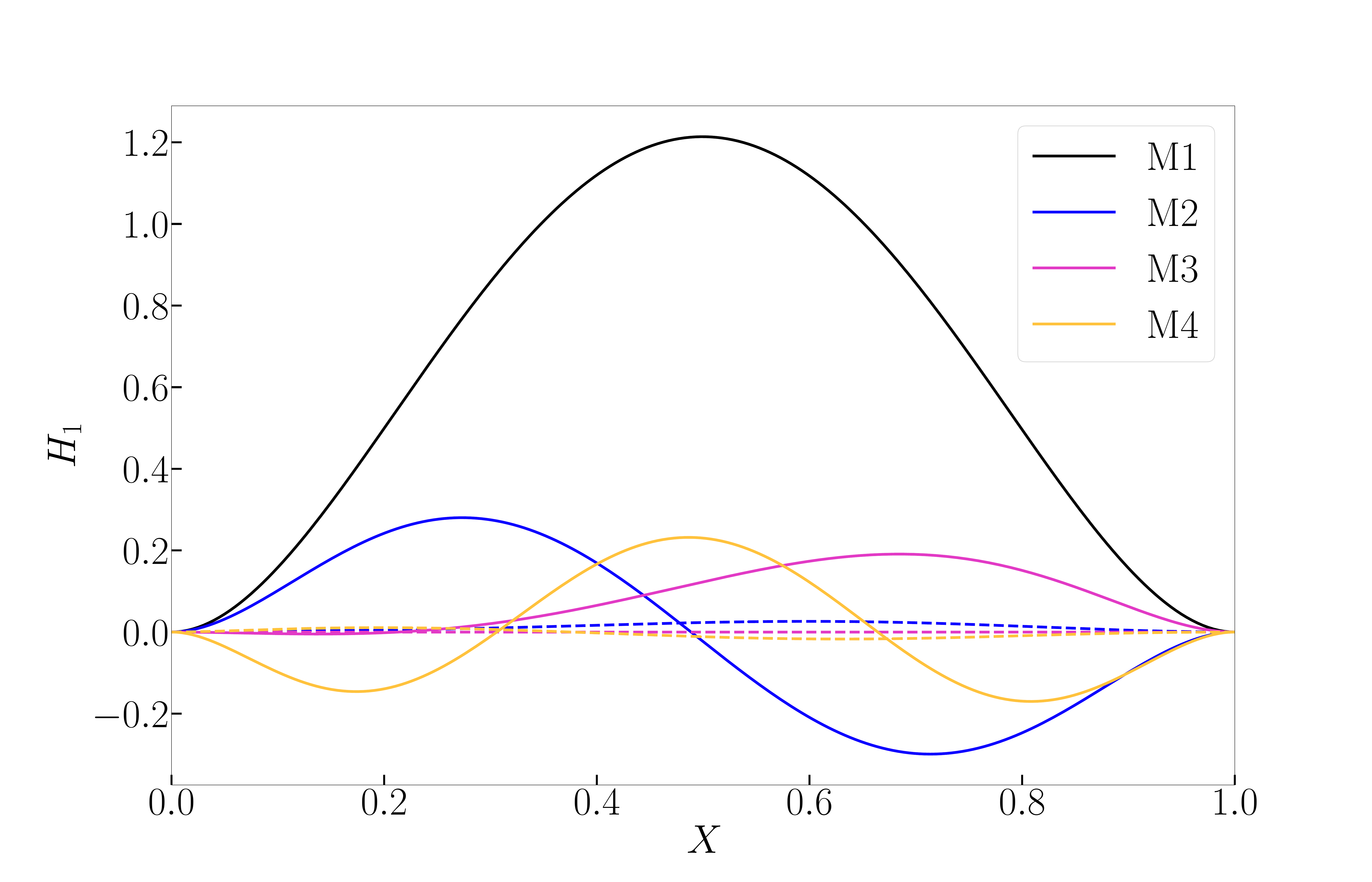}}
`    \caption{Eigenfunctions $Q_1(X)$ and $H_1(X)$ for pure bending cases. `M1' to `M4' denote mode 1 through 4, i.e., the first four eigenmodes with distinct $\imag(\sigma)$. Solid curves represent the real parts of the eigenfunctions, while dashed curves represent the imaginary parts. The normalization of the eigenfunctions is arbitrary, thus the relative magnitude of the eigenfunctions of the different modes is not important, so they are scaled to fit the plots; only their qualitative features are being highlighted here.}
    \label{fig:eigenfunc_0alphal}
\end{figure}

{As shown in Fig.~\ref{fig:eigenspectra}, the eigenspectra are discrete and symmetric about the imaginary axis. They resemble a ``seagull'' shape. As we zoom into the first 20 eigenvalues, the shapes are ``seagull''-like again but upside down. The case of $Re=10$ in Fig.~\subref*{fig:re1000tension} is an exception in that the eigenvalues form a small hole in the middle of the complex plane. We believe that this change can be attributed to the strong inertial effects in the flow in this case. 

Furthermore, for higher-order eigenvalues (i.e., larger $|\sigma|$), their real part grows much more rapidly than their imaginary part. Nevertheless, as shown in Fig.~\ref{fig:eigenspectra}, $\imag(\sigma)$ appears to be plateauing for large $|\real(\sigma)|$. The presence of a large number of eigenvalues with large $|\real(\sigma)|$ suggests that there are corresponding eigenmodes that are highly oscillatory. This observation highlights the \emph{stiffness} of the unsteady FSI problem and motivates the development of the fully-implicit finite-difference scheme with under-relaxation, used in Sect.~\ref{sec:inflate} and as described in Appendix~\ref{app:compute}. Since our transient simulation always reach a steady state, and no eigenvalues with $\imag(\sigma)>0$ have been identified via the Chebyshev method, we are led to conclude that the steady-state deformation is linearly stable to small perturbations, or at least to relatively low-frequency perturbations. A detailed analysis could be performed in future work to understand the asymptotic behavior of the eigenspectra, and to completely address the stability of the system to high-frequency disturbances.}

\begin{figure}
    \subfloat[$Re=0.5$: $Q_1(X)$. \label{Re50Q1}]{\includegraphics[width=0.5\textwidth]{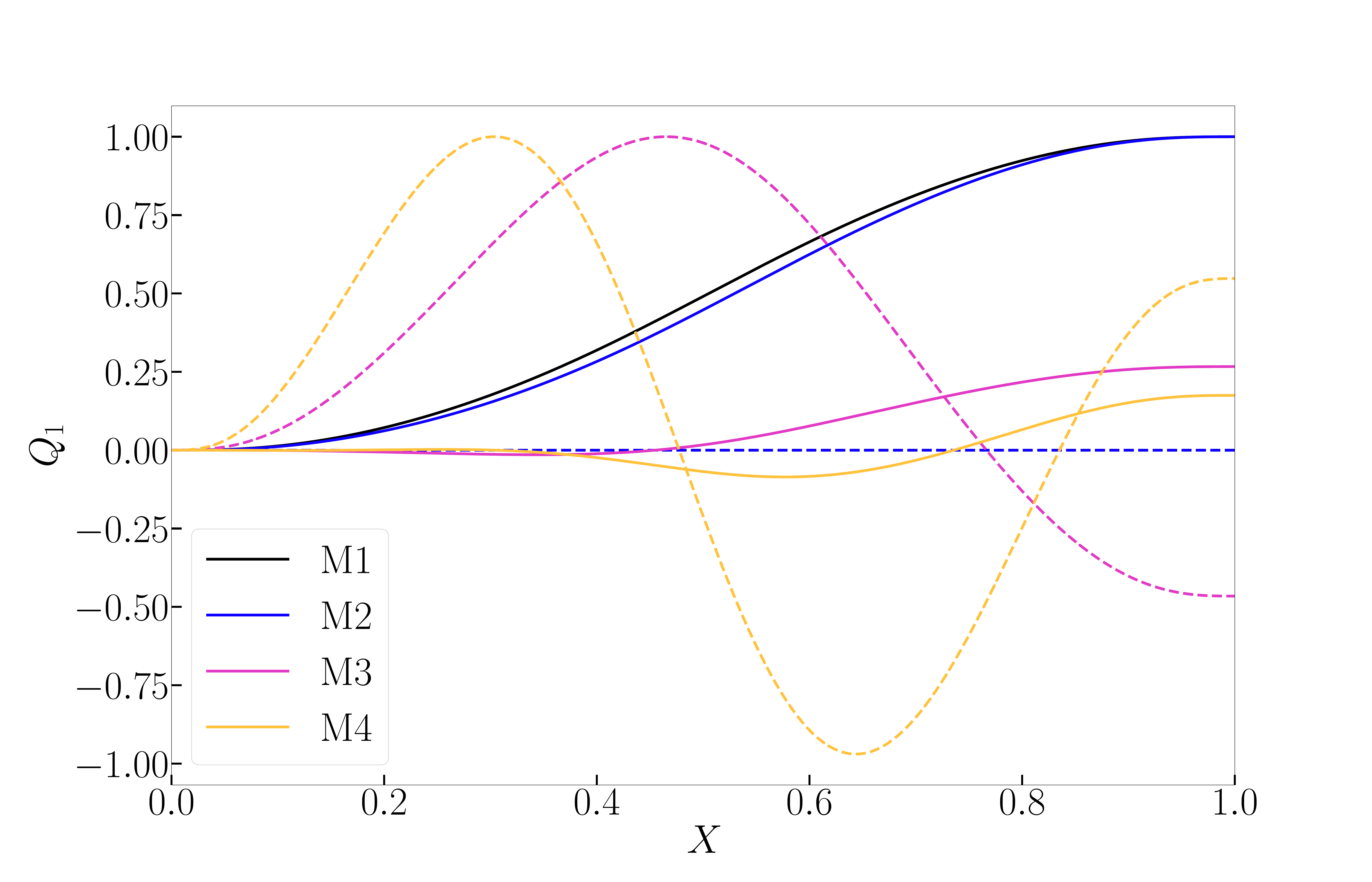}}
    \hfill
    \subfloat[$Re=0.5$: $H_1(X)$. \label{Re50H1}]{\includegraphics[width=0.5\textwidth]{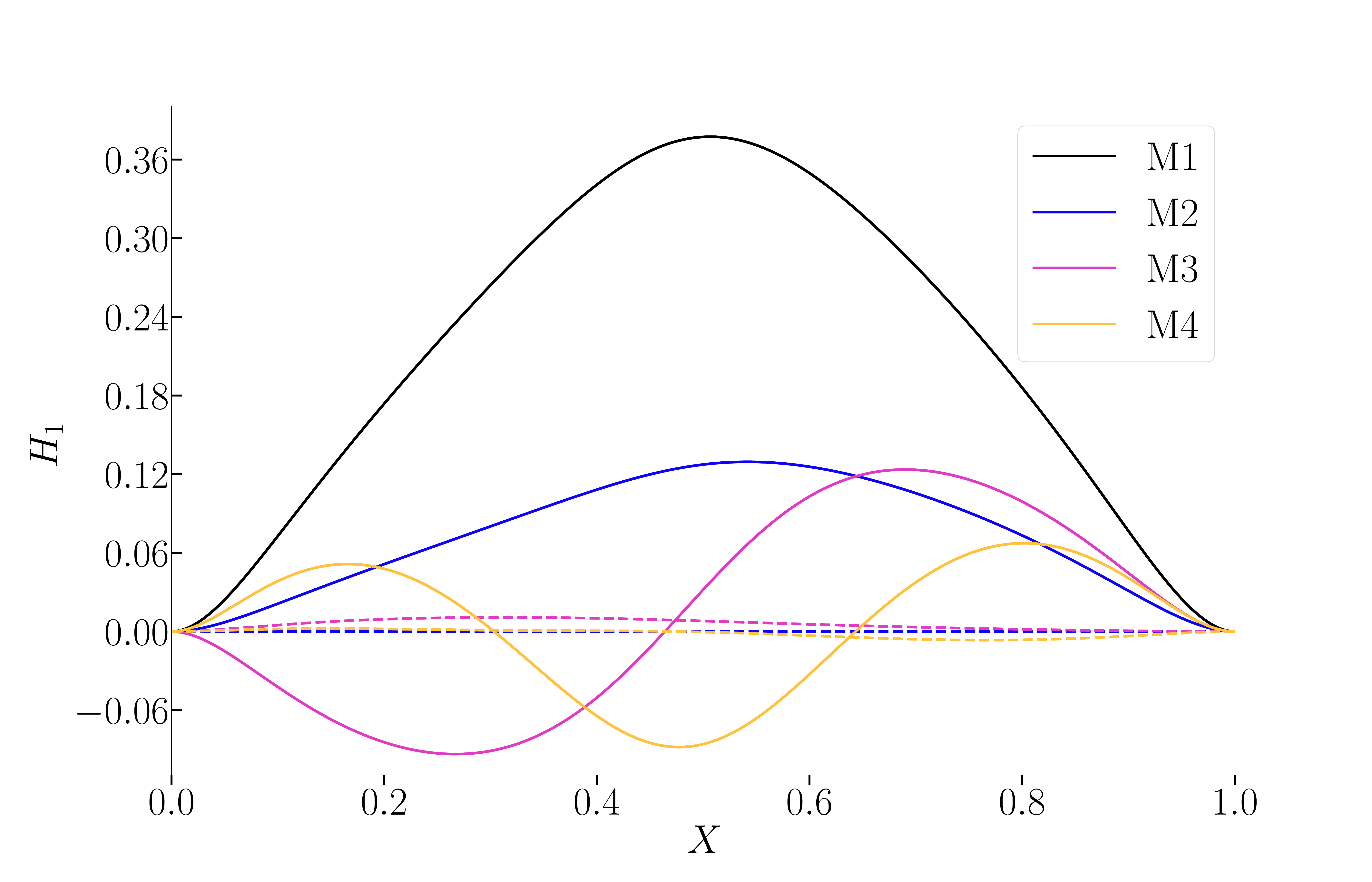}}
    \hfill
    \subfloat[$Re=10$: $Q_1(X)$. \label{Re1000Q1}]{\includegraphics[width=0.5\textwidth]{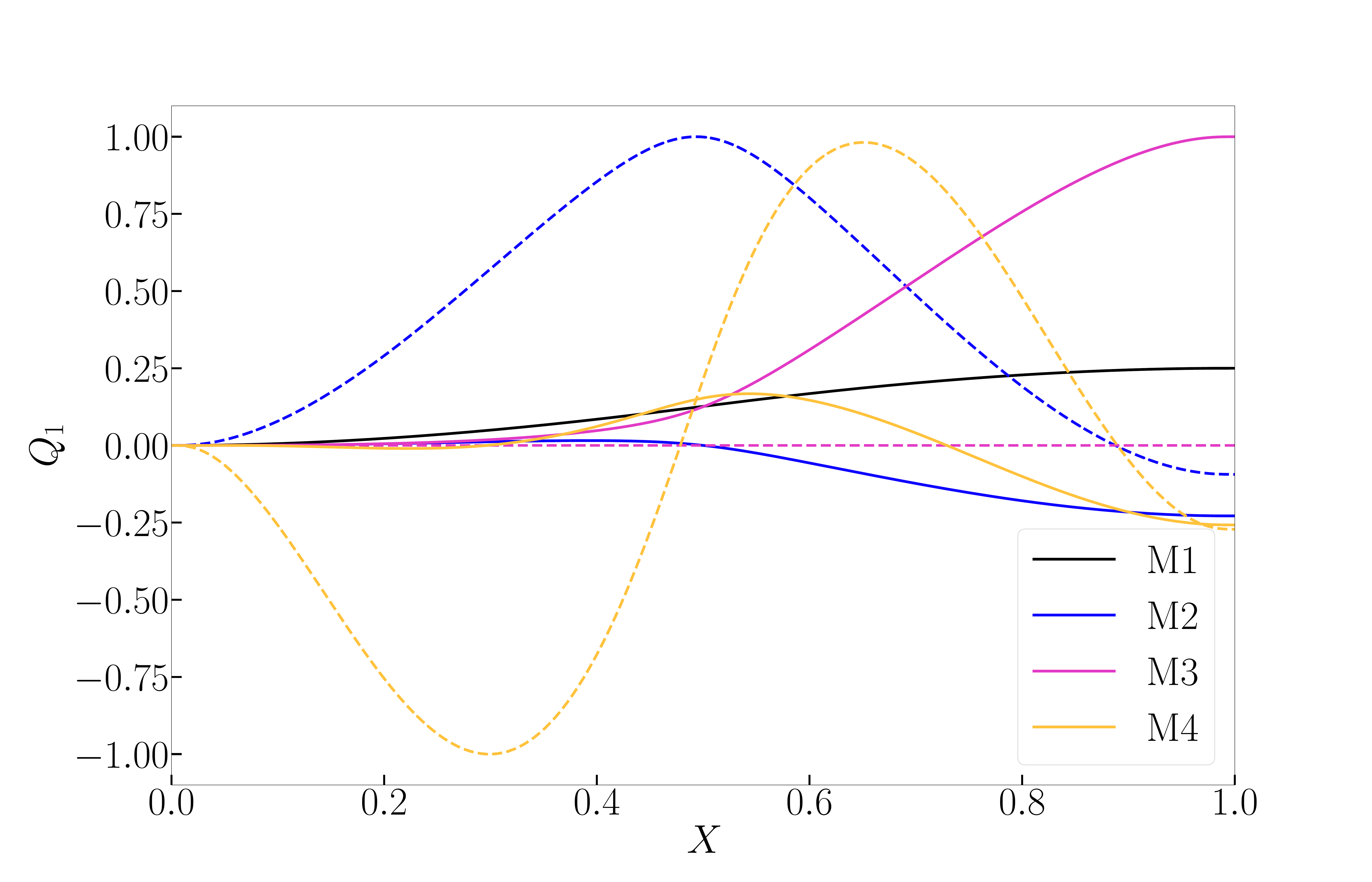}}
    \hfill
    \subfloat[$Re=10$: $H_1(X)$. \label{Re180H1}]{\includegraphics[width=0.5\textwidth]{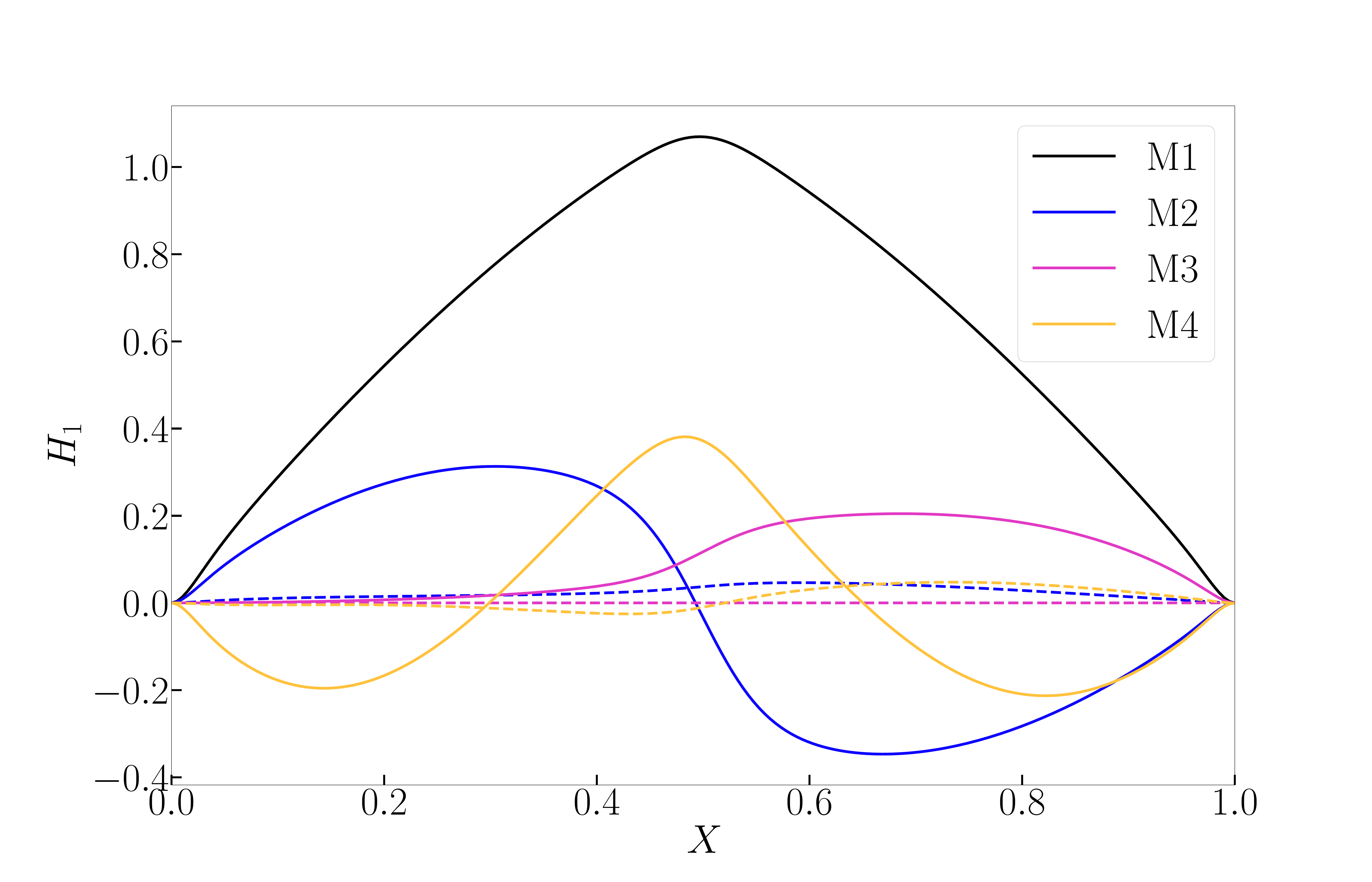}}
    \caption{Eigenfunctions $Q_1(X)$ and $H_1(X)$ for cases with bending and tension. Again, `M1' to `M4' denote mode 1 through 4, i.e., the first four eigenmodes with distinct $\imag(\sigma)$. Solid curves represent the real parts of the eigenfunction, while dashed curves represent the imaginary parts. The normalization of the eigenfunctions is arbitrary; only their qualitative features are being highlighted here.}
    \label{fig:eigenfunc}
\end{figure}

{Another interesting observation regarding the eigenspectra is that there are always two purely imaginary eigenvalues. For the cases in Fig.~\subref*{fig:re50bending} and \subref*{fig:re50tension}, these eigenvalues appear as the first and second eigenvalues, while in Figs.~\subref*{fig:re180bending} and \subref*{fig:re1000tension}, these are the first and the third eigenvalues. It can be shown that the corresponding eigenfunctions are purely real. In relation to this observation, note that Butler et al.~\cite{BBRV19} used an alternative way to calculate the eigenvalue and eigenfunctions. Translated to our setting, we can introduce $\widetilde{\sigma}=-\mathrm{i}\sigma$ and substitute it into Eqs.~\eqref{eq:linearized} to exclude any complex-valued solutions. Then, the system is linearly unstable if $\widetilde{\sigma}>0$. The new set of equations can be viewed as a boundary value problem for $Q_1(X)$ and $H_1(X)$ with $\widetilde{\sigma}$ as an unknown parameter. The idea from \cite{BBRV19} is to then use, e.g., SciPy's {\tt solve\_bvp} to solve the reformulated problem, provided that a proper initial guess for $\widetilde{\sigma}$ is given. However, this method can only provide a \emph{single}, real eigenpair at a time (which corresponds to the case of purely imaginary $\sigma$ in our model). The eigenvalue calculated in this way is sensitive to the initial guess. Nevertheless, this approach can provide an independent validation of our stability calculation by the Chebyshev method. If a positive eigenvalue is returned by {\tt solve\_bvp}, then we would immediately conclude that the system is linearly unstable. However, it is important to note that the opposite is \emph{not} true. If the eigenvalue returned is negative, then the result is inconclusive as we do not know whether this is the eigenvalue with smallest $|\imag(\sigma)|$, and we cannot make a definitive statement about the stability of the system. Finally, in Fig.~\ref{fig:mode0&1_Re50}, we compare the results from the Chebyshev pseudospectral  method (using different $N$) with the results of the formulation based on \cite{BBRV19} using {\tt solve\_bvp}, for the first two modes shown in Fig.~\subref*{fig:re50tension}. As evidenced by Fig.~\ref{fig:mode0&1_Re50}, the eigenvalues and eigenfunctions from both methods agree completely. Thus, we are confident in the accuracy of the eigenspectra in Fig.~\ref{fig:eigenspectra}, which were computed by the Chebyshev method.}

{Next, we show the eigenfunctions corresponding to the first four eigenvalues with distinct imaginary parts for each case shown in Fig.~\ref{fig:eigenspectra}. As discussed above, the eigenfunctions corresponding to the eigenvalues with the same imaginary part but opposite real part are conjugate pairs and thus not interesting to show here. Figure~\ref{fig:eigenfunc_0alphal} shows the eigenfunctions for the two cases in Figs.~\subref*{fig:re50bending} and \subref*{fig:re180bending}, while Fig.~\ref{fig:eigenfunc} corresponds to the cases in Figs.~\subref*{fig:re50tension} and \subref*{fig:re1000tension}. One common feature of these plots is that, for the two modes with purely imaginary eigenvalues, $\real(H_1)$ has only one maximum (hump), while the other two modes display two and three humps, respectively. Wavelike shapes are also observed in $\imag(Q_1)$ for the fourth mode. Of course, it is expected that there will be more humps in the eigenfunctions of higher-order modes. This observation is the reason for increasing the number of Gauss--Lobatto points to properly resolve the oscillatory nature of the higher-order eigenfunctions. Finally, unsurprisingly, inertial effects lead to a distinct shape of $H_1(X)$ for $Re=10$, compared to the other three cases.} 

\begin{figure}
    \subfloat[First mode: $\sigma = -1.3547\mathrm{i}$.~\label{stab_trans_mode0}]{\includegraphics[width=0.5\textwidth]{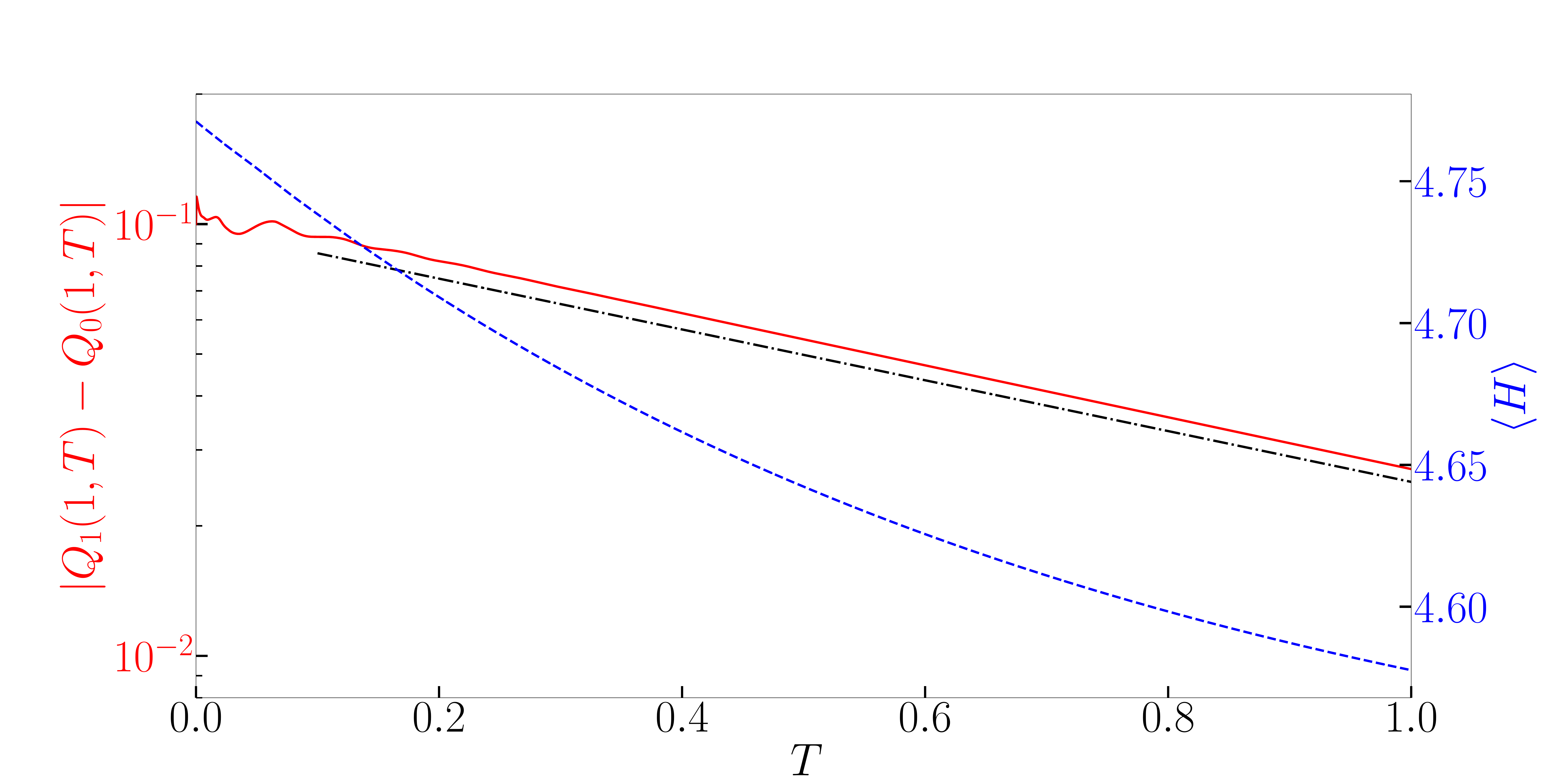}}
    \hfill
    \subfloat[Second mode: $\sigma = 31.2167-2.4504\mathrm{i}$.~\label{stab_trans_mode1}]{\includegraphics[width=0.5\textwidth]{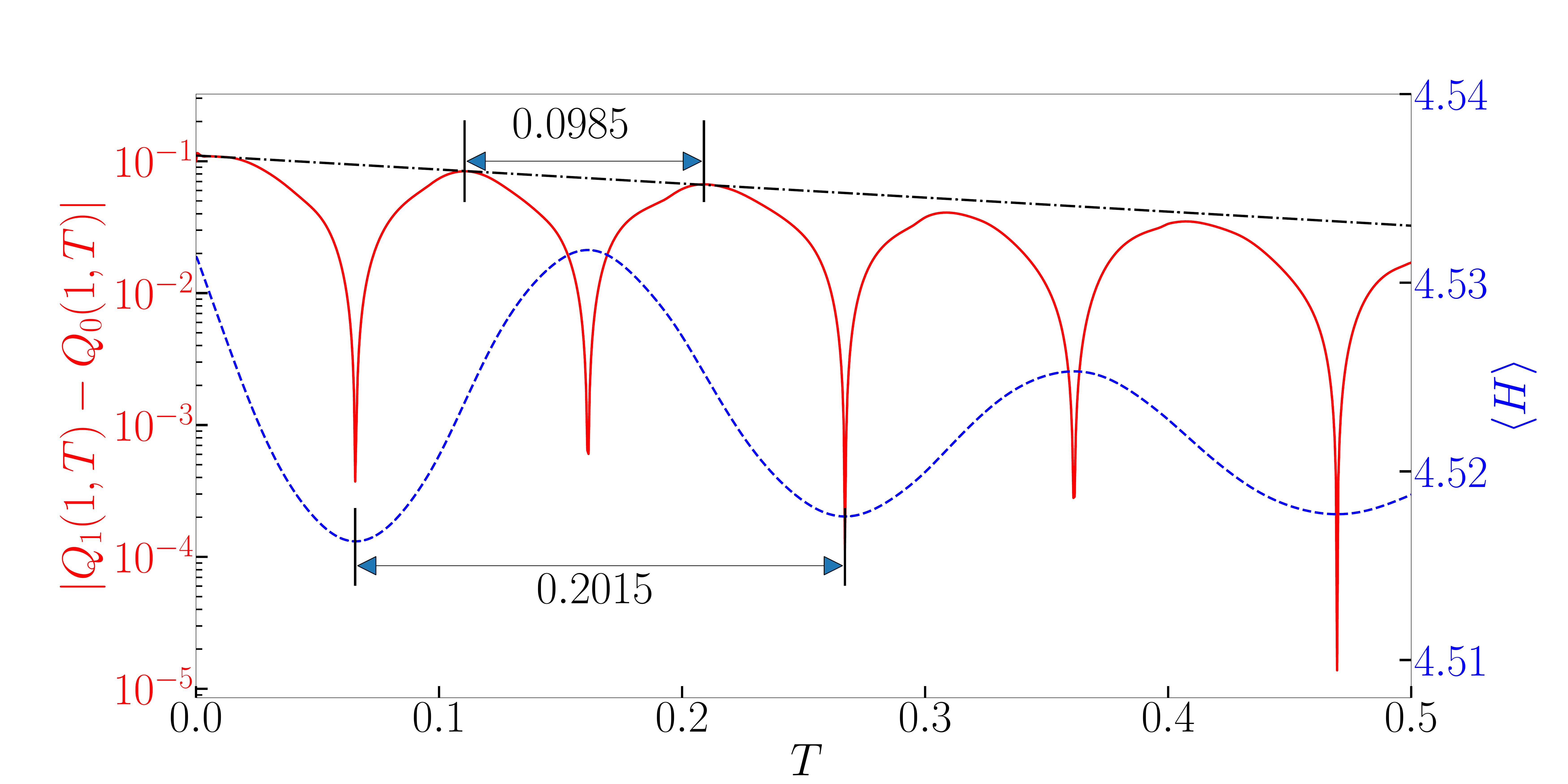}}
    \caption{The time histories of the difference of the instantaneous outlet flow rate from the base state, i.e., $|Q(1,T)-Q_0(1,T)|$, and the axially-average deformed channel height, $\langle H \rangle$, after substituting the eigenfunctions of the first and second mode for $Re=10$ into the initial perturbations respectively (see Eqs.~\eqref{eq:lin_perturb} and taking $T=0$). The slope of the dot-dashed lines represents the imaginary part of the corresponding eigenvalues from the linear stability calculation.}
    \label{fig:stab_trans}
\end{figure}

{The final validation of this linear analysis is to compare the predicted growth of perturbations to the time-evolution of the nonlinear problem. To this end, we take the initial condition to be  Eqs.~\eqref{eq:lin_perturb} (at $T=0$), where $Q_1$ and $H_1$ given by the eigenfunctions of the linear problem computed above. Then, $Q(X,T)$ and $H(X,T)$ should evolve according to Eqs.~\eqref{eq:lin_perturb} with the corresponding eigenvalue setting the time dependence. For example, fixing $\delta=0.1$ and taking $Q(X,0)=1+\delta Q_1(X)$ and $H(X,0)=H_0(X)+\delta H_1(X)$, where $\{Q_1,H_1\}$ is a linear eigenmode for $Re=10$, as the initial conditions for the transient simulation, Fig.~\ref{fig:stab_trans} shows the time histories of the outlet flow rate and the deformed channel's height. Since the eigenvalue of the first mode is purely imaginary, Fig.~\subref*{stab_trans_mode0} shows that the deviation of the outlet flow rate from the base state, $|Q(1,T)-Q_0(1,T)|$, as well as the average deformed channel height, $\langle H\rangle$, both decay without oscillations. Importantly, the decay rate of the fully-nonlinear simulation agrees with $\imag(\sigma)$. As for the results of the second mode, shown in Fig.~\subref*{stab_trans_mode1}, both $Q$ and $\langle H\rangle$ oscillate in time because $\real (\sigma)\neq 0$. In this example, $\real (\sigma)$ represents the temporal period of the eigenmode, which is $\approx 2\pi/31.2167=0.2013$ based on the solution by the Chebyshev pseudospectral method. Clearly, the oscillation period observed in the transient simulation is very close to this predicted value. Furthermore, the decay rate of the amplitudes of $|Q(1,T)-Q_0(1,T)|$ agrees with $\imag(\sigma)$.}

\section{Conclusion}
\label{sec:concl}

In this paper, we derived a one-dimensional (1D) model for unsteady viscous fluid--structure interactions (FSIs) at finite Reynolds number, starting from a two-dimensional (2D) Cartesian geometry in which an initially rectangular fluid domain contains a Newtonian fluid obeying the Navier--Stokes equations, while the top boundary of the geometry is a beam of finite thickness that supports both bending and nonlinear tension. The parameter space of this model was shown to consist of a Reynolds number $Re$, a Strouhal number $St$, a dimensionless elasticity modulus $\Sigma$, the channel aspect ratio $\epsilon$, a dimensionless compliance (or, FSI/coupling) parameter $\beta=Re/\Sigma$, and a dimensionless nonlinear tension $\alpha$. Fixing the microchannel's geometry, we explored the physical effects of $\alpha$, $Re$ (which necessarily involves changing $St$),  and $\Sigma$ on unsteady FSI dynamics in this system.

Specifically, with our reduced-order 1D model in hand, we characterized the hydrodynamics and the deformation at steady state through representative quantities: an axially-averaged hydrodynamic pressure, $\langle P \rangle$, and the maximum deformation of the top wall, $H_{\mathrm{max}}$. We derived different scaling regimes for $\langle P \rangle$ and $H_{\mathrm{max}}$ with respect to $Re$ and $\Sigma$ at steady state. Since our model allows $Re=\mathcal{O}(1)$, inertia and viscosity are two competing forces in the flow. Depending on which of these two forces dominates, the pressure distribution within the channel is markedly different, which, in turn, leads to two scaling regimes for $\langle P \rangle$. On the other hand, the deformation of the solid is the outcome of the competition between bending and tension in the beam that represents the channel's top wall. By considering  different force balances (in the fluid and in the solid), we found four scaling regimes for $H_{\mathrm{max}}$. Each of these predicted scaling regimes was verified across a wide range of $Re$ and $\Sigma$ values by numerical solutions of the steady-state problem. {In future mathematical work, it would be worthwhile to revisit these scalings with an asymptotic analysis of the governing equations, in the limiting cases of extreme values of the parameters, including identifying any singular perturbations and boundary layer structures.}

To highlight the key aspects of scaling the governing equations in the viscous (lubrication) limit, we then addressed some exemplar unsteady coupled flow--deformation behaviors through numerical simulations via our implicit finite-difference scheme. Specifically, at ``moderate'' $Re$, which in the present context we take to be $Re=\mathcal{O}(1)$, and in the presence of nonlinear tension, an intermediate almost-stable state, which resembles a beam's buckling mode, exists. Nevertheless, the intermediate ``buckled'' state observed is not one of the solutions admitted by the steady-state problem, therefore this is a distinct, purely transient, effect. Overall, complex transients were observed in all unsteady regimes considered because $St=\mathcal{O}(1)$, meaning the flow and deformation are tightly coupled. This stiff transient response was not considered in many previous works in which either $St\to0$ or $St\to\infty$.

{The stiffness of the FSI system is also evidenced by the multiple highly oscillatory modes observed in the linear stability analysis. With the Chebyshev pseudospectral method, we were able to resolve the linearized problem's eigenspectra, which resemble a ``seagull'' shape. The high-frequency oscillations are indicated by the rapid increase of the real parts of the eigenvalues. The corresponding eigenfunctions of theses higher mode display more maxima (humps). Notably, fluid  inertial effects were shown to be significantly affect to change the eigenspectra and eigenfunctions. However, for all the examples discussed herein, whether tension of the top wall was included or not, and spanning ``low'' $Re$ to ``moderate'' $Re$, the imaginary parts of the eigenvalues (i.e., the growth rates of linear normal modes) remained negative, indicating that the steady state of our model microscale FSI system is linearly stable to perturbations. Since the flat steady state, $Q_0(X)=1$ and $H_0(X)=1$, is not a solution of our model FSI system, the stability problem considered herein is different from previous constant-tension-dominated systems \cite{XBJ13,XBJ14}. Specifically, in our problem we did not find multiple neutral modes, which can lead to various bifurcation scenarios in the spectrum. Furthermore, we have kept the upstream flux fixed, as is most common in microfluidic systems, thus precluding any instabilities that could be induced by sufficiently vigorous oscillations \cite{XBJ13}. In future work, it would be of interest to address the possibilities of such instabilities in soft-walled microchannels, by replacing the fixed flux upstream boundary condition in our model with a prescribed pressure drop.}

The present work is a complementary line of inquiry to Pedley's collapsible tubes research program (see, e.g., \cite{HJ03,PPP15}) in which reduced-order (typically, 1D) models have been derived for physiological FSIs under the boundary-layer (high $Re$) scaling of the Navier--Stokes equations. The main difference between the latter and our present work is that we have scaled the Navier--Stokes equations in the lubrication limit relevant to microfluidics, which changes the relative ``importance'' of various flow effects on the coupled FSI problem. Our work is also distinct from previous FSI models in which \emph{inviscid} flow is coupled to a nonlinear beam with stretching and rotation \cite{KAIS09,EIKC14}, a nonlinear von K\'arm\'an plate \cite{CDLW16}, or an elastic tube \cite{GBGP18}. Nevertheless, it would be worthwhile generalizing the mathematical techniques and stability results obtained in \cite{KAIS09,CDLW16,GBGP18} to the present \emph{viscous} context. Microscale unsteady FSI with non-Newtonian fluid rheology, e.g., shear-rate-dependent viscosity \cite{BBG17,ADC18}, is another avenue of future research. {Furthermore, it would also be of interest to explore other actuation mechanisms for this system, such as electrostatic forcing, with applications to MEMS resonators \cite{NY04}.} It might also be prudent to take capillary effects into account during \emph{unsteady} microscale FSI, building upon the steady case \cite{AS15} and prior work on \emph{elastocapillarity} \cite{BRB18}. 

Finally, there are a number of distinguished limits of our model that would be of interest to analyze in future work. These limiting models are discussed in \cite{Inamdar2018}. The most salient limits worth pointing out here, in particular in relation to previous microchannel FSI studies, are $St \rightarrow 0$ and $St \rightarrow \infty$. These two cases correspond to physical situations in which either the solid time scale or the fluid time scale, respectively, dominates the FSI dynamics. Consequently, in each of these limits, the unsteady effects in the other medium are negligible. Since either the solid mechanics strongly affects the fluid mechanics or vice versa, in the $St \rightarrow 0$ and $St \rightarrow \infty$ limits respectively, then in each limit one of the mechanical problems is ``subjugated'' to the other, leading to something akin to weakly-coupled \emph{one-way} FSI. A related discussion on the comparison between fluid and solid time scales can be found in~\cite{MEG17}.

\acknowledgments
We thank V.\ Anand, F.\ Municchi and T.\ C.\ Shidhore each for a critical reading of the manuscript and helpful suggestions. We also thank the anonymous referees for incisive remarks that have improved the manuscript significantly. This research was supported, in part, by the US National Science Foundation under grant No.\ CBET-1705637. 

\bibliography{Mendeley_References_ICC,other_refs}

\begin{thebibliography}{87}%
\makeatletter
\providecommand \@ifxundefined [1]{%
 \@ifx{#1\undefined}
}%
\providecommand \@ifnum [1]{%
 \ifnum #1\expandafter \@firstoftwo
 \else \expandafter \@secondoftwo
 \fi
}%
\providecommand \@ifx [1]{%
 \ifx #1\expandafter \@firstoftwo
 \else \expandafter \@secondoftwo
 \fi
}%
\providecommand \natexlab [1]{#1}%
\providecommand \enquote  [1]{``#1''}%
\providecommand \bibnamefont  [1]{#1}%
\providecommand \bibfnamefont [1]{#1}%
\providecommand \citenamefont [1]{#1}%
\providecommand \href@noop [0]{\@secondoftwo}%
\providecommand \href [0]{\begingroup \@sanitize@url \@href}%
\providecommand \@href[1]{\@@startlink{#1}\@@href}%
\providecommand \@@href[1]{\endgroup#1\@@endlink}%
\providecommand \@sanitize@url [0]{\catcode `\\12\catcode `\$12\catcode
  `\&12\catcode `\#12\catcode `\^12\catcode `\_12\catcode `\%12\relax}%
\providecommand \@@startlink[1]{}%
\providecommand \@@endlink[0]{}%
\providecommand \url  [0]{\begingroup\@sanitize@url \@url }%
\providecommand \@url [1]{\endgroup\@href {#1}{\urlprefix }}%
\providecommand \urlprefix  [0]{URL }%
\providecommand \Eprint [0]{\href }%
\providecommand \doibase [0]{https://doi.org/}%
\providecommand \selectlanguage [0]{\@gobble}%
\providecommand \bibinfo  [0]{\@secondoftwo}%
\providecommand \bibfield  [0]{\@secondoftwo}%
\providecommand \translation [1]{[#1]}%
\providecommand \BibitemOpen [0]{}%
\providecommand \bibitemStop [0]{}%
\providecommand \bibitemNoStop [0]{.\EOS\space}%
\providecommand \EOS [0]{\spacefactor3000\relax}%
\providecommand \BibitemShut  [1]{\csname bibitem#1\endcsname}%
\let\auto@bib@innerbib\@empty
\bibitem [{\citenamefont {Nguyen}\ and\ \citenamefont {Wereley}(2006)}]{NW06}%
  \BibitemOpen
  \bibfield  {author} {\bibinfo {author} {\bibfnamefont {N.-T.}\ \bibnamefont
  {Nguyen}}\ and\ \bibinfo {author} {\bibfnamefont {S.~T.}\ \bibnamefont
  {Wereley}},\ }\href@noop {} {\emph {\bibinfo {title} {{Fundamentals and
  Applications of Microfluidics}}}},\ \bibinfo {edition} {2nd}\ ed.,\
  Integrated Microsystems Series\ (\bibinfo  {publisher} {Artech House},\
  \bibinfo {address} {Norwood, MA},\ \bibinfo {year} {2006})\BibitemShut
  {NoStop}%
\bibitem [{\citenamefont {Reyes}\ \emph {et~al.}(2002)\citenamefont {Reyes},
  \citenamefont {Iossifidis}, \citenamefont {Auroux},\ and\ \citenamefont
  {Manz}}]{RDIAM02}%
  \BibitemOpen
  \bibfield  {author} {\bibinfo {author} {\bibfnamefont {D.~R.}\ \bibnamefont
  {Reyes}}, \bibinfo {author} {\bibfnamefont {D.}~\bibnamefont {Iossifidis}},
  \bibinfo {author} {\bibfnamefont {P.-A.}\ \bibnamefont {Auroux}},\ and\
  \bibinfo {author} {\bibfnamefont {A.}~\bibnamefont {Manz}},\ }\bibfield
  {title} {\bibinfo {title} {{Micro total analysis systems. 1. Introduction,
  theory, and technology}},\ }\href {https://doi.org/10.1021/ac0202435}
  {\bibfield  {journal} {\bibinfo  {journal} {Anal. Chem.}\ }\textbf {\bibinfo
  {volume} {74}},\ \bibinfo {pages} {2623} (\bibinfo {year}
  {2002})}\BibitemShut {NoStop}%
\bibitem [{\citenamefont {Auroux}\ \emph {et~al.}(2002)\citenamefont {Auroux},
  \citenamefont {Iossifidis}, \citenamefont {Reyes},\ and\ \citenamefont
  {Manz}}]{AIRM02}%
  \BibitemOpen
  \bibfield  {author} {\bibinfo {author} {\bibfnamefont {P.-A.}\ \bibnamefont
  {Auroux}}, \bibinfo {author} {\bibfnamefont {D.}~\bibnamefont {Iossifidis}},
  \bibinfo {author} {\bibfnamefont {D.~R.}\ \bibnamefont {Reyes}},\ and\
  \bibinfo {author} {\bibfnamefont {A.}~\bibnamefont {Manz}},\ }\bibfield
  {title} {\bibinfo {title} {{Micro total analysis systems. 2. Analytical
  standard operations and applications}},\ }\href
  {https://doi.org/10.1021/ac020239t} {\bibfield  {journal} {\bibinfo
  {journal} {Anal. Chem.}\ }\textbf {\bibinfo {volume} {74}},\ \bibinfo {pages}
  {2637} (\bibinfo {year} {2002})}\BibitemShut {NoStop}%
\bibitem [{\citenamefont {Bienvenue}\ \emph {et~al.}(2004)\citenamefont
  {Bienvenue}, \citenamefont {Karlinsey}, \citenamefont {Landers},\ and\
  \citenamefont {Ferrance}}]{BKLF04}%
  \BibitemOpen
  \bibfield  {author} {\bibinfo {author} {\bibfnamefont {J.~M.}\ \bibnamefont
  {Bienvenue}}, \bibinfo {author} {\bibfnamefont {J.}~\bibnamefont
  {Karlinsey}}, \bibinfo {author} {\bibfnamefont {J.~P.}\ \bibnamefont
  {Landers}},\ and\ \bibinfo {author} {\bibfnamefont {J.~P.}\ \bibnamefont
  {Ferrance}},\ }\bibfield  {title} {\bibinfo {title} {{Clinical Applications
  of Microfluidic Devices}},\ }in\ \href
  {https://doi.org/10.1201/9780203912980} {\emph {\bibinfo {booktitle}
  {Electrokinetic Phenomena: Principles and Applications in Analytical
  Chemistry and Microchip Technology}}},\ \bibinfo {editor} {edited by\
  \bibinfo {editor} {\bibfnamefont {A.~S.}\ \bibnamefont {Rathore}}\ and\
  \bibinfo {editor} {\bibfnamefont {A.}~\bibnamefont {Guttman}}}\ (\bibinfo
  {publisher} {Marcel Dekker, Inc.},\ \bibinfo {address} {New York},\ \bibinfo
  {year} {2004})\ Chap.~\bibinfo {chapter} {15}, pp.\ \bibinfo {pages}
  {427--469}\BibitemShut {NoStop}%
\bibitem [{\citenamefont {Chakraborty}(2013)}]{C13_book}%
  \BibitemOpen
  \bibinfo {editor} {\bibfnamefont {S.}~\bibnamefont {Chakraborty}},\ ed.,\
  \href@noop {} {\emph {\bibinfo {title} {{Microfluidics and Microscale
  Transport Processes}}}},\ IIT Kharagpur Research Monograph Series\ (\bibinfo
  {publisher} {CRC Press},\ \bibinfo {address} {Boca Raton, FL},\ \bibinfo
  {year} {2013})\BibitemShut {NoStop}%
\bibitem [{\citenamefont {Huh}\ \emph {et~al.}(2010)\citenamefont {Huh},
  \citenamefont {Matthews}, \citenamefont {Mammoto}, \citenamefont
  {Montoya-Zavala}, \citenamefont {Hsin},\ and\ \citenamefont
  {Ingber}}]{Huhetal10}%
  \BibitemOpen
  \bibfield  {author} {\bibinfo {author} {\bibfnamefont {D.}~\bibnamefont
  {Huh}}, \bibinfo {author} {\bibfnamefont {B.~D.}\ \bibnamefont {Matthews}},
  \bibinfo {author} {\bibfnamefont {A.}~\bibnamefont {Mammoto}}, \bibinfo
  {author} {\bibfnamefont {M.}~\bibnamefont {Montoya-Zavala}}, \bibinfo
  {author} {\bibfnamefont {H.~Y.}\ \bibnamefont {Hsin}},\ and\ \bibinfo
  {author} {\bibfnamefont {D.~E.}\ \bibnamefont {Ingber}},\ }\bibfield  {title}
  {\bibinfo {title} {{Reconstituting organ-level lung functions on a chip}},\
  }\href {https://doi.org/10.1126/science.1188302} {\bibfield  {journal}
  {\bibinfo  {journal} {Science}\ }\textbf {\bibinfo {volume} {328}},\ \bibinfo
  {pages} {1662} (\bibinfo {year} {2010})}\BibitemShut {NoStop}%
\bibitem [{\citenamefont {Karan}\ \emph {et~al.}(2018)\citenamefont {Karan},
  \citenamefont {Chakraborty},\ and\ \citenamefont {Chakraborty}}]{KCC18}%
  \BibitemOpen
  \bibfield  {author} {\bibinfo {author} {\bibfnamefont {P.}~\bibnamefont
  {Karan}}, \bibinfo {author} {\bibfnamefont {J.}~\bibnamefont {Chakraborty}},\
  and\ \bibinfo {author} {\bibfnamefont {S.}~\bibnamefont {Chakraborty}},\
  }\bibfield  {title} {\bibinfo {title} {{Small-scale flow with deformable
  boundaries}},\ }\href {https://doi.org/10.1007/s41745-018-0073-5} {\bibfield
  {journal} {\bibinfo  {journal} {J. Indian Inst. Sci.}\ }\textbf {\bibinfo
  {volume} {98}},\ \bibinfo {pages} {159} (\bibinfo {year} {2018})}\BibitemShut
  {NoStop}%
\bibitem [{\citenamefont {Hosoi}\ and\ \citenamefont {Mahadevan}(2004)}]{HM04}%
  \BibitemOpen
  \bibfield  {author} {\bibinfo {author} {\bibfnamefont {A.~E.}\ \bibnamefont
  {Hosoi}}\ and\ \bibinfo {author} {\bibfnamefont {L.}~\bibnamefont
  {Mahadevan}},\ }\bibfield  {title} {\bibinfo {title} {{Peeling, healing, and
  bursting in a lubricated elastic sheet}},\ }\href
  {https://doi.org/10.1103/PhysRevLett.93.137802} {\bibfield  {journal}
  {\bibinfo  {journal} {Phys. Rev. Lett.}\ }\textbf {\bibinfo {volume} {93}},\
  \bibinfo {pages} {137802} (\bibinfo {year} {2004})}\BibitemShut {NoStop}%
\bibitem [{\citenamefont {Hewitt}\ \emph {et~al.}(2015)\citenamefont {Hewitt},
  \citenamefont {Balmforth},\ and\ \citenamefont {De~Bruyn}}]{HBDB14}%
  \BibitemOpen
  \bibfield  {author} {\bibinfo {author} {\bibfnamefont {I.~J.}\ \bibnamefont
  {Hewitt}}, \bibinfo {author} {\bibfnamefont {N.~J.}\ \bibnamefont
  {Balmforth}},\ and\ \bibinfo {author} {\bibfnamefont {J.~R.}\ \bibnamefont
  {De~Bruyn}},\ }\bibfield  {title} {\bibinfo {title} {{Elastic-plated gravity
  currents}},\ }\href {https://doi.org/10.1017/S0956792514000291} {\bibfield
  {journal} {\bibinfo  {journal} {Eur. J. Appl. Math.}\ }\textbf {\bibinfo
  {volume} {26}},\ \bibinfo {pages} {1} (\bibinfo {year} {2015})}\BibitemShut
  {NoStop}%
\bibitem [{\citenamefont {Ho}\ and\ \citenamefont {Tai}(1998)}]{HT98}%
  \BibitemOpen
  \bibfield  {author} {\bibinfo {author} {\bibfnamefont {C.-M.}\ \bibnamefont
  {Ho}}\ and\ \bibinfo {author} {\bibfnamefont {Y.-C.}\ \bibnamefont {Tai}},\
  }\bibfield  {title} {\bibinfo {title} {{Micro-electro-mechanical-systems
  (MEMS) and fluid flows}},\ }\href
  {https://doi.org/10.1146/annurev.fluid.30.1.579} {\bibfield  {journal}
  {\bibinfo  {journal} {Annu. Rev. Fluid Mech.}\ }\textbf {\bibinfo {volume}
  {30}},\ \bibinfo {pages} {579} (\bibinfo {year} {1998})}\BibitemShut
  {NoStop}%
\bibitem [{\citenamefont {Juel}\ \emph {et~al.}(2018)\citenamefont {Juel},
  \citenamefont {Pihler-Puzovi{\'{c}}},\ and\ \citenamefont {Heil}}]{JPPH18}%
  \BibitemOpen
  \bibfield  {author} {\bibinfo {author} {\bibfnamefont {A.}~\bibnamefont
  {Juel}}, \bibinfo {author} {\bibfnamefont {D.}~\bibnamefont
  {Pihler-Puzovi{\'{c}}}},\ and\ \bibinfo {author} {\bibfnamefont
  {M.}~\bibnamefont {Heil}},\ }\bibfield  {title} {\bibinfo {title}
  {{Instabilities in blistering}},\ }\href
  {https://doi.org/10.1146/annurev-fluid-122316-045106} {\bibfield  {journal}
  {\bibinfo  {journal} {Annu. Rev. Fluid Mech.}\ }\textbf {\bibinfo {volume}
  {50}},\ \bibinfo {pages} {691} (\bibinfo {year} {2018})}\BibitemShut
  {NoStop}%
\bibitem [{\citenamefont {Gomez}\ \emph {et~al.}(2017)\citenamefont {Gomez},
  \citenamefont {Moulton},\ and\ \citenamefont {Vella}}]{GMV17}%
  \BibitemOpen
  \bibfield  {author} {\bibinfo {author} {\bibfnamefont {M.}~\bibnamefont
  {Gomez}}, \bibinfo {author} {\bibfnamefont {D.~E.}\ \bibnamefont {Moulton}},\
  and\ \bibinfo {author} {\bibfnamefont {D.}~\bibnamefont {Vella}},\ }\bibfield
   {title} {\bibinfo {title} {{Passive control of viscous flow via elastic
  snap-through}},\ }\href {https://doi.org/10.1103/PhysRevLett.119.144502}
  {\bibfield  {journal} {\bibinfo  {journal} {Phys. Rev. Lett.}\ }\textbf
  {\bibinfo {volume} {119}},\ \bibinfo {pages} {144502} (\bibinfo {year}
  {2017})}\BibitemShut {NoStop}%
\bibitem [{\citenamefont {Borcia}\ \emph {et~al.}(2018)\citenamefont {Borcia},
  \citenamefont {Bestehorn}, \citenamefont {Uhlig}, \citenamefont {Gaudet},\
  and\ \citenamefont {Schenk}}]{BBUGS18}%
  \BibitemOpen
  \bibfield  {author} {\bibinfo {author} {\bibfnamefont {R.}~\bibnamefont
  {Borcia}}, \bibinfo {author} {\bibfnamefont {M.}~\bibnamefont {Bestehorn}},
  \bibinfo {author} {\bibfnamefont {S.}~\bibnamefont {Uhlig}}, \bibinfo
  {author} {\bibfnamefont {M.}~\bibnamefont {Gaudet}},\ and\ \bibinfo {author}
  {\bibfnamefont {H.}~\bibnamefont {Schenk}},\ }\bibfield  {title} {\bibinfo
  {title} {{Liquid pumping induced by transverse forced vibrations of an
  elastic beam: A lubrication approach}},\ }\href
  {https://doi.org/10.1103/PhysRevFluids.3.084202} {\bibfield  {journal}
  {\bibinfo  {journal} {Phys. Rev. Fluids}\ }\textbf {\bibinfo {volume} {3}},\
  \bibinfo {pages} {084202} (\bibinfo {year} {2018})}\BibitemShut {NoStop}%
\bibitem [{\citenamefont {Matia}\ \emph {et~al.}(2017)\citenamefont {Matia},
  \citenamefont {Elimelech},\ and\ \citenamefont {Gat}}]{MEG17}%
  \BibitemOpen
  \bibfield  {author} {\bibinfo {author} {\bibfnamefont {Y.}~\bibnamefont
  {Matia}}, \bibinfo {author} {\bibfnamefont {T.}~\bibnamefont {Elimelech}},\
  and\ \bibinfo {author} {\bibfnamefont {A.~D.}\ \bibnamefont {Gat}},\
  }\bibfield  {title} {\bibinfo {title} {{Leveraging internal viscous flow to
  extend the capabilities of beam-shaped soft robotic actuators}},\ }\href
  {https://doi.org/10.1089/soro.2016.0048} {\bibfield  {journal} {\bibinfo
  {journal} {Soft Robotics}\ }\textbf {\bibinfo {volume} {4}},\ \bibinfo
  {pages} {126} (\bibinfo {year} {2017})}\BibitemShut {NoStop}%
\bibitem [{\citenamefont {Boyko}\ \emph {et~al.}(2019)\citenamefont {Boyko},
  \citenamefont {Eshel}, \citenamefont {Gommed}, \citenamefont {Gat},\ and\
  \citenamefont {Bercovici}}]{BGB18}%
  \BibitemOpen
  \bibfield  {author} {\bibinfo {author} {\bibfnamefont {E.}~\bibnamefont
  {Boyko}}, \bibinfo {author} {\bibfnamefont {R.}~\bibnamefont {Eshel}},
  \bibinfo {author} {\bibfnamefont {K.}~\bibnamefont {Gommed}}, \bibinfo
  {author} {\bibfnamefont {A.~D.}\ \bibnamefont {Gat}},\ and\ \bibinfo {author}
  {\bibfnamefont {M.}~\bibnamefont {Bercovici}},\ }\bibfield  {title} {\bibinfo
  {title} {{Elastohydrodynamics of a pre-stretched finite elastic sheet
  lubricated by a thin viscous film with application to microfluidic soft
  actuators}},\ }\href {https://doi.org/10.1017/jfm.2018.967} {\bibfield
  {journal} {\bibinfo  {journal} {J. Fluid Mech.}\ }\textbf {\bibinfo {volume}
  {862}},\ \bibinfo {pages} {732} (\bibinfo {year} {2019})}\BibitemShut
  {NoStop}%
\bibitem [{\citenamefont {Salem}\ \emph {et~al.}(2020)\citenamefont {Salem},
  \citenamefont {Gamus}, \citenamefont {Or},\ and\ \citenamefont
  {Gat}}]{SGOG20}%
  \BibitemOpen
  \bibfield  {author} {\bibinfo {author} {\bibfnamefont {L.}~\bibnamefont
  {Salem}}, \bibinfo {author} {\bibfnamefont {B.}~\bibnamefont {Gamus}},
  \bibinfo {author} {\bibfnamefont {Y.}~\bibnamefont {Or}},\ and\ \bibinfo
  {author} {\bibfnamefont {A.~D.}\ \bibnamefont {Gat}},\ }\bibfield  {title}
  {\bibinfo {title} {{Leveraging Viscous Peeling to Create and Activate Soft
  Actuators and Microfluidic Devices}},\ }\href
  {https://doi.org/10.1089/soro.2019.0005} {\bibfield  {journal} {\bibinfo
  {journal} {Soft Robotics}\ }\textbf {\bibinfo {volume} {7}},\ \bibinfo
  {pages} {76} (\bibinfo {year} {2020})}\BibitemShut {NoStop}%
\bibitem [{\citenamefont {Xia}\ and\ \citenamefont {Whitesides}(1998)}]{XW98}%
  \BibitemOpen
  \bibfield  {author} {\bibinfo {author} {\bibfnamefont {Y.}~\bibnamefont
  {Xia}}\ and\ \bibinfo {author} {\bibfnamefont {G.~M.}\ \bibnamefont
  {Whitesides}},\ }\bibfield  {title} {\bibinfo {title} {{Soft lithography}},\
  }\href {https://doi.org/10.1146/annurev.matsci.28.1.153} {\bibfield
  {journal} {\bibinfo  {journal} {Annu. Rev. Mater. Sci.}\ }\textbf {\bibinfo
  {volume} {28}},\ \bibinfo {pages} {153} (\bibinfo {year} {1998})}\BibitemShut
  {NoStop}%
\bibitem [{\citenamefont {Therriault}\ \emph {et~al.}(2003)\citenamefont
  {Therriault}, \citenamefont {White},\ and\ \citenamefont {Lewis}}]{TWL03}%
  \BibitemOpen
  \bibfield  {author} {\bibinfo {author} {\bibfnamefont {D.}~\bibnamefont
  {Therriault}}, \bibinfo {author} {\bibfnamefont {S.~R.}\ \bibnamefont
  {White}},\ and\ \bibinfo {author} {\bibfnamefont {J.~A.}\ \bibnamefont
  {Lewis}},\ }\bibfield  {title} {\bibinfo {title} {{Chaotic mixing in
  three-dimensional microvascular networks fabricated by direct-write
  assembly}},\ }\href {https://doi.org/10.1038/nmat863} {\bibfield  {journal}
  {\bibinfo  {journal} {Nat. Mat.}\ }\textbf {\bibinfo {volume} {2}},\ \bibinfo
  {pages} {265} (\bibinfo {year} {2003})}\BibitemShut {NoStop}%
\bibitem [{\citenamefont {Kitson}\ \emph {et~al.}(2012)\citenamefont {Kitson},
  \citenamefont {Rosnes}, \citenamefont {Sans}, \citenamefont {Dragone},\ and\
  \citenamefont {Cronin}}]{KRSDC12}%
  \BibitemOpen
  \bibfield  {author} {\bibinfo {author} {\bibfnamefont {P.~J.}\ \bibnamefont
  {Kitson}}, \bibinfo {author} {\bibfnamefont {M.~H.}\ \bibnamefont {Rosnes}},
  \bibinfo {author} {\bibfnamefont {V.}~\bibnamefont {Sans}}, \bibinfo {author}
  {\bibfnamefont {V.}~\bibnamefont {Dragone}},\ and\ \bibinfo {author}
  {\bibfnamefont {L.}~\bibnamefont {Cronin}},\ }\bibfield  {title} {\bibinfo
  {title} {{Configurable 3D-Printed millifluidic and microfluidic `lab on a
  chip' reactionware devices}},\ }\href {https://doi.org/10.1039/c2lc40761b}
  {\bibfield  {journal} {\bibinfo  {journal} {Lab Chip}\ }\textbf {\bibinfo
  {volume} {12}},\ \bibinfo {pages} {3267} (\bibinfo {year}
  {2012})}\BibitemShut {NoStop}%
\bibitem [{\citenamefont {Su}\ \emph {et~al.}(2016)\citenamefont {Su},
  \citenamefont {Cook}, \citenamefont {Fang},\ and\ \citenamefont
  {Tentzeris}}]{SCFT16}%
  \BibitemOpen
  \bibfield  {author} {\bibinfo {author} {\bibfnamefont {W.}~\bibnamefont
  {Su}}, \bibinfo {author} {\bibfnamefont {B.~S.}\ \bibnamefont {Cook}},
  \bibinfo {author} {\bibfnamefont {Y.}~\bibnamefont {Fang}},\ and\ \bibinfo
  {author} {\bibfnamefont {M.~M.}\ \bibnamefont {Tentzeris}},\ }\bibfield
  {title} {\bibinfo {title} {{Fully inkjet-printed microfluidics: A solution to
  low-cost rapid three-dimensional microfluidics fabrication with numerous
  electrical and sensing applications}},\ }\href
  {https://doi.org/10.1038/srep35111} {\bibfield  {journal} {\bibinfo
  {journal} {Sci. Rep.}\ }\textbf {\bibinfo {volume} {6}},\ \bibinfo {pages}
  {35111} (\bibinfo {year} {2016})}\BibitemShut {NoStop}%
\bibitem [{\citenamefont {Bruus}(2008)}]{B08}%
  \BibitemOpen
  \bibfield  {author} {\bibinfo {author} {\bibfnamefont {H.}~\bibnamefont
  {Bruus}},\ }\href@noop {} {\emph {\bibinfo {title} {{Theoretical
  Microfluidics}}}},\ Oxford Master Series in Condensed Matter Physics\
  (\bibinfo  {publisher} {Oxford University Press},\ \bibinfo {address}
  {Oxford, UK},\ \bibinfo {year} {2008})\BibitemShut {NoStop}%
\bibitem [{\citenamefont {McDonald}\ and\ \citenamefont
  {Whitesides}(2002)}]{MW02}%
  \BibitemOpen
  \bibfield  {author} {\bibinfo {author} {\bibfnamefont {J.~C.}\ \bibnamefont
  {McDonald}}\ and\ \bibinfo {author} {\bibfnamefont {G.~M.}\ \bibnamefont
  {Whitesides}},\ }\bibfield  {title} {\bibinfo {title}
  {{Poly(dimethylsiloxane) as a material for fabricating microfluidic
  devices}},\ }\href {https://doi.org/10.1021/ar010110q} {\bibfield  {journal}
  {\bibinfo  {journal} {Acc. Chem. Res.}\ }\textbf {\bibinfo {volume} {35}},\
  \bibinfo {pages} {491} (\bibinfo {year} {2002})}\BibitemShut {NoStop}%
\bibitem [{\citenamefont {Friend}\ and\ \citenamefont {Yeo}(2010)}]{FY10}%
  \BibitemOpen
  \bibfield  {author} {\bibinfo {author} {\bibfnamefont {J.}~\bibnamefont
  {Friend}}\ and\ \bibinfo {author} {\bibfnamefont {L.}~\bibnamefont {Yeo}},\
  }\bibfield  {title} {\bibinfo {title} {{Fabrication of microfluidic devices
  using polydimethylsiloxane}},\ }\href {https://doi.org/10.1063/1.3259624}
  {\bibfield  {journal} {\bibinfo  {journal} {Biomicrofluidics}\ }\textbf
  {\bibinfo {volume} {4}},\ \bibinfo {pages} {026502} (\bibinfo {year}
  {2010})}\BibitemShut {NoStop}%
\bibitem [{\citenamefont {Johnston}\ \emph {et~al.}(2014)\citenamefont
  {Johnston}, \citenamefont {McCluskey}, \citenamefont {Tan},\ and\
  \citenamefont {Tracey}}]{JMTT14}%
  \BibitemOpen
  \bibfield  {author} {\bibinfo {author} {\bibfnamefont {I.~D.}\ \bibnamefont
  {Johnston}}, \bibinfo {author} {\bibfnamefont {D.~K.}\ \bibnamefont
  {McCluskey}}, \bibinfo {author} {\bibfnamefont {C.~K.~L.}\ \bibnamefont
  {Tan}},\ and\ \bibinfo {author} {\bibfnamefont {M.~C.}\ \bibnamefont
  {Tracey}},\ }\bibfield  {title} {\bibinfo {title} {{Mechanical
  characterization of bulk Sylgard 184 for microfluidics and
  microengineering}},\ }\href {https://doi.org/10.1088/0960-1317/24/3/035017}
  {\bibfield  {journal} {\bibinfo  {journal} {J. Micromech. Microeng.}\
  }\textbf {\bibinfo {volume} {24}},\ \bibinfo {pages} {35017} (\bibinfo {year}
  {2014})}\BibitemShut {NoStop}%
\bibitem [{\citenamefont {Verma}\ and\ \citenamefont {Kumaran}(2013)}]{VK13}%
  \BibitemOpen
  \bibfield  {author} {\bibinfo {author} {\bibfnamefont {M.~K.~S.}\
  \bibnamefont {Verma}}\ and\ \bibinfo {author} {\bibfnamefont
  {V.}~\bibnamefont {Kumaran}},\ }\bibfield  {title} {\bibinfo {title} {{A
  multifold reduction in the transition Reynolds number, and ultra-fast mixing,
  in a micro-channel due to a dynamical instability induced by a soft wall}},\
  }\href {https://doi.org/10.1017/jfm.2013.264} {\bibfield  {journal} {\bibinfo
   {journal} {J. Fluid Mech.}\ }\textbf {\bibinfo {volume} {727}},\ \bibinfo
  {pages} {407} (\bibinfo {year} {2013})}\BibitemShut {NoStop}%
\bibitem [{\citenamefont {Squires}\ and\ \citenamefont {Quake}(2005)}]{SQ05}%
  \BibitemOpen
  \bibfield  {author} {\bibinfo {author} {\bibfnamefont {T.~M.}\ \bibnamefont
  {Squires}}\ and\ \bibinfo {author} {\bibfnamefont {S.~R.}\ \bibnamefont
  {Quake}},\ }\bibfield  {title} {\bibinfo {title} {{Microfluidics: Fluid
  physics at the nanoliter scale}},\ }\href
  {https://doi.org/10.1103/RevModPhys.77.977} {\bibfield  {journal} {\bibinfo
  {journal} {Rev. Mod. Phys.}\ }\textbf {\bibinfo {volume} {77}},\ \bibinfo
  {pages} {977} (\bibinfo {year} {2005})}\BibitemShut {NoStop}%
\bibitem [{\citenamefont {Stone}\ \emph {et~al.}(2004)\citenamefont {Stone},
  \citenamefont {Stroock},\ and\ \citenamefont {Ajdari}}]{SSA04}%
  \BibitemOpen
  \bibfield  {author} {\bibinfo {author} {\bibfnamefont {H.~A.}\ \bibnamefont
  {Stone}}, \bibinfo {author} {\bibfnamefont {A.~D.}\ \bibnamefont {Stroock}},\
  and\ \bibinfo {author} {\bibfnamefont {A.}~\bibnamefont {Ajdari}},\
  }\bibfield  {title} {\bibinfo {title} {{Engineering flows in small devices:
  Microfluidics toward a Lab-on-a-Chip}},\ }\href
  {https://doi.org/10.1146/annurev.fluid.36.050802.122124} {\bibfield
  {journal} {\bibinfo  {journal} {Annu. Rev. Fluid Mech.}\ }\textbf {\bibinfo
  {volume} {36}},\ \bibinfo {pages} {381} (\bibinfo {year} {2004})}\BibitemShut
  {NoStop}%
\bibitem [{\citenamefont {Panton}(2013)}]{panton}%
  \BibitemOpen
  \bibfield  {author} {\bibinfo {author} {\bibfnamefont {R.~L.}\ \bibnamefont
  {Panton}},\ }\href {https://doi.org/10.1002/9781118713075} {\emph {\bibinfo
  {title} {{Incompressible Flow}}}},\ \bibinfo {edition} {4th}\ ed.\ (\bibinfo
  {publisher} {John Wiley {\&} Sons},\ \bibinfo {address} {Hoboken, NJ},\
  \bibinfo {year} {2013})\BibitemShut {NoStop}%
\bibitem [{\citenamefont {Holden}\ \emph {et~al.}(2003)\citenamefont {Holden},
  \citenamefont {Kumar}, \citenamefont {Beskok},\ and\ \citenamefont
  {Cremer}}]{HKBC03}%
  \BibitemOpen
  \bibfield  {author} {\bibinfo {author} {\bibfnamefont {M.~A.}\ \bibnamefont
  {Holden}}, \bibinfo {author} {\bibfnamefont {S.}~\bibnamefont {Kumar}},
  \bibinfo {author} {\bibfnamefont {A.}~\bibnamefont {Beskok}},\ and\ \bibinfo
  {author} {\bibfnamefont {P.~S.}\ \bibnamefont {Cremer}},\ }\bibfield  {title}
  {\bibinfo {title} {{Microfluidic diffusion diluter: bulging of PDMS
  microchannels under pressure-driven flow}},\ }\href
  {https://doi.org/10.1088/0960-1317/13/3/309} {\bibfield  {journal} {\bibinfo
  {journal} {J. Micromech. Microeng.}\ }\textbf {\bibinfo {volume} {13}},\
  \bibinfo {pages} {412} (\bibinfo {year} {2003})}\BibitemShut {NoStop}%
\bibitem [{\citenamefont {Gervais}\ \emph {et~al.}(2006)\citenamefont
  {Gervais}, \citenamefont {El-Ali}, \citenamefont {G{\"{u}}nther},\ and\
  \citenamefont {Jensen}}]{GEGJ06}%
  \BibitemOpen
  \bibfield  {author} {\bibinfo {author} {\bibfnamefont {T.}~\bibnamefont
  {Gervais}}, \bibinfo {author} {\bibfnamefont {J.}~\bibnamefont {El-Ali}},
  \bibinfo {author} {\bibfnamefont {A.}~\bibnamefont {G{\"{u}}nther}},\ and\
  \bibinfo {author} {\bibfnamefont {K.~F.}\ \bibnamefont {Jensen}},\ }\bibfield
   {title} {\bibinfo {title} {{Flow-induced deformation of shallow microfluidic
  channels}},\ }\href {https://doi.org/10.1039/b513524a} {\bibfield  {journal}
  {\bibinfo  {journal} {Lab Chip}\ }\textbf {\bibinfo {volume} {6}},\ \bibinfo
  {pages} {500} (\bibinfo {year} {2006})}\BibitemShut {NoStop}%
\bibitem [{\citenamefont {Raj}\ \emph {et~al.}(2017)\citenamefont {Raj},
  \citenamefont {DasGupta},\ and\ \citenamefont {Chakraborty}}]{RDC17}%
  \BibitemOpen
  \bibfield  {author} {\bibinfo {author} {\bibfnamefont {M.~K.}\ \bibnamefont
  {Raj}}, \bibinfo {author} {\bibfnamefont {S.}~\bibnamefont {DasGupta}},\ and\
  \bibinfo {author} {\bibfnamefont {S.}~\bibnamefont {Chakraborty}},\
  }\bibfield  {title} {\bibinfo {title} {{Hydrodynamics in deformable
  microchannels}},\ }\href {https://doi.org/10.1007/s10404-017-1908-5}
  {\bibfield  {journal} {\bibinfo  {journal} {Microfluid. Nanofluid.}\ }\textbf
  {\bibinfo {volume} {21}},\ \bibinfo {pages} {70} (\bibinfo {year}
  {2017})}\BibitemShut {NoStop}%
\bibitem [{\citenamefont {Christov}\ \emph {et~al.}(2018)\citenamefont
  {Christov}, \citenamefont {Cognet}, \citenamefont {Shidhore},\ and\
  \citenamefont {Stone}}]{CCSS17}%
  \BibitemOpen
  \bibfield  {author} {\bibinfo {author} {\bibfnamefont {I.~C.}\ \bibnamefont
  {Christov}}, \bibinfo {author} {\bibfnamefont {V.}~\bibnamefont {Cognet}},
  \bibinfo {author} {\bibfnamefont {T.~C.}\ \bibnamefont {Shidhore}},\ and\
  \bibinfo {author} {\bibfnamefont {H.~A.}\ \bibnamefont {Stone}},\ }\bibfield
  {title} {\bibinfo {title} {{Flow rate--pressure drop relation for deformable
  shallow microfluidic channels}},\ }\href
  {https://doi.org/10.1017/jfm.2018.30} {\bibfield  {journal} {\bibinfo
  {journal} {J. Fluid Mech.}\ }\textbf {\bibinfo {volume} {814}},\ \bibinfo
  {pages} {267} (\bibinfo {year} {2018})}\BibitemShut {NoStop}%
\bibitem [{Note1()}]{Note1}%
  \BibitemOpen
  \bibinfo {note} {See, e.g., Fung's illustration \cite [Figure~3.4:2]{F97} for
  a schematic visual representation of this FSI feed-back loop in hemoelastic
  system.}\BibitemShut {Stop}%
\bibitem [{\citenamefont {Srinivas}\ and\ \citenamefont
  {Kumaran}(2017)}]{SK17}%
  \BibitemOpen
  \bibfield  {author} {\bibinfo {author} {\bibfnamefont {S.~S.}\ \bibnamefont
  {Srinivas}}\ and\ \bibinfo {author} {\bibfnamefont {V.}~\bibnamefont
  {Kumaran}},\ }\bibfield  {title} {\bibinfo {title} {{Effect of
  viscoelasticity on the soft-wall transition and turbulence in a
  microchannel}},\ }\href {https://doi.org/10.1017/jfm.2016.839} {\bibfield
  {journal} {\bibinfo  {journal} {J. Fluid Mech.}\ }\textbf {\bibinfo {volume}
  {812}},\ \bibinfo {pages} {1076} (\bibinfo {year} {2017})}\BibitemShut
  {NoStop}%
\bibitem [{\citenamefont {Verma}\ and\ \citenamefont {Kumaran}(2015)}]{VK15}%
  \BibitemOpen
  \bibfield  {author} {\bibinfo {author} {\bibfnamefont {M.~K.~S.}\
  \bibnamefont {Verma}}\ and\ \bibinfo {author} {\bibfnamefont
  {V.}~\bibnamefont {Kumaran}},\ }\bibfield  {title} {\bibinfo {title}
  {{Stability of the flow in a soft tube deformed due to an applied pressure
  gradient}},\ }\href {https://doi.org/10.1103/PhysRevE.91.043001} {\bibfield
  {journal} {\bibinfo  {journal} {Phys. Rev. E}\ }\textbf {\bibinfo {volume}
  {91}},\ \bibinfo {pages} {043001} (\bibinfo {year} {2015})}\BibitemShut
  {NoStop}%
\bibitem [{\citenamefont {Fung}(1997)}]{F97}%
  \BibitemOpen
  \bibfield  {author} {\bibinfo {author} {\bibfnamefont {Y.~C.}\ \bibnamefont
  {Fung}},\ }\href {https://doi.org/10.1007/978-1-4757-2696-1} {\emph {\bibinfo
  {title} {{Biomechanics: Circulation}}}},\ \bibinfo {edition} {2nd}\ ed.\
  (\bibinfo  {publisher} {Springer-Verlag},\ \bibinfo {address} {New York,
  NY},\ \bibinfo {year} {1997})\BibitemShut {NoStop}%
\bibitem [{\citenamefont {Pedley}(1980)}]{P80}%
  \BibitemOpen
  \bibfield  {author} {\bibinfo {author} {\bibfnamefont {T.~J.}\ \bibnamefont
  {Pedley}},\ }\href@noop {} {\emph {\bibinfo {title} {{The Fluid Mechanics of
  Large Blood Vessels}}}}\ (\bibinfo  {publisher} {Cambridge University
  Press},\ \bibinfo {address} {Cambridge},\ \bibinfo {year} {1980})\BibitemShut
  {NoStop}%
\bibitem [{\citenamefont {Grotberg}\ and\ \citenamefont {Jensen}(2004)}]{GJ04}%
  \BibitemOpen
  \bibfield  {author} {\bibinfo {author} {\bibfnamefont {J.~B.}\ \bibnamefont
  {Grotberg}}\ and\ \bibinfo {author} {\bibfnamefont {O.~E.}\ \bibnamefont
  {Jensen}},\ }\bibfield  {title} {\bibinfo {title} {{Biofluid mechanics in
  flexible tubes}},\ }\href
  {https://doi.org/10.1146/annurev.fluid.36.050802.121918} {\bibfield
  {journal} {\bibinfo  {journal} {Annu. Rev. Fluid Mech.}\ }\textbf {\bibinfo
  {volume} {36}},\ \bibinfo {pages} {121} (\bibinfo {year} {2004})}\BibitemShut
  {NoStop}%
\bibitem [{\citenamefont {Lind}\ \emph {et~al.}(2017)\citenamefont {Lind},
  \citenamefont {Busbee}, \citenamefont {Valentine}, \citenamefont
  {Pasqualini}, \citenamefont {Yuan}, \citenamefont {Yadid}, \citenamefont
  {Park}, \citenamefont {Kotikian}, \citenamefont {Nesmith}, \citenamefont
  {Campbell}, \citenamefont {Vlassak}, \citenamefont {Lewis},\ and\
  \citenamefont {Parker}}]{Lindetal17}%
  \BibitemOpen
  \bibfield  {author} {\bibinfo {author} {\bibfnamefont {J.~U.}\ \bibnamefont
  {Lind}}, \bibinfo {author} {\bibfnamefont {T.~A.}\ \bibnamefont {Busbee}},
  \bibinfo {author} {\bibfnamefont {A.~D.}\ \bibnamefont {Valentine}}, \bibinfo
  {author} {\bibfnamefont {F.~S.}\ \bibnamefont {Pasqualini}}, \bibinfo
  {author} {\bibfnamefont {H.}~\bibnamefont {Yuan}}, \bibinfo {author}
  {\bibfnamefont {M.}~\bibnamefont {Yadid}}, \bibinfo {author} {\bibfnamefont
  {S.~J.}\ \bibnamefont {Park}}, \bibinfo {author} {\bibfnamefont
  {A.}~\bibnamefont {Kotikian}}, \bibinfo {author} {\bibfnamefont {A.~P.}\
  \bibnamefont {Nesmith}}, \bibinfo {author} {\bibfnamefont {P.~H.}\
  \bibnamefont {Campbell}}, \bibinfo {author} {\bibfnamefont {J.~J.}\
  \bibnamefont {Vlassak}}, \bibinfo {author} {\bibfnamefont {J.~A.}\
  \bibnamefont {Lewis}},\ and\ \bibinfo {author} {\bibfnamefont {K.~K.}\
  \bibnamefont {Parker}},\ }\bibfield  {title} {\bibinfo {title} {{Instrumented
  cardiac microphysiological devices via multimaterial three-dimensional
  printing}},\ }\href {https://doi.org/10.1038/nmat4782} {\bibfield  {journal}
  {\bibinfo  {journal} {Nat. Mat.}\ }\textbf {\bibinfo {volume} {16}},\
  \bibinfo {pages} {303} (\bibinfo {year} {2017})}\BibitemShut {NoStop}%
\bibitem [{\citenamefont {Dendukuri}\ \emph {et~al.}(2007)\citenamefont
  {Dendukuri}, \citenamefont {Gu}, \citenamefont {Pregibon}, \citenamefont
  {Hatton},\ and\ \citenamefont {Doyle}}]{DGPHD07}%
  \BibitemOpen
  \bibfield  {author} {\bibinfo {author} {\bibfnamefont {D.}~\bibnamefont
  {Dendukuri}}, \bibinfo {author} {\bibfnamefont {S.~S.}\ \bibnamefont {Gu}},
  \bibinfo {author} {\bibfnamefont {D.~C.}\ \bibnamefont {Pregibon}}, \bibinfo
  {author} {\bibfnamefont {T.~A.}\ \bibnamefont {Hatton}},\ and\ \bibinfo
  {author} {\bibfnamefont {P.~S.}\ \bibnamefont {Doyle}},\ }\bibfield  {title}
  {\bibinfo {title} {{Stop-flow lithography in a microfluidic device}},\ }\href
  {https://doi.org/10.1039/b703457a} {\bibfield  {journal} {\bibinfo  {journal}
  {Lab Chip}\ }\textbf {\bibinfo {volume} {7}},\ \bibinfo {pages} {818}
  (\bibinfo {year} {2007})}\BibitemShut {NoStop}%
\bibitem [{\citenamefont {Mukherjee}\ \emph {et~al.}(2013)\citenamefont
  {Mukherjee}, \citenamefont {Chakraborty},\ and\ \citenamefont
  {Chakraborty}}]{MCC13}%
  \BibitemOpen
  \bibfield  {author} {\bibinfo {author} {\bibfnamefont {U.}~\bibnamefont
  {Mukherjee}}, \bibinfo {author} {\bibfnamefont {J.}~\bibnamefont
  {Chakraborty}},\ and\ \bibinfo {author} {\bibfnamefont {S.}~\bibnamefont
  {Chakraborty}},\ }\bibfield  {title} {\bibinfo {title} {{Relaxation
  characteristics of a compliant microfluidic channel under electroosmotic
  flow}},\ }\href {https://doi.org/10.1039/c2sm27247d} {\bibfield  {journal}
  {\bibinfo  {journal} {Soft Matter}\ }\textbf {\bibinfo {volume} {9}},\
  \bibinfo {pages} {1562} (\bibinfo {year} {2013})}\BibitemShut {NoStop}%
\bibitem [{\citenamefont {Naik}\ \emph {et~al.}(2017)\citenamefont {Naik},
  \citenamefont {Chakraborty},\ and\ \citenamefont {Chakraborty}}]{NCC17}%
  \BibitemOpen
  \bibfield  {author} {\bibinfo {author} {\bibfnamefont {K.~G.}\ \bibnamefont
  {Naik}}, \bibinfo {author} {\bibfnamefont {S.}~\bibnamefont {Chakraborty}},\
  and\ \bibinfo {author} {\bibfnamefont {J.}~\bibnamefont {Chakraborty}},\
  }\bibfield  {title} {\bibinfo {title} {{Finite size effects of ionic species
  sensitively determine load bearing capacities of lubricated systems under
  combined influence of electrokinetics and surface compliance}},\ }\href
  {https://doi.org/10.1039/C7SM01423F} {\bibfield  {journal} {\bibinfo
  {journal} {Soft Matter}\ }\textbf {\bibinfo {volume} {13}},\ \bibinfo {pages}
  {6422} (\bibinfo {year} {2017})}\BibitemShut {NoStop}%
\bibitem [{\citenamefont {Mart{\'{i}}nez-Calvo}\ \emph
  {et~al.}(2020)\citenamefont {Mart{\'{i}}nez-Calvo}, \citenamefont {Sevilla},
  \citenamefont {Peng},\ and\ \citenamefont {Stone}}]{MCSPS19}%
  \BibitemOpen
  \bibfield  {author} {\bibinfo {author} {\bibfnamefont {A.}~\bibnamefont
  {Mart{\'{i}}nez-Calvo}}, \bibinfo {author} {\bibfnamefont {A.}~\bibnamefont
  {Sevilla}}, \bibinfo {author} {\bibfnamefont {G.~G.}\ \bibnamefont {Peng}},\
  and\ \bibinfo {author} {\bibfnamefont {H.~A.}\ \bibnamefont {Stone}},\
  }\bibfield  {title} {\bibinfo {title} {{Start-up flow in shallow deformable
  microchannels}},\ }\href {https://doi.org/10.1017/jfm.2019.994} {\bibfield
  {journal} {\bibinfo  {journal} {J. Fluid Mech.}\ }\textbf {\bibinfo {volume}
  {885}},\ \bibinfo {pages} {A25} (\bibinfo {year} {2020})}\BibitemShut
  {NoStop}%
\bibitem [{\citenamefont {Riley}\ \emph {et~al.}(1988)\citenamefont {Riley},
  \citenamefont {Gad-el Hak},\ and\ \citenamefont {Metcalfe}}]{RGHM88}%
  \BibitemOpen
  \bibfield  {author} {\bibinfo {author} {\bibfnamefont {J.~J.}\ \bibnamefont
  {Riley}}, \bibinfo {author} {\bibfnamefont {M.}~\bibnamefont {Gad-el Hak}},\
  and\ \bibinfo {author} {\bibfnamefont {R.~W.}\ \bibnamefont {Metcalfe}},\
  }\bibfield  {title} {\bibinfo {title} {{Complaint coatings}},\ }\href
  {https://doi.org/10.1146/annurev.fl.20.010188.002141} {\bibfield  {journal}
  {\bibinfo  {journal} {Annu. Rev. Fluid Mech.}\ }\textbf {\bibinfo {volume}
  {20}},\ \bibinfo {pages} {393} (\bibinfo {year} {1988})}\BibitemShut
  {NoStop}%
\bibitem [{\citenamefont {Gad-el Hak}(1996)}]{G96}%
  \BibitemOpen
  \bibfield  {author} {\bibinfo {author} {\bibfnamefont {M.}~\bibnamefont
  {Gad-el Hak}},\ }\bibfield  {title} {\bibinfo {title} {{Compliant coatings: A
  decade of progress}},\ }\href {https://doi.org/10.1115/1.3101966} {\bibfield
  {journal} {\bibinfo  {journal} {Appl. Mech. Rev.}\ }\textbf {\bibinfo
  {volume} {49}},\ \bibinfo {pages} {S147} (\bibinfo {year}
  {1996})}\BibitemShut {NoStop}%
\bibitem [{Pai(2003)}]{Pai03}%
  \BibitemOpen
  \bibfield  {title} {\bibinfo {title} {{Plates in axial flow}},\ }in\ \href
  {https://doi.org/10.1016/S1874-5652(04)80006-1} {\emph {\bibinfo {booktitle}
  {Fluid-Structure Interactions}}},\ Vol.~\bibinfo {volume} {2},\ \bibinfo
  {editor} {edited by\ \bibinfo {editor} {\bibfnamefont {M.~P.}\ \bibnamefont
  {Pa{\"{i}}doussis}}}\ (\bibinfo  {publisher} {Academic Press},\ \bibinfo
  {year} {2003})\ Chap.~\bibinfo {chapter} {10}, pp.\ \bibinfo {pages}
  {1137--1220}\BibitemShut {NoStop}%
\bibitem [{\citenamefont {Luo}\ and\ \citenamefont {Pedley}(1998)}]{LP98}%
  \BibitemOpen
  \bibfield  {author} {\bibinfo {author} {\bibfnamefont {X.~Y.}\ \bibnamefont
  {Luo}}\ and\ \bibinfo {author} {\bibfnamefont {T.~J.}\ \bibnamefont
  {Pedley}},\ }\bibfield  {title} {\bibinfo {title} {{The effects of wall
  inertia on flow in a two-dimensional collapsible channel}},\ }\href
  {https://doi.org/10.1017/S0022112098001062} {\bibfield  {journal} {\bibinfo
  {journal} {J. Fluid Mech.}\ }\textbf {\bibinfo {volume} {363}},\ \bibinfo
  {pages} {253} (\bibinfo {year} {1998})}\BibitemShut {NoStop}%
\bibitem [{\citenamefont {Cai}\ and\ \citenamefont {Luo}(2003)}]{CL03}%
  \BibitemOpen
  \bibfield  {author} {\bibinfo {author} {\bibfnamefont {Z.~X.}\ \bibnamefont
  {Cai}}\ and\ \bibinfo {author} {\bibfnamefont {X.~Y.}\ \bibnamefont {Luo}},\
  }\bibfield  {title} {\bibinfo {title} {{A fluid--beam model for flow in a
  collapsible channel}},\ }\href
  {https://doi.org/10.1016/S0889-9746(02)00112-3} {\bibfield  {journal}
  {\bibinfo  {journal} {J. Fluids Struct.}\ }\textbf {\bibinfo {volume} {17}},\
  \bibinfo {pages} {125} (\bibinfo {year} {2003})}\BibitemShut {NoStop}%
\bibitem [{\citenamefont {Xu}\ \emph {et~al.}(2013)\citenamefont {Xu},
  \citenamefont {Billingham},\ and\ \citenamefont {Jensen}}]{XBJ13}%
  \BibitemOpen
  \bibfield  {author} {\bibinfo {author} {\bibfnamefont {F.}~\bibnamefont
  {Xu}}, \bibinfo {author} {\bibfnamefont {J.}~\bibnamefont {Billingham}},\
  and\ \bibinfo {author} {\bibfnamefont {O.~E.}\ \bibnamefont {Jensen}},\
  }\bibfield  {title} {\bibinfo {title} {{Divergence-driven oscillations in a
  flexible-channel flow with fixed upstream flux}},\ }\href
  {https://doi.org/10.1017/jfm.2013.97} {\bibfield  {journal} {\bibinfo
  {journal} {J. Fluid Mech.}\ }\textbf {\bibinfo {volume} {723}},\ \bibinfo
  {pages} {706} (\bibinfo {year} {2013})}\BibitemShut {NoStop}%
\bibitem [{\citenamefont {Duprat}\ and\ \citenamefont {Stone}(2016)}]{DS16}%
  \BibitemOpen
  \bibinfo {editor} {\bibfnamefont {C.}~\bibnamefont {Duprat}}\ and\ \bibinfo
  {editor} {\bibfnamefont {H.~A.}\ \bibnamefont {Stone}},\ eds.,\ \href
  {https://doi.org/10.1039/9781782628491} {\emph {\bibinfo {title}
  {{Fluid--Structure Interactions in Low-Reynolds-Number Flows}}}}\ (\bibinfo
  {publisher} {The Royal Society of Chemistry},\ \bibinfo {address} {Cambridge,
  UK},\ \bibinfo {year} {2016})\BibitemShut {NoStop}%
\bibitem [{\citenamefont {Ottino}(1989)}]{O89}%
  \BibitemOpen
  \bibfield  {author} {\bibinfo {author} {\bibfnamefont {J.~M.}\ \bibnamefont
  {Ottino}},\ }\href {https://doi.org/10.2277/0521368782} {\emph {\bibinfo
  {title} {{The Kinematics of Mixing: Stretching, Chaos, and Transport}}}},\
  \bibinfo {series} {Cambridge Texts in Applied Mathematics}, Vol.~\bibinfo
  {volume} {3}\ (\bibinfo  {publisher} {Cambridge University Press},\ \bibinfo
  {address} {Cambridge, UK},\ \bibinfo {year} {1989})\BibitemShut {NoStop}%
\bibitem [{\citenamefont {Ottino}\ and\ \citenamefont {Wiggins}(2004)}]{OW04}%
  \BibitemOpen
  \bibfield  {author} {\bibinfo {author} {\bibfnamefont {J.~M.}\ \bibnamefont
  {Ottino}}\ and\ \bibinfo {author} {\bibfnamefont {S.}~\bibnamefont
  {Wiggins}},\ }\bibfield  {title} {\bibinfo {title} {{Introduction: mixing in
  microfluidics}},\ }\href {https://doi.org/10.1098/rsta.2003.1355} {\bibfield
  {journal} {\bibinfo  {journal} {Phil. Trans. R. Soc. A}\ }\textbf {\bibinfo
  {volume} {362}},\ \bibinfo {pages} {923} (\bibinfo {year}
  {2004})}\BibitemShut {NoStop}%
\bibitem [{Note2()}]{Note2}%
  \BibitemOpen
  \bibinfo {note} {{A less general unsteady 1D model was derived in \cite
  {BBUGS18}, in parallel to the present work. The latter model does not
  consider fluid inertia, stretching of the elastic wall, the inflationary
  nonlinear dynamics, or the stability of the deformed channel
  shape.}}\BibitemShut {Stop}%
\bibitem [{\citenamefont {Reddy}(2007)}]{reddy07}%
  \BibitemOpen
  \bibfield  {author} {\bibinfo {author} {\bibfnamefont {J.~N.}\ \bibnamefont
  {Reddy}},\ }\href@noop {} {\emph {\bibinfo {title} {{Theory and Analysis of
  Elastic Plates and Shells}}}},\ \bibinfo {edition} {2nd}\ ed.\ (\bibinfo
  {publisher} {CRC Press, an imprint of Taylor {\&} Francis Group},\ \bibinfo
  {address} {Boca Raton, FL},\ \bibinfo {year} {2007})\BibitemShut {NoStop}%
\bibitem [{\citenamefont {Kodio}\ \emph {et~al.}(2017)\citenamefont {Kodio},
  \citenamefont {Griffiths},\ and\ \citenamefont {Vella}}]{KGV17}%
  \BibitemOpen
  \bibfield  {author} {\bibinfo {author} {\bibfnamefont {O.}~\bibnamefont
  {Kodio}}, \bibinfo {author} {\bibfnamefont {I.~M.}\ \bibnamefont
  {Griffiths}},\ and\ \bibinfo {author} {\bibfnamefont {D.}~\bibnamefont
  {Vella}},\ }\bibfield  {title} {\bibinfo {title} {{Lubricated wrinkles:
  Imposed constraints affect the dynamics of wrinkle coarsening}},\ }\href
  {https://doi.org/10.1103/PhysRevFluids.2.014202} {\bibfield  {journal}
  {\bibinfo  {journal} {Phys. Rev. Fluids}\ }\textbf {\bibinfo {volume} {2}},\
  \bibinfo {pages} {014202} (\bibinfo {year} {2017})}\BibitemShut {NoStop}%
\bibitem [{\citenamefont {Stewart}\ \emph {et~al.}(2009)\citenamefont
  {Stewart}, \citenamefont {Waters},\ and\ \citenamefont {Jensen}}]{SWJ09}%
  \BibitemOpen
  \bibfield  {author} {\bibinfo {author} {\bibfnamefont {P.~S.}\ \bibnamefont
  {Stewart}}, \bibinfo {author} {\bibfnamefont {S.~L.}\ \bibnamefont
  {Waters}},\ and\ \bibinfo {author} {\bibfnamefont {O.~E.}\ \bibnamefont
  {Jensen}},\ }\bibfield  {title} {\bibinfo {title} {{Local and global
  instabilities of flow in a flexible-walled channel}},\ }\href
  {https://doi.org/10.1016/j.euromechflu.2009.03.002} {\bibfield  {journal}
  {\bibinfo  {journal} {Eur. J. Mech. B/Fluids}\ }\textbf {\bibinfo {volume}
  {28}},\ \bibinfo {pages} {541} (\bibinfo {year} {2009})}\BibitemShut
  {NoStop}%
\bibitem [{Note3()}]{Note3}%
  \BibitemOpen
  \bibinfo {note} {We use the notation $\Sigma ^*$, which coincides with the
  notation in \cite {VK13} and the references therein, because our definition
  is similar to the dimensionless solid mechanics parameter
  therein.}\BibitemShut {Stop}%
\bibitem [{\citenamefont {Sollier}\ \emph {et~al.}(2011)\citenamefont
  {Sollier}, \citenamefont {Murray}, \citenamefont {Maoddi},\ and\
  \citenamefont {Di~Carlo}}]{SMMC11}%
  \BibitemOpen
  \bibfield  {author} {\bibinfo {author} {\bibfnamefont {E.}~\bibnamefont
  {Sollier}}, \bibinfo {author} {\bibfnamefont {C.}~\bibnamefont {Murray}},
  \bibinfo {author} {\bibfnamefont {P.}~\bibnamefont {Maoddi}},\ and\ \bibinfo
  {author} {\bibfnamefont {D.}~\bibnamefont {Di~Carlo}},\ }\bibfield  {title}
  {\bibinfo {title} {{Rapid prototyping polymers for microfluidic devices and
  high pressure injections}},\ }\href {https://doi.org/10.1039/c1lc20514e}
  {\bibfield  {journal} {\bibinfo  {journal} {Lab Chip}\ }\textbf {\bibinfo
  {volume} {11}},\ \bibinfo {pages} {3752} (\bibinfo {year}
  {2011})}\BibitemShut {NoStop}%
\bibitem [{\citenamefont {Stewart}(2017)}]{S17}%
  \BibitemOpen
  \bibfield  {author} {\bibinfo {author} {\bibfnamefont {P.~S.}\ \bibnamefont
  {Stewart}},\ }\bibfield  {title} {\bibinfo {title} {{Instabilities in
  flexible channel flow with large external pressure}},\ }\href
  {https://doi.org/10.1017/jfm.2017.404} {\bibfield  {journal} {\bibinfo
  {journal} {J. Fluid Mech.}\ }\textbf {\bibinfo {volume} {825}},\ \bibinfo
  {pages} {922} (\bibinfo {year} {2017})}\BibitemShut {NoStop}%
\bibitem [{\citenamefont {Virtanen}\ \emph {et~al.}(2020)\citenamefont
  {Virtanen}, \citenamefont {Gommers}, \citenamefont {Oliphant}, \citenamefont
  {Haberland}, \citenamefont {Reddy}, \citenamefont {Cournapeau}, \citenamefont
  {Burovski}, \citenamefont {Peterson}, \citenamefont {Weckesser},
  \citenamefont {Bright}, \citenamefont {van~der Walt}, \citenamefont {Brett},
  \citenamefont {Wilson}, \citenamefont {Millman}, \citenamefont {Mayorov},
  \citenamefont {Nelson}, \citenamefont {Jones}, \citenamefont {Kern},
  \citenamefont {Larson}, \citenamefont {Carey}, \citenamefont {Polat},
  \citenamefont {Feng}, \citenamefont {Moore}, \citenamefont {VanderPlas},
  \citenamefont {Laxalde}, \citenamefont {Perktold}, \citenamefont {Cimrman},
  \citenamefont {Henriksen}, \citenamefont {Quintero}, \citenamefont {Harris},
  \citenamefont {Archibald}, \citenamefont {Ribeiro}, \citenamefont
  {Pedregosa},\ and\ \citenamefont {van Mulbregt}}]{SciPy}%
  \BibitemOpen
  \bibfield  {author} {\bibinfo {author} {\bibfnamefont {P.}~\bibnamefont
  {Virtanen}}, \bibinfo {author} {\bibfnamefont {R.}~\bibnamefont {Gommers}},
  \bibinfo {author} {\bibfnamefont {T.~E.}\ \bibnamefont {Oliphant}}, \bibinfo
  {author} {\bibfnamefont {M.}~\bibnamefont {Haberland}}, \bibinfo {author}
  {\bibfnamefont {T.}~\bibnamefont {Reddy}}, \bibinfo {author} {\bibfnamefont
  {D.}~\bibnamefont {Cournapeau}}, \bibinfo {author} {\bibfnamefont
  {E.}~\bibnamefont {Burovski}}, \bibinfo {author} {\bibfnamefont
  {P.}~\bibnamefont {Peterson}}, \bibinfo {author} {\bibfnamefont
  {W.}~\bibnamefont {Weckesser}}, \bibinfo {author} {\bibfnamefont
  {J.}~\bibnamefont {Bright}}, \bibinfo {author} {\bibfnamefont {S.~J.}\
  \bibnamefont {van~der Walt}}, \bibinfo {author} {\bibfnamefont
  {M.}~\bibnamefont {Brett}}, \bibinfo {author} {\bibfnamefont
  {J.}~\bibnamefont {Wilson}}, \bibinfo {author} {\bibfnamefont {K.~J.}\
  \bibnamefont {Millman}}, \bibinfo {author} {\bibfnamefont {N.}~\bibnamefont
  {Mayorov}}, \bibinfo {author} {\bibfnamefont {A.~R.~J.}\ \bibnamefont
  {Nelson}}, \bibinfo {author} {\bibfnamefont {E.}~\bibnamefont {Jones}},
  \bibinfo {author} {\bibfnamefont {R.}~\bibnamefont {Kern}}, \bibinfo {author}
  {\bibfnamefont {E.}~\bibnamefont {Larson}}, \bibinfo {author} {\bibfnamefont
  {C.~J.}\ \bibnamefont {Carey}}, \bibinfo {author} {\bibfnamefont
  {I.}~\bibnamefont {Polat}}, \bibinfo {author} {\bibfnamefont
  {Y.}~\bibnamefont {Feng}}, \bibinfo {author} {\bibfnamefont {E.~W.}\
  \bibnamefont {Moore}}, \bibinfo {author} {\bibfnamefont {J.}~\bibnamefont
  {VanderPlas}}, \bibinfo {author} {\bibfnamefont {D.}~\bibnamefont {Laxalde}},
  \bibinfo {author} {\bibfnamefont {J.}~\bibnamefont {Perktold}}, \bibinfo
  {author} {\bibfnamefont {R.}~\bibnamefont {Cimrman}}, \bibinfo {author}
  {\bibfnamefont {I.}~\bibnamefont {Henriksen}}, \bibinfo {author}
  {\bibfnamefont {E.~A.}\ \bibnamefont {Quintero}}, \bibinfo {author}
  {\bibfnamefont {C.~R.}\ \bibnamefont {Harris}}, \bibinfo {author}
  {\bibfnamefont {A.~M.}\ \bibnamefont {Archibald}}, \bibinfo {author}
  {\bibfnamefont {A.~H.}\ \bibnamefont {Ribeiro}}, \bibinfo {author}
  {\bibfnamefont {F.}~\bibnamefont {Pedregosa}},\ and\ \bibinfo {author}
  {\bibfnamefont {P.}~\bibnamefont {van Mulbregt}},\ }\bibfield  {title}
  {\bibinfo {title} {{SciPy 1.0: fundamental algorithms for scientific
  computing in Python}},\ }\href {https://doi.org/10.1038/s41592-019-0686-2}
  {\bibfield  {journal} {\bibinfo  {journal} {Nature Methods}\ }\textbf
  {\bibinfo {volume} {17}},\ \bibinfo {pages} {261} (\bibinfo {year}
  {2020})}\BibitemShut {NoStop}%
\bibitem [{Note4()}]{Note4}%
  \BibitemOpen
  \bibinfo {note} {See the Supplemental Material at [URL will be inserted by
  publisher] for the video {\protect \tt notensionRe50.mp4} showing the time
  evolution of the shape of the microchannel $\alpha =0$ and
  $Re=0.5$.}\BibitemShut {Stop}%
\bibitem [{Note5()}]{Note5}%
  \BibitemOpen
  \bibinfo {note} {See the Supplemental Material at [URL will be inserted by
  publisher] for the video {\protect \tt notensionRe180.mp4} showing the time
  evolution of the shape of the microchannel $\alpha =0$ and
  $Re=1.8$.}\BibitemShut {Stop}%
\bibitem [{Note6()}]{Note6}%
  \BibitemOpen
  \bibinfo {note} {See the Supplemental Material at [URL will be inserted by
  publisher] for the video {\protect \tt tensionRe50.mp4} showing the time
  evolution of the shape of the microchannel $\alpha \protect \ne 0$ and
  $Re=0.5$.}\BibitemShut {Stop}%
\bibitem [{Note7()}]{Note7}%
  \BibitemOpen
  \bibinfo {note} {See the Supplemental Material at [URL will be inserted by
  publisher] for the video {\protect \tt tensionRe1000.mp4} showing the time
  evolution of the shape of the microchannel $\alpha \protect \ne 0$ and
  $Re=10$.}\BibitemShut {Stop}%
\bibitem [{\citenamefont {Schmid}\ and\ \citenamefont
  {Henningson}(2001)}]{SH01}%
  \BibitemOpen
  \bibfield  {author} {\bibinfo {author} {\bibfnamefont {P.~J.}\ \bibnamefont
  {Schmid}}\ and\ \bibinfo {author} {\bibfnamefont {D.~S.}\ \bibnamefont
  {Henningson}},\ }\href {https://doi.org/10.1007/978-1-4613-0185-1} {\emph
  {\bibinfo {title} {{Stability and Transition in Shear Flows}}}},\ \bibinfo
  {series} {Applied Mathematical Sciences}, Vol.\ \bibinfo {volume} {142}\
  (\bibinfo  {publisher} {Springer},\ \bibinfo {address} {New York, NY},\
  \bibinfo {year} {2001})\BibitemShut {NoStop}%
\bibitem [{\citenamefont {Boyd}(2000)}]{Boyd00}%
  \BibitemOpen
  \bibfield  {author} {\bibinfo {author} {\bibfnamefont {J.~P.}\ \bibnamefont
  {Boyd}},\ }\href@noop {} {\emph {\bibinfo {title} {{Chebyshev and Fourier
  Spectral Methods}}}},\ \bibinfo {edition} {2nd}\ ed.\ (\bibinfo  {publisher}
  {Dover Publications},\ \bibinfo {address} {Mineola, NY},\ \bibinfo {year}
  {2000})\BibitemShut {NoStop}%
\bibitem [{\citenamefont {Davis}\ and\ \citenamefont {Troian}(2003)}]{DT03}%
  \BibitemOpen
  \bibfield  {author} {\bibinfo {author} {\bibfnamefont {J.~M.}\ \bibnamefont
  {Davis}}\ and\ \bibinfo {author} {\bibfnamefont {S.~M.}\ \bibnamefont
  {Troian}},\ }\bibfield  {title} {\bibinfo {title} {{On a generalized approach
  to the linear stability of spatially nonuniform thin film flows}},\ }\href
  {https://doi.org/10.1063/1.1564094} {\bibfield  {journal} {\bibinfo
  {journal} {Phys. Fluids}\ }\textbf {\bibinfo {volume} {15}},\ \bibinfo
  {pages} {1344} (\bibinfo {year} {2003})}\BibitemShut {NoStop}%
\bibitem [{\citenamefont {Symon}\ \emph {et~al.}(2018)\citenamefont {Symon},
  \citenamefont {Rosenberg}, \citenamefont {Dawson},\ and\ \citenamefont
  {McKeon}}]{SRDM18}%
  \BibitemOpen
  \bibfield  {author} {\bibinfo {author} {\bibfnamefont {S.}~\bibnamefont
  {Symon}}, \bibinfo {author} {\bibfnamefont {K.}~\bibnamefont {Rosenberg}},
  \bibinfo {author} {\bibfnamefont {S.~T.~M.}\ \bibnamefont {Dawson}},\ and\
  \bibinfo {author} {\bibfnamefont {B.~J.}\ \bibnamefont {McKeon}},\ }\bibfield
   {title} {\bibinfo {title} {{On non-normality and classification of
  amplification mechanisms in stability and resolvent analysis}},\ }\href
  {https://doi.org/10.1103/PhysRevFluids.3.053902} {\bibfield  {journal}
  {\bibinfo  {journal} {Phys. Rev. Fluids}\ }\textbf {\bibinfo {volume} {3}},\
  \bibinfo {pages} {053902} (\bibinfo {year} {2018})}\BibitemShut {NoStop}%
\bibitem [{\citenamefont {Butler}\ \emph {et~al.}(2019)\citenamefont {Butler},
  \citenamefont {Box}, \citenamefont {Robert},\ and\ \citenamefont
  {Vella}}]{BBRV19}%
  \BibitemOpen
  \bibfield  {author} {\bibinfo {author} {\bibfnamefont {M.}~\bibnamefont
  {Butler}}, \bibinfo {author} {\bibfnamefont {F.}~\bibnamefont {Box}},
  \bibinfo {author} {\bibfnamefont {T.}~\bibnamefont {Robert}},\ and\ \bibinfo
  {author} {\bibfnamefont {D.}~\bibnamefont {Vella}},\ }\bibfield  {title}
  {\bibinfo {title} {{Elasto-capillary adhesion: Effect of deformability on
  adhesion strength and detachment}},\ }\href
  {https://doi.org/10.1103/PhysRevFluids.4.033601} {\bibfield  {journal}
  {\bibinfo  {journal} {Phys. Rev. Fluids}\ }\textbf {\bibinfo {volume} {4}},\
  \bibinfo {pages} {033601} (\bibinfo {year} {2019})}\BibitemShut {NoStop}%
\bibitem [{\citenamefont {Xu}\ \emph {et~al.}(2014)\citenamefont {Xu},
  \citenamefont {Billingham},\ and\ \citenamefont {Jensen}}]{XBJ14}%
  \BibitemOpen
  \bibfield  {author} {\bibinfo {author} {\bibfnamefont {F.}~\bibnamefont
  {Xu}}, \bibinfo {author} {\bibfnamefont {J.}~\bibnamefont {Billingham}},\
  and\ \bibinfo {author} {\bibfnamefont {O.~E.}\ \bibnamefont {Jensen}},\
  }\bibfield  {title} {\bibinfo {title} {{Resonance-driven oscillations in a
  flexible-channel flow with fixed upstream flux and a long downstream rigid
  segment}},\ }\href {https://doi.org/10.1017/jfm.2014.136} {\bibfield
  {journal} {\bibinfo  {journal} {J. Fluid Mech.}\ }\textbf {\bibinfo {volume}
  {746}},\ \bibinfo {pages} {368} (\bibinfo {year} {2014})}\BibitemShut
  {NoStop}%
\bibitem [{\citenamefont {Heil}\ and\ \citenamefont {Jensen}(2003)}]{HJ03}%
  \BibitemOpen
  \bibfield  {author} {\bibinfo {author} {\bibfnamefont {M.}~\bibnamefont
  {Heil}}\ and\ \bibinfo {author} {\bibfnamefont {O.~E.}\ \bibnamefont
  {Jensen}},\ }\bibfield  {title} {\bibinfo {title} {{Flows in deformable tubes
  and channels: Theoretical models and biological applications}},\ }in\ \href
  {https://doi.org/10.1007/978-94-017-0415-1{\_}2} {\emph {\bibinfo {booktitle}
  {Flow Past Highly Compliant Boundaries and in Collapsible Tubes}}},\ \bibinfo
  {series} {Fluid Mechanics and Its Applications}, Vol.~\bibinfo {volume}
  {72},\ \bibinfo {editor} {edited by\ \bibinfo {editor} {\bibfnamefont
  {P.~W.}\ \bibnamefont {Carpenter}}\ and\ \bibinfo {editor} {\bibfnamefont
  {T.~J.}\ \bibnamefont {Pedley}}}\ (\bibinfo  {publisher} {Springer},\
  \bibinfo {year} {2003})\ pp.\ \bibinfo {pages} {15--49}\BibitemShut {NoStop}%
\bibitem [{\citenamefont {Pedley}\ and\ \citenamefont
  {Pihler-Puzovi{\'{c}}}(2015)}]{PPP15}%
  \BibitemOpen
  \bibfield  {author} {\bibinfo {author} {\bibfnamefont {T.~J.}\ \bibnamefont
  {Pedley}}\ and\ \bibinfo {author} {\bibfnamefont {D.}~\bibnamefont
  {Pihler-Puzovi{\'{c}}}},\ }\bibfield  {title} {\bibinfo {title} {{Flow and
  oscillations in collapsible tubes: Physiological applications and
  low-dimensional models}},\ }\href {https://doi.org/10.1007/s12046-015-0363-9}
  {\bibfield  {journal} {\bibinfo  {journal} {S{\={a}}dh{\={a}}na: J. Indian
  Acad. Sci.}\ }\textbf {\bibinfo {volume} {40}},\ \bibinfo {pages} {891}
  (\bibinfo {year} {2015})}\BibitemShut {NoStop}%
\bibitem [{\citenamefont {Kaya}\ \emph {et~al.}(2009)\citenamefont {Kaya},
  \citenamefont {Aulisa}, \citenamefont {Ibragimov},\ and\ \citenamefont
  {Seshaiyer}}]{KAIS09}%
  \BibitemOpen
  \bibfield  {author} {\bibinfo {author} {\bibfnamefont {E.}~\bibnamefont
  {Kaya}}, \bibinfo {author} {\bibfnamefont {E.}~\bibnamefont {Aulisa}},
  \bibinfo {author} {\bibfnamefont {A.}~\bibnamefont {Ibragimov}},\ and\
  \bibinfo {author} {\bibfnamefont {P.}~\bibnamefont {Seshaiyer}},\ }\bibfield
  {title} {\bibinfo {title} {{A stability estimate for fluid structure
  interaction problem with non-linear beam}},\ }in\ \href
  {https://doi.org/10.3934/proc.2009.2009.424} {\emph {\bibinfo {booktitle}
  {Proceedings of the 7th AIMS International Conference}}},\ \bibinfo {series
  and number} {Dynamical Systems and Differential Equations (DCDS)},\ \bibinfo
  {editor} {edited by\ \bibinfo {editor} {\bibfnamefont {X.}~\bibnamefont
  {Hou}}, \bibinfo {editor} {\bibfnamefont {X.}~\bibnamefont {Lu}}, \bibinfo
  {editor} {\bibfnamefont {A.}~\bibnamefont {Miranville}}, \bibinfo {editor}
  {\bibfnamefont {J.}~\bibnamefont {Su}},\ and\ \bibinfo {editor}
  {\bibfnamefont {J.}~\bibnamefont {Zhu}}}\ (\bibinfo {address} {Arlington,
  Texas, USA},\ \bibinfo {year} {2009})\ pp.\ \bibinfo {pages}
  {424--432}\BibitemShut {NoStop}%
\bibitem [{\citenamefont {Aulisa}\ \emph {et~al.}(2014)\citenamefont {Aulisa},
  \citenamefont {Ibragimov},\ and\ \citenamefont {Kaya-Cekin}}]{EIKC14}%
  \BibitemOpen
  \bibfield  {author} {\bibinfo {author} {\bibfnamefont {E.}~\bibnamefont
  {Aulisa}}, \bibinfo {author} {\bibfnamefont {A.}~\bibnamefont {Ibragimov}},\
  and\ \bibinfo {author} {\bibfnamefont {E.~Y.}\ \bibnamefont {Kaya-Cekin}},\
  }\bibfield  {title} {\bibinfo {title} {{Fluid structure interaction problem
  with changing thickness beam and slightly compressible fluid}},\ }\href
  {https://doi.org/10.3934/dcdss.2014.7.1133} {\bibfield  {journal} {\bibinfo
  {journal} {Discr. Contin. Dynam. Syst. Ser. S}\ }\textbf {\bibinfo {volume}
  {7}},\ \bibinfo {pages} {1133} (\bibinfo {year} {2014})}\BibitemShut
  {NoStop}%
\bibitem [{\citenamefont {Chueshov}\ \emph {et~al.}(2016)\citenamefont
  {Chueshov}, \citenamefont {Dowell}, \citenamefont {Lasiecka},\ and\
  \citenamefont {Webster}}]{CDLW16}%
  \BibitemOpen
  \bibfield  {author} {\bibinfo {author} {\bibfnamefont {I.}~\bibnamefont
  {Chueshov}}, \bibinfo {author} {\bibfnamefont {E.~H.}\ \bibnamefont
  {Dowell}}, \bibinfo {author} {\bibfnamefont {I.}~\bibnamefont {Lasiecka}},\
  and\ \bibinfo {author} {\bibfnamefont {J.~T.}\ \bibnamefont {Webster}},\
  }\bibfield  {title} {\bibinfo {title} {{Nonlinear elastic plate in a flow of
  gas: Recent results and conjectures}},\ }\href
  {https://doi.org/10.1007/s00245-016-9349-1} {\bibfield  {journal} {\bibinfo
  {journal} {Appl. Math. Opt.}\ }\textbf {\bibinfo {volume} {73}},\ \bibinfo
  {pages} {475} (\bibinfo {year} {2016})}\BibitemShut {NoStop}%
\bibitem [{\citenamefont {Gay-Balmaz}\ \emph {et~al.}(2018)\citenamefont
  {Gay-Balmaz}, \citenamefont {Georgievskii},\ and\ \citenamefont
  {Putkaradze}}]{GBGP18}%
  \BibitemOpen
  \bibfield  {author} {\bibinfo {author} {\bibfnamefont {F.}~\bibnamefont
  {Gay-Balmaz}}, \bibinfo {author} {\bibfnamefont {D.}~\bibnamefont
  {Georgievskii}},\ and\ \bibinfo {author} {\bibfnamefont {V.}~\bibnamefont
  {Putkaradze}},\ }\bibfield  {title} {\bibinfo {title} {{Stability of helical
  tubes conveying fluid}},\ }\href
  {https://doi.org/10.1016/j.jfluidstructs.2017.12.020} {\bibfield  {journal}
  {\bibinfo  {journal} {J. Fluids Struct.}\ }\textbf {\bibinfo {volume} {78}},\
  \bibinfo {pages} {146} (\bibinfo {year} {2018})}\BibitemShut {NoStop}%
\bibitem [{\citenamefont {Boyko}\ \emph {et~al.}(2017)\citenamefont {Boyko},
  \citenamefont {Bercovici},\ and\ \citenamefont {Gat}}]{BBG17}%
  \BibitemOpen
  \bibfield  {author} {\bibinfo {author} {\bibfnamefont {E.}~\bibnamefont
  {Boyko}}, \bibinfo {author} {\bibfnamefont {M.}~\bibnamefont {Bercovici}},\
  and\ \bibinfo {author} {\bibfnamefont {A.~D.}\ \bibnamefont {Gat}},\
  }\bibfield  {title} {\bibinfo {title} {{Viscous-elastic dynamics of power-law
  fluids within an elastic cylinder}},\ }\href
  {https://doi.org/10.1103/PhysRevFluids.2.073301} {\bibfield  {journal}
  {\bibinfo  {journal} {Phys. Rev. Fluids}\ }\textbf {\bibinfo {volume} {2}},\
  \bibinfo {pages} {073301} (\bibinfo {year} {2017})}\BibitemShut {NoStop}%
\bibitem [{\citenamefont {Anand}\ \emph {et~al.}(2019)\citenamefont {Anand},
  \citenamefont {David~JR},\ and\ \citenamefont {Christov}}]{ADC18}%
  \BibitemOpen
  \bibfield  {author} {\bibinfo {author} {\bibfnamefont {V.}~\bibnamefont
  {Anand}}, \bibinfo {author} {\bibfnamefont {J.}~\bibnamefont {David~JR}},\
  and\ \bibinfo {author} {\bibfnamefont {I.~C.}\ \bibnamefont {Christov}},\
  }\bibfield  {title} {\bibinfo {title} {{Non-Newtonian fluid--structure
  interactions: Static response of a microchannel due to internal flow of a
  power-law fluid}},\ }\href {https://doi.org/10.1016/j.jnnfm.2018.12.008}
  {\bibfield  {journal} {\bibinfo  {journal} {J. Non-Newtonian Fluid Mech.}\
  }\textbf {\bibinfo {volume} {264}},\ \bibinfo {pages} {62} (\bibinfo {year}
  {2019})}\BibitemShut {NoStop}%
\bibitem [{\citenamefont {Nayfeh}\ and\ \citenamefont {Younis}(2004)}]{NY04}%
  \BibitemOpen
  \bibfield  {author} {\bibinfo {author} {\bibfnamefont {A.~H.}\ \bibnamefont
  {Nayfeh}}\ and\ \bibinfo {author} {\bibfnamefont {M.~I.}\ \bibnamefont
  {Younis}},\ }\bibfield  {title} {\bibinfo {title} {{A new approach to the
  modeling and simulation of flexible microstructures under the effect of
  squeeze-film damping}},\ }\href {https://doi.org/10.1088/0960-1317/14/2/002}
  {\bibfield  {journal} {\bibinfo  {journal} {J. Micromech. Microeng.}\
  }\textbf {\bibinfo {volume} {14}},\ \bibinfo {pages} {170} (\bibinfo {year}
  {2004})}\BibitemShut {NoStop}%
\bibitem [{\citenamefont {Anoop}\ and\ \citenamefont {Sen}(2015)}]{AS15}%
  \BibitemOpen
  \bibfield  {author} {\bibinfo {author} {\bibfnamefont {R.}~\bibnamefont
  {Anoop}}\ and\ \bibinfo {author} {\bibfnamefont {A.~K.}\ \bibnamefont
  {Sen}},\ }\bibfield  {title} {\bibinfo {title} {{Capillary flow enhancement
  in rectangular polymer microchannels with a deformable wall}},\ }\href
  {https://doi.org/10.1103/PhysRevE.92.013024} {\bibfield  {journal} {\bibinfo
  {journal} {Phys. Rev. E}\ }\textbf {\bibinfo {volume} {92}},\ \bibinfo
  {pages} {013024} (\bibinfo {year} {2015})}\BibitemShut {NoStop}%
\bibitem [{\citenamefont {Bico}\ \emph {et~al.}(2018)\citenamefont {Bico},
  \citenamefont {Reyssat},\ and\ \citenamefont {Roman}}]{BRB18}%
  \BibitemOpen
  \bibfield  {author} {\bibinfo {author} {\bibfnamefont {J.}~\bibnamefont
  {Bico}}, \bibinfo {author} {\bibfnamefont {E.}~\bibnamefont {Reyssat}},\ and\
  \bibinfo {author} {\bibfnamefont {B.}~\bibnamefont {Roman}},\ }\bibfield
  {title} {\bibinfo {title} {{Elastocapillarity: When surface tension deforms
  elastic solids}},\ }\href
  {https://doi.org/10.1146/annurev-fluid-122316-050130} {\bibfield  {journal}
  {\bibinfo  {journal} {Annu. Rev. Fluid Mech.}\ }\textbf {\bibinfo {volume}
  {50}},\ \bibinfo {pages} {629} (\bibinfo {year} {2018})}\BibitemShut
  {NoStop}%
\bibitem [{\citenamefont {Inamdar}(2018)}]{Inamdar2018}%
  \BibitemOpen
  \bibfield  {author} {\bibinfo {author} {\bibfnamefont {T.~C.}\ \bibnamefont
  {Inamdar}},\ }\emph {\bibinfo {title} {Unsteady Fluid-structure Interactions
  in Soft-walled Microchannels}},\ \href
  {https://docs.lib.purdue.edu/dissertations/AAI10793067/} {Master's thesis},\
  \bibinfo  {school} {Purdue University}, \bibinfo {address} {West Lafayette,
  Indiana} (\bibinfo {year} {2018})\BibitemShut {NoStop}%
\bibitem [{\citenamefont {Hou}\ \emph {et~al.}(2012)\citenamefont {Hou},
  \citenamefont {Wang},\ and\ \citenamefont {Layton}}]{HWL12}%
  \BibitemOpen
  \bibfield  {author} {\bibinfo {author} {\bibfnamefont {G.}~\bibnamefont
  {Hou}}, \bibinfo {author} {\bibfnamefont {J.}~\bibnamefont {Wang}},\ and\
  \bibinfo {author} {\bibfnamefont {A.}~\bibnamefont {Layton}},\ }\bibfield
  {title} {\bibinfo {title} {{Numerical methods for fluid-structure interaction
  --- A review}},\ }\href {https://doi.org/10.4208/cicp.291210.290411s}
  {\bibfield  {journal} {\bibinfo  {journal} {Commun. Comput. Phys.}\ }\textbf
  {\bibinfo {volume} {12}},\ \bibinfo {pages} {337} (\bibinfo {year}
  {2012})}\BibitemShut {NoStop}%
\bibitem [{\citenamefont {Ferziger}\ and\ \citenamefont
  {Peri{\'{c}}}(2002)}]{F02}%
  \BibitemOpen
  \bibfield  {author} {\bibinfo {author} {\bibfnamefont {J.~H.}\ \bibnamefont
  {Ferziger}}\ and\ \bibinfo {author} {\bibfnamefont {M.}~\bibnamefont
  {Peri{\'{c}}}},\ }\href {https://doi.org/10.1007/978-3-642-56026-2} {\emph
  {\bibinfo {title} {{Computational Methods for Fluid Dynamics}}}},\ \bibinfo
  {edition} {3rd}\ ed.\ (\bibinfo  {publisher} {Springer-Verlag},\ \bibinfo
  {address} {New York},\ \bibinfo {year} {2002})\BibitemShut {NoStop}%
\bibitem [{\citenamefont {Roache}(1998)}]{R98}%
  \BibitemOpen
  \bibfield  {author} {\bibinfo {author} {\bibfnamefont {P.~J.}\ \bibnamefont
  {Roache}},\ }\href@noop {} {\emph {\bibinfo {title} {{Verification and
  Validation in Computational Science and Engineering}}}}\ (\bibinfo
  {publisher} {Hermosa Publishers},\ \bibinfo {address} {Socorro, New Mexico},\
  \bibinfo {year} {1998})\BibitemShut {NoStop}%
\bibitem [{\citenamefont {Huang}\ and\ \citenamefont {Sloan}(1994)}]{HS94}%
  \BibitemOpen
  \bibfield  {author} {\bibinfo {author} {\bibfnamefont {W.}~\bibnamefont
  {Huang}}\ and\ \bibinfo {author} {\bibfnamefont {D.~M.}\ \bibnamefont
  {Sloan}},\ }\bibfield  {title} {\bibinfo {title} {{The pseudospectral method
  for solving differential eigenvalue problems}},\ }\href
  {https://doi.org/10.1006/JCPH.1994.1073} {\bibfield  {journal} {\bibinfo
  {journal} {J. Comput. Phys.}\ }\textbf {\bibinfo {volume} {111}},\ \bibinfo
  {pages} {399} (\bibinfo {year} {1994})}\BibitemShut {NoStop}%
\bibitem [{\citenamefont {Shen}\ \emph {et~al.}(2011)\citenamefont {Shen},
  \citenamefont {Tang},\ and\ \citenamefont {Wang}}]{STW11}%
  \BibitemOpen
  \bibfield  {author} {\bibinfo {author} {\bibfnamefont {J.}~\bibnamefont
  {Shen}}, \bibinfo {author} {\bibfnamefont {T.}~\bibnamefont {Tang}},\ and\
  \bibinfo {author} {\bibfnamefont {L.-L.}\ \bibnamefont {Wang}},\ }\href
  {https://doi.org/10.1007/978-3-540-71041-7} {\emph {\bibinfo {title}
  {Spectral Methods Algorithms, Analysis and Applications}}},\ \bibinfo
  {series} {Springer Series in Computational Mathematics}, Vol.~\bibinfo
  {volume} {41}\ (\bibinfo  {publisher} {Springer-Verlag},\ \bibinfo {address}
  {Berlin/Heidelberg},\ \bibinfo {year} {2011})\BibitemShut {NoStop}%
\end{thebibliography}%


\appendix

\section{Numerical scheme for the coupled FSI equations and implementation details}
\label{app:compute}

To summarize, Eq.~\eqref{solidmodel3} is the time-stepping equation of the FSI, while Eqs.~\eqref{betaeq2}, \eqref{cont7} and \eqref{xmom8} are used to calculate the pressure $P$ iteratively. The construction of accurate and stable method for such two-way coupled FSIs is generally a difficult problem (see, e.g., \cite{HWL12} and the references therein). The scheme described below is based on a segregated solver, which was implemented in Python, making use of the routines from SciPy \cite{SciPy}. In this section, for brevity, we denote the displacement $U_Y$ by $U$, without fear of confusion.

\begin{enumerate}
\item Create mesh (uniform grid of $N$ points on the domain $[0,1]$ with nodes $X_j = j\Delta X$, $\Delta X=j/(N-1)$, $j=0,\hdots,N-1$) and the appropriate sparse matrices corresponding to the spatial derivative operators.

\item Initialize by setting ($\forall j$): $Q^\star = Q^{n+1} = Q^n = Q^{n-1} = 1$; $H^\star = H^{n+1} = H^n = H^{n-1} = 1$; $U^\star = U^{n+1} = U^n = U^{n-1} = 0$; $P = 0$. A superscript of $n$, $(n-1)$ and $\star$ denotes values at the current time step, the previous time step, and at the intermediate (sub-time-step) iteration stage, respectively.

\item Start the time loop from $n=0$ and preform $N_T$ time steps with fixed time step $\Delta T$, to the final selected time.

\item Start inner iteration loop to determine $Q^\star$ and resolve the nonlinearity.

\item Solve for pressure from Eq.~\eqref{xmom8} by 
discretizing the time and space derivatives and rewriting the equation as a definite integral for the pressure:
\begin{equation}\label{peq_discrete}
P(X,T)-P(1,T) = \int_1^{X} \Bigg[ -\frac{Re St}{H^\star}\left(\frac{4Q^\star - 4Q^n + Q^{n-1}}{2\Delta T}\right) 
- \frac{6}{5}\frac{Re}{H^\star} \left\{\bm{D}_1\left[\frac{(Q^\star)^2}{H^\star}\right]\right\} -\frac{12Q^\star}{(H^\star)^3} \Bigg] \mathrm{d}X,
\end{equation}
which is subject to the outlet boundary condition: $P(1,T)=0$. Here, $\bm{D}_1$ is a discrete spatial operator corresponding to $\partial/\partial X$ constructed with second-order central-difference schemes (CDS) \cite{F02}.

A custom integration loop is used to obtain $P(X_j,T)$. Specifically, the values of the integral over every grid cell, i.e., $\int_{X_j}^{X_{j+1}} \cdots \, \mathrm{d}X$, which we term a ``sub-integral,'' is pre-calculated. Then, to evaluate $P(X_j,T)$ at a particular node $j$, the appropriate set of sub-integrals are added together. This results in significant computational cost savings as compared to using a built-in integration routine that recalculates the whole integral for each $j$.

\item Solve for $U^\star$ via Eq.~\eqref{solidmodel3}, which is discretized fully-implicitly with second-order CDS for the second-order time derivative, yielding a standard matrix equation formulation: 
\begin{equation}
\left\{ \bm{D}_4 -\alpha (\bm{D}_1[U^\star])^2 \bm{D}_2+ \frac{1}{(\Delta T)^2}\bm{I} \right\} U^\star = -\frac{1}{(\Delta T)^2}\left(-2U^n+U^{n-1}\right) + P.
\label{eq:Axb}
\end{equation}

$\bm{D}_4$, $\bm{D}_2$, $\bm{D}_1$ are the discrete fourth, second and first spatial differentiation matrices constructed using second-order CDS, appropriately modified for the present boundary conditions. $\bm{I}$ is the $N\times N$ identity matrix. Equation~\eqref{eq:Axb} is solved using {\tt spsolve} form the linear algebra libraries in SciPy \cite{SciPy}. 

$U^\star$ is a temporary variable used during the internal (sub-time-step) iterations to get a fully-implicit scheme for stable loading of the beam. If necessary, $U^\star$ can be under-relaxed:
\begin{equation}\label{relaxeq}
U^\star=\omega U^\star+(1-\omega)U^{n+1},
\end{equation}
where $\omega\in(0,1]$ is the \emph{relaxation factor}.

\item Update $H^\star$ from $U^
\star$ via Eq.~\eqref{betaeq2}.

\item Find $Q^\star$ by solving Eq.~\eqref{cont7}, which can again be re-cast as a definite integral from $0$ to $X$, and after discretizing the time derivative becomes:
\begin{equation}\label{qeq_discrete}
Q^\star(X,T) - Q(0,T) = - St\int_0^X \left(\frac{3H^\star - 4H^n + H^{n-1}}{2\Delta T}\right) \mathrm{d}X,
\end{equation}
were $Q(0,T)=1$ is the flow rate at the inlet given by the boundary condition. 

\item Update the inner iteration. To ensure that the both the solid and fluid solutions have independently converged, the residual is calculated as:
\begin{equation}\label{residual}
r=\max \left\{ \frac{\max_j |U^\star-U^{n+1}|}{\max_j |U^{n+1}|},\frac{\max_j |P-P^{n+1}|}{\max_j |P^{n+1}|} \right\},
\end{equation}
and a tolerance of $10^{-6}$ is used for the residual convergence criterion after testing tolerance values ranging from $10^{-4}$ to $10^{-8}$. As we never use values stored in the $(n+1)$ variables, they can be used as containers for old values of the $\star$ variables, which simplifies the definition of $r$ above. 

Iterating on $U^\star$ causes the values of $Q^\star$ and $H^\star$ to change, which results in changes in $P$. So, the beam bending equation is our main time-stepping equation and $P$ is our nonlinear loading that requires internal iterations to obtain a stable scheme. While iterating on $P$, all of $U^n$, $U^{n-1}$, $H^n$, $H^{n-1}$, $Q^n$ and $Q^{n-1}$ are known and constant. Set ($\forall j$): $U^{n+1}=U^\star$, $H^{n+1}=H^\star$, $Q^{n+1}=Q^\star$.

\item Update (external) time-stepping loop by setting ($\forall j$):
$U^{n-1}=U^n$, $U^n=U^{n+1}$; $H^{n-1}=H^n$, $H^n=H^{n+1}$; $Q^{n-1}=Q^n$, $Q^n=Q^{n+1}$.

\end{enumerate}

The discrete spatial operators were verified to be second-order accurate by applying them to smooth test functions. The second-order accuracy of the whole time-stepping algorithm was verified by setting up test cases for the solid mechanics problem (no flow) and the coupled-problem (prescribed wall motion generating flow) for which exact solutions were calculated via the \emph{method of manufactured solutions} \cite{R98}. {A mesh size and time step independence study was carried out before selecting $\Delta T$; for further details, see \cite{Inamdar2018}.}

For homogeneous boundary conditions, the same stencil can be used for the boundary points (as that used away from the boundary points on the grid) when defining $\bm{D}_1$, $\bm{D}_2$ and $\bm{D}_4$. Specifically, this re-use of the stencil can be accomplished by omitting weights that falls on the boundary points' neighbors outside the boundary. For example, in case of $\bm{D}_2$ formed by CDS, the first row will have $-2/(\Delta X)^2$, $1/(\Delta X)^2$ as its entries in first and second column, omitting the weight $1/(\Delta X)^2$ that would be applied to the left-hand ($j=-1$) neighbor.

\section{Chebyshev pseudospectral methods for the generalized eigenvalue problem}
\label{app:cheb_stability}

We use the Chebyshev pseudospectral method \cite{SH01,Boyd00} to compute the spectrum (eigenvalues and eigenfunctions) of the system defined by Eq.~\eqref{stability_matrix} and the boundary conditions in Eqs.~\eqref{stability_bc}. The numerical method was implemented in Python using routines from SciPy \cite{SciPy}. Since the Chebyshev pseudospectral method is derived for the domain $[-1,+1]$, we use the change of variables $\widetilde{X}\equiv2X-1$ to transform the computational domain from $\left\{X|X\in[0,1]\right\}$ to $\{\widetilde{X}|\widetilde{X}\in[-1,1]\}$. Then, $\mathrm{d}^m/\mathrm{d}X^m =2^m\mathrm{d}^m/\mathrm{d}\widetilde{X}^m $ and we denote $\widetilde{Q}(\widetilde{X})=Q_1(X)$ and $\widetilde{H}(\widetilde{X})=H_1(X)$, dropping the ``$1$'' subscript for simplicity. Note that the non-constant coefficients in Eq.~\eqref{stability_matrix}, which involve the steady-state shape of the microchannel $H_0(X)$ and its derivatives, have been precomputed using SciPy's {\tt{solve\_bvp}} and are known at this stage.

We introduce the Gauss--Lobatto points:
    \begin{equation}\label{Gauss-Lobatto}
        \widetilde{X}_j=-\cos\left(\frac{j\pi}{N}\right),\qquad j=0,1,\hdots,N.
    \end{equation}
Generally speaking, the aim of the Chebyshev pseudospectral method is to find a high-order polynomial, valid on the whole domain, to approximate the exact solution of the problem. In order to determine the coefficients of the polynomial, it is required that the differential equations be satisfied at the interior points, i.e., at $\widetilde{X}=\widetilde{X}_j$ with $j=1,\hdots,N-1$, while the boundary conditions are imposed at $\widetilde{X}=\widetilde{X}_0\equiv-1$ and $\widetilde{X}=\widetilde{X}_N\equiv+1$. For our problem, recall that there are three boundary conditions for $\widetilde{Q}$ and five boundary conditions for $\widetilde{H}$ [see Eqs.~\eqref{stability_bc}]. Therefore, we need to (and, indeed, can) uniquely determine a polynomial of order $N+1$ for $\widetilde{Q}$ and a polynomial of order $N+3$ for $\widetilde{H}$.

To this end, we follow Huang and Sloan \cite{HS94}, but we use different modified bases to construct polynomials for the eigenfunctions $\widetilde{Q}$ and $\widetilde{H}$. Actually, the modified bases in \cite{HS94} [see Eq.~(3.3) therein] do not apply to our problem because, here, the boundary conditions in  Eqs.~\eqref{stability_bc} involve higher derivatives at $\widetilde{X}_N=+1$ and do not meet the requirements for boundary conditions in \cite[Eq.~(3.1)]{HS94}. Instead, we construct the polynomials for eigenfunctions as follows:
\begin{subequations}\label{modified_bases}
\begin{align}\label{Q1_poly}
    \widetilde{Q}(\widetilde{X}) &\approx (1+\widetilde{X})\sum_{j=1}^N\frac{Q_j}{1+\widetilde{X}_j}l_j(\widetilde{X}),\\
\label{H1_poly}
    \widetilde{H}(\widetilde{X}) &\approx (1+\widetilde{X})(1-\widetilde{X})\sum_{j=1}^{N-1}\frac{H_j}{(1+\widetilde{X}_j)(1-\widetilde{X}_j)}l_j(\widetilde{X})+(1+\widetilde{X})(1-\widetilde{X})^2H_Nl_N(\widetilde{X}),
\end{align}
\end{subequations}
where $Q_j$ and $H_j$ are the function values at the collocation points. Here, $l_k(\widetilde{X})$ denotes the $k$-th basis Lagrange interpolating polynomial, defined as
\begin{equation}\label{lagrange}
    l_k(\widetilde{X})=\prod_{i=0,i\neq k}^N\frac{\widetilde{X}-\widetilde{X}_i}{\widetilde{X}_k-\widetilde{X}_i}, \qquad l_k(\widetilde{X}_j)=\delta_{kj},
\end{equation}
where $\delta_{kj}$ is the Kronecker delta symbol. With this property of the Lagrange interpolating polynomial, it follows that $\widetilde{Q}(\widetilde{X}_j)=Q_j$ $j=1,2,\hdots,N$, while $\widetilde{H}(\widetilde{X}_j)=H_j$ for $j=1,2,\hdots,N-1$. Note $H_N\neq H(\widetilde{X}_N)$. In fact, we already know from Eq.~\eqref{H1_bc} that $H(\widetilde{X}_N)=0$ because of the clamped boundary condition. However, $H_N$ is still needed, simply functioned as a coefficient, to satisfy the very last condition in Eq.~\eqref{H1_bc}. More importantly, it is easily verified that  Eqs.~\eqref{modified_bases} satisfy all the boundary conditions for $\widetilde{Q}$ and $\widetilde{H}$ except $\mathrm{d}\widetilde{Q}/\mathrm{d}\widetilde{X}=0$ and $\mathrm{d}^4\widetilde{H}/\mathrm{d}\widetilde{X}^4=0$. These two boundary conditions need to be imposed explicitly as extra two rows in the discretized matrix corresponding to the system in Eq.~\eqref{stability_matrix}.

Next, we substitute Eqs.~\eqref{modified_bases} into Eq.~\eqref{stability_matrix} and discretize it by requiring that Eq.~\eqref{stability_matrix} be satisfied at the interior collocation points,  i.e., $\widetilde{X}=\widetilde{X}_j$ with $j=1,2,\hdots,N-1$. At the same time, we impose the two unsatisfied boundary conditions at $\widetilde{X}_N=+1$ and add them into the discretized system: $\widehat{\bm{A}}\widehat{\bm{\psi}}=\sigma\widehat{\bm{B}}\widehat{\bm{\psi}}$, where $\widehat{\bm{\psi}}=[Q_1,Q_2,\hdots,Q_N,H_1,H_2,\hdots,H_N]^T$. Both $\widehat{\bm{A}}$ and $\widehat{\bm{B}}$ are $2N\times2N$ block matrices. Specifically, the matrix $\widehat{\bm{B}}$ is singular because the two homogeneous boundary conditions, imposed as its $N$-th and $2N$-th rows, respectively. These BCs do not involve the eigenvalue $\sigma$, which makes these two rows of $\widehat{\bm{B}}$ each equal to the zero vector.

One of the most important details that must be taken care of to obtain the discretized eigenvalue problem is dealing with the differentiation of Eqs.~\eqref{modified_bases} at the collocation points. Fortunately, derivatives of the Lagrange interpolating polynomials at the Gauss--Lobatto points have explicit representations \cite{STW11}. Let $\bm{D}$ denote the first-order differentiation matrix of the Lagrange interpolating polynomial basis at the Gauss--Lobatto points, then
\begin{equation}\label{lagrange_diff}
    D_{kj}=\left.\frac{\mathrm{d} l_j}{\mathrm{d} \widetilde{X}}\right|_{\widetilde{X}=\widetilde{X}_k}=\begin{cases}
    \displaystyle-\frac{2N^2+1}{6},&\qquad k=j=0,\\[3mm]
    \displaystyle\frac{\widetilde{c}_k}{\widetilde{c}_j}\frac{(-1)^{k+j}}{\widetilde{X}_k-\widetilde{X}_j},&\qquad k\ne j,\;\;0\le k,j\le N,\\[4mm]
    \displaystyle-\frac{\widetilde{X}_k}{2(1-\widetilde{X}_k^2)},&\qquad k= j,\;\;1\le k,j\le N-1,\\[5mm]
    \displaystyle\frac{2N^2+1}{6},&\qquad k=j=N,
    \end{cases}
\end{equation}
where $\widetilde{c}_0=\widetilde{c}_N=2$ and $\widetilde{c}_j=1$ for $1\le j\le N-1$. Furthermore, if we denote the $m$-th order differentiation matrix for the Lagrange interpolating polynomial basis as $\bm{D}^m$, the higher-order differentiation matrices can be obtained through matrix multiplication of the lower-order ones. For example, $\bm{D}^2=\bm{D}\times \bm{D}$ and $\bm{D}^3=\bm{D}\times \bm{D}^2=\bm{D}\times \bm{D} \times \bm{D}$. However, what we really need is the differentiation matrix with respect to the modified polynomials in Eqs.~\eqref{Q1_poly} and \eqref{H1_poly}. Similarly, we denote the first-order differentiation matrix with respect to $\widetilde{Q}$ and $\widetilde{H}$ as $\widetilde{\bm{D}}_Q$ and $\widetilde{\bm{D}}_H$, respectively, and the $m$-th higher order differentiation matrices as $\widetilde{\bm{D}}_Q^m$ and $\widetilde{\bm{D}}_H^m$. Clearly, $\widetilde{\bm{D}}_Q$, $\widetilde{\bm{D}}_Q$, $\widetilde{\bm{D}}_Q^m$, and $\widetilde{\bm{D}}_H^m$ should be modified from Eq.~\eqref{lagrange_diff}. The complete expressions are quite lengthy, thus we do not include them here. Instead, we just write down $\widetilde{\bm{D}}_Q$ as an example:  
\begin{equation}\label{DQ}
    \widetilde{\bm{D}}_Q = \left[\bm{I}+\diag\big((1+\widetilde{X}_k),1\le k\le N\big)\times \bm{D}\right] \times \diag\left(1/(1+\widetilde{X}_k),1\le k\le N\right),
\end{equation}
where $\bm{I}$ is the identity matrix and $\diag(\cdot,1\le k\le N)$ denotes an $N\times N$ diagonal matrix with entries given by the first input. Furthermore, it is important to note that the modification of the Lagrange interpolating polynomial no longer allows us to build the higher-order differentiation matrix via  multiplication of the lower-order differentiation matrices.

Before solving the discretized eigenvalue problem $\widehat{\bm{A}}\widehat{\bm{\psi}}=\sigma\widehat{\bm{B}}\widehat{\bm{\psi}}$, a preconditioner is needed to reduce the condition number of  $\widehat{\bm{A}}$. By analogy to \cite{HS94}, we find that the following preconditioner successfully reduces the condition number of $\widehat{\bm{A}}$ by four orders of magnitude:
\begin{equation}
    \widehat{\bm{S}}=\diag\left(\frac{1}{1+\widetilde{X}_k},1\le k\le N; (1+\widetilde{X}_k)^2(1-\widetilde{X}_k)^2,1\le k\le N-1; 0.1 \right).
\end{equation}
Here, the semicolons denote concatenation of elements along the diagonals of the $2N\times 2N$ matrix $\widehat{\bm{S}}$. Therefore, the eigenvalue problem that we need to solve numerically is actually $\widehat{\bm{S}}\widehat{\bm{A}}\widehat{\bm{\psi}}=\sigma\widehat{\bm{S}}\widehat{\bm{B}}\widehat{\bm{\psi}}$.

Finally, for the computational results reported in the main text above, we invert the matrix $\widehat{\bm{S}}\widehat{\bm{A}}$ and solve the regular eigenvalue problem, $\widehat{\bm{A}}^{-1}\widehat{\bm{S}}^{-1}\widehat{\bm{S}}\widehat{\bm{B}}\widehat{\bm{\psi}}=\sigma^{-1}\widehat{\bm{\psi}}$, numerically. Note that $\widehat{\bm{S}}\widehat{\bm{B}}$ is not invertible because $\widehat{\bm{B}}$ is singular. SciPy's routine {\tt{eig}} is used to obtain the spectrum (eigenvalues and eigenfunctions) of this system.

\end{document}